\documentclass[final,1p,times,twocolumn]{elsarticle}
\usepackage{graphicx,amssymb}

\journal{ Advances in Colloid and Interface Science}

\newcommand{\Eb}{\mbox{\bf E}}
\newcommand{\Jb}{\mbox{\bf J}}
\newcommand{\ub}{\mbox{\bf u}}
\newcommand{\rb}{\mbox{\bf r}}
\newcommand{\Ub}{\mbox{\bf U}}
\newcommand{\Fb}{\mbox{\bf F}}
\newcommand{\fb}{\mbox{\bf f}}
\newcommand{\Tb}{\mbox{\bf T}}

\newcommand{\Ib}{\mbox{\bf I}}
\newcommand{\Mb}{\mbox{\bf M}}
\newcommand{\del}{\mbox{\boldmath{$\nabla$}}}
\newcommand{\nhat}{\hat{n}}

\begin{document}

\begin{frontmatter}

\title{
Towards an understanding of induced-charge electrokinetics \\ at large applied voltages in concentrated solutions }

\author[mitche,mit,espci]{Martin Z. Bazant\corref{cor1}}
\ead{bazant@mit.edu}
\author[mit]{Mustafa Sabri Kilic}
\author[olin]{Brian D. Storey}
\author[espci]{Armand Ajdari}
\address[mitche]{Department of Chemical Engineering, Massachusetts Institute of Technology, Cambridge, MA 02139}
\address[mit]{Department of Mathematics, Massachusetts Institute of Technology, Cambridge, MA 02139}
\address[olin]{Franklin W. Olin College of Engineering, Needham, MA 02492}
\address[espci]{CNRS UMR Gulliver 7083, ESPCI, 10 rue Vauquelin, 75005 Paris, France}
\cortext[cor1]{Corresponding author}


\date{\today}

\begin{abstract}

  The venerable theory of electrokinetic phenomena rests on the hypothesis of a dilute solution of point-like ions in quasi-equilibrium with a weakly charged surface, whose potential relative to the bulk is of order the thermal voltage ($kT/e \approx 25$  mV at room temperature). In nonlinear electrokinetic phenomena, such as AC or induced-charge 
  electro-osmosis (ACEO, ICEO) and induced-charge electrophoresis (ICEP), several Volts
  $\approx 100\, kT/e$ are applied to polarizable surfaces in microscopic geometries, and the resulting electric fields and  induced surface charges are large enough to violate the assumptions of the classical theory. In this  article, we review the experimental and theoretical literatures, highlight discrepancies between theory and experiment, introduce possible modifications of the theory, and analyze their consequences. 
  We argue that, in response to a  large applied voltage, the 
  ``compact layer'' and ``shear plane'' effectively advance into the liquid, due to the crowding of counter-ions. Using simple continuum models, we predict two general trends at
  large voltages: (i)  ionic crowding against a blocking surface expands the diffuse double layer and
  thus decreases its differential capacitance, and (ii) a charge-induced 
  viscosity increase near the surface reduces the electro-osmotic mobility; each trend is enhanced by dielectric saturation. The first effect is able to predict high-frequency flow reversal in ACEO pumps, while the second may explain the decay of ICEO flow with increasing salt concentration. Through
  several colloidal examples, such as ICEP of an
  uncharged metal sphere in an asymmetric electrolyte, we show that nonlinear  
  electrokinetic phenomena are generally ion-specific. Similar theoretical issues arise in nanofluidics (due to 
  confinement) and ionic liquids (due to the lack of solvent), so the paper 
  concludes with a general framework of modified electrokinetic equations for finite-sized ions.

\end{abstract}

\begin{keyword}
Nonlinear electrokinetics, microfluidics, induced-charge electro-osmosis, electrophoresis, AC electro-osmosis, 
concentrated solution, modified Poisson-Boltzmann theory, steric effects, hard-sphere liquid, lattice-gas, viscoelectric effect, solvation, ionic liquids, non-equilibrium thermodynamics
\end{keyword}
\end{frontmatter}


\section{ Introduction}

\subsection{ Nonlinear ``induced-charge" electrokinetic phenomena }

\begin{figure}
\begin{center}
\includegraphics[width=4in]{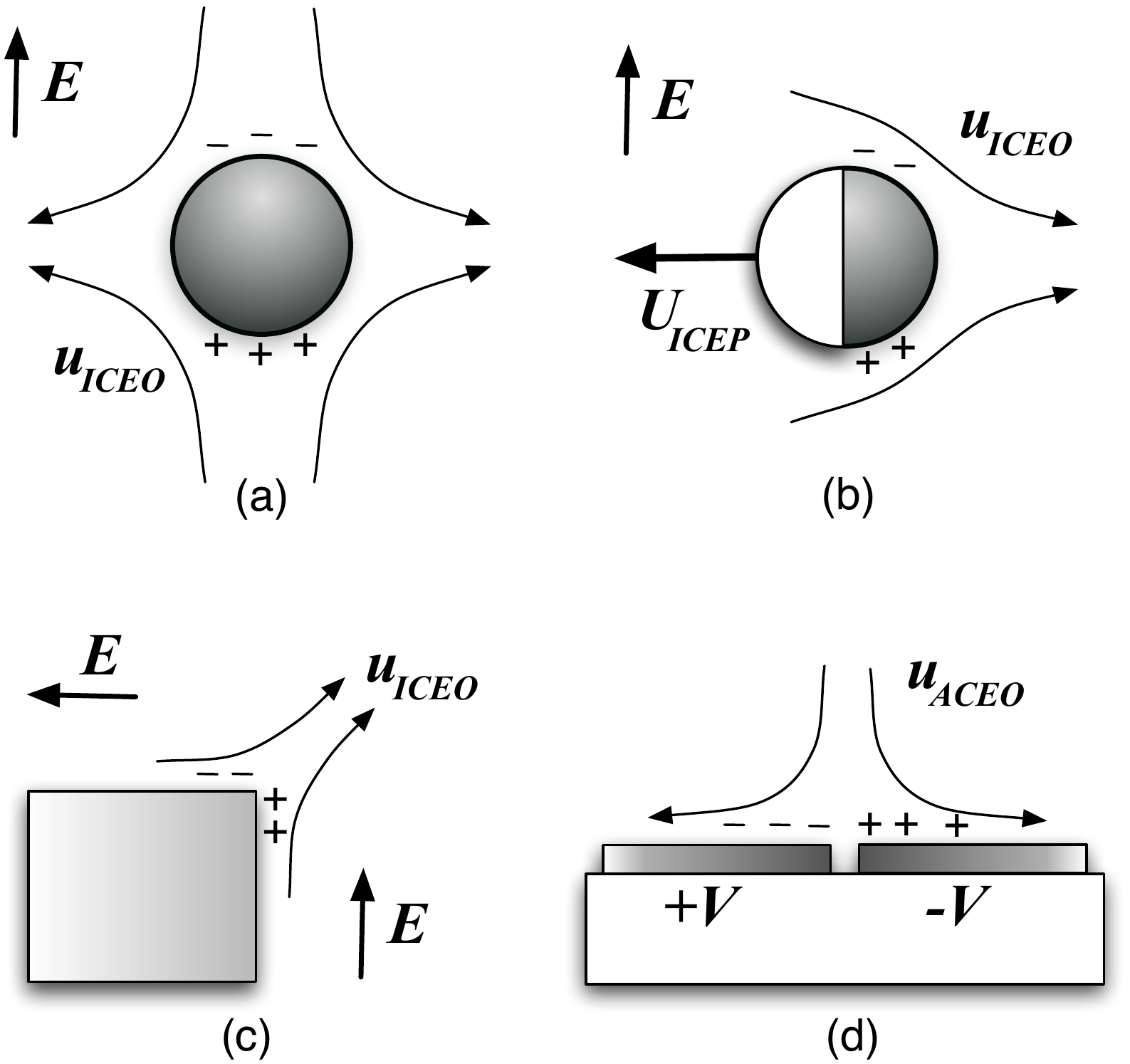}
\caption{ \label{fig:examples}
Examples of nonlinear electrokinetic phenomena, driven by induced charge ($+$, $-$)in the diffuse part of the electrochemical double layer at polarizable, blocking surfaces, subject to an applied electric field $E$ or voltage $V$. (a) Induced-charge electro-osmosis (ICEO) around a metal post~\cite{iceo2004a,iceo2004b,levitan2005,murtsovkin1996},
(b) induced-charge electrophoresis (ICEP) of a metal/insulator Janus particle~\cite{squires2006,gangwal2008},
(c) a nonlinear electrokinetic jet of ICEO flow at a sharp corner in a dielectric microchannel~\cite{thamida2002,yossifon2006},
and (d) AC electro-osmosis (ACEO) over a pair of microelectrodes~\cite{ramos1999,ajdari2000}.
}\end{center}
\end{figure}

Due to favorable scaling with miniaturization, electrokinetic phenomena are finding many new applications in
microfluidics~\cite{stone2004,squires2005,laser2004}, but often in new 
situations that raise fundamental theoretical questions. The classical
theory of electrokinetics, dating back to Helmholtz and
Smoluchowski a century ago~\cite{lyklema2003}, was developed for the effective {\it linear} hydrodynamic slip driven by an electric field past a surface in chemical {\it equilibrium} with the solution, whose double-layer
voltage is of order the thermal voltage, ${kT}/{e} = 25$ mV, and
approximately constant
~\cite{hunter_book,lyklema_book_vol2,russel_book,anderson1989,dukhin1995,delgado2007}. The
discovery of AC electro-osmotic flow (ACEO) over micro-electrodes~\cite{ramos1999,ajdari2000,encyclopedia_ACEO}
has shifted attention to a new {\it  nonlinear} regime~\cite{encyclopedia_nonlinear}, where the {\it induced}
double-layer voltage is typically several Volts $\approx 100\, kT/e$,
oscillating at frequencies up to 100 kHz, and nonuniform at the micron
scale. Related  phenomena of induced-charge electro-osmosis (ICEO)
~\cite{iceo2004a,iceo2004b,squires2009review} also occur around polarizable
particles~\cite{murtsovkin1996,encyclopedia_polarizable} and
microstructures~\cite{levitan2005,harnett2008} (in AC or DC fields), as well as driven biological membranes~\cite{lacoste2009}. Due to broken symmetries in ICEO flow, 
asymmetric colloidal particles undergo nonlinear, induced-charge electrophoresis
(ICEP)~\cite{iceo2004a,squires2006,yariv2005,gangwal2008}.  Some of these fundamental nonlinear electrokinetic phenomena are illustrated in  Fig. ~\ref{fig:examples}. 

A ``Standard Model" (outlined below) has emerged to describe a wide variety of induced-charge electrokinetic phenomena,  but some crucial aspects remain unexplained. In their pioneering work  25 years ago in the USSR, which went unnoticed in the West until recently~\cite{iceo2004a,iceo2004b}, V. A. Murtsovkin, 
A. S. Dukhin and collaborators  first 
predicted quadrupolar flow (which we call ``ICEO'') around a
polarizable sphere in a uniform electric field ~\cite{gamayunov1986} and observed
the phenomenon using mercury drops~\cite{murtsovkin1991} and metal
particles~\cite{gamayunov1992}, although the flow was sometimes in the
opposite direction to the theory, as discussed below. (See Ref.~\cite{murtsovkin1996} for a review.) 
More recently, in microfluidics, Ramos et al. observed and modeled
ACEO flows over a pair of micro-electrodes, and the theory
over-predicted the observed velocity by an order of
magnitude~\cite{ramos1999,green2000b,gonzalez2000,green2002}. Around
the same time, Ajdari used a similar model to predict ACEO pumping by
asymmetric electrode arrays~\cite{ajdari2000}, which was 
demonstrated using planar electrodes of unequal widths and
gaps~\cite{brown2000,ramos2003,studer2002,mpholo2003,studer2004,debesset2004},
but the model cannot predict experimentally observed
flow reversal at high frequency and loss of flow at high salt
concentration~\cite{studer2004,urbanski2006,microTAS2007}, even if
extended to large voltages~\cite{olesen2006,olesen_thesis}. The same 
model has also been used to predict faster three-dimensional ACEO pump
geometries~\cite{bazant2006},  prior to experimental
verification~\cite{urbanski2006,urbanski2007,weiss2008,hilber2008,chien-chih}, but again the data
departs from the theory at large voltages. Discrepancies
between theory and experiments, including flow reversal, also arise in traveling-wave
electro-osmosis (TWEO) for electrode arrays applying a wave-like
four-phase voltage
pulse~\cite{cahill2004,ramos2005,garcia2006,ramos2007}. Recent
observations of ICEO flow around metal
microstructures~\cite{levitan2005,harnett2008}, ICEP rotation of metal
rods~\cite{rose2007}, ICEP translation of metallo-dielectric
particles~\cite{gangwal2008} have likewise confirmed qualitative
theoretical predictions
~\cite{iceo2004a,iceo2004b,squires2006,saintillan2006,wall_preprint}, 
while exhibiting the same poorly understood decay of the velocity at
high salt concentration. We conclude that there are 
fundamental gaps in our understanding of induced-charge electrokinetic phenomena.

In this article, we review recent experimental and theoretical work in this growing area of nonlinear electrokinetics, as well as some possibly relevant literatures from other fields, and we consider a number of possible modifications of classical electrokinetic theory. Some of these ideas are new, while others were proposed long ago by O. Stern, J. J. Bikerman, J. Lyklema, and others, and effectively forgotten.  We build the case that at least some failures of the Standard Model can be attributed to the breakdown of the dilute-solution
approximation at large induced voltages. Using simple models, we
predict two general effects associated with counterion crowding --
decay of the double-layer capacitance and reduction of the
electro-osmotic mobility -- which begin to explain the experimental
data.  Our models, although incomplete, also imply generic new
ion-specific nonlinear electrokinetic phenomena at large voltages
related to atomic-level details of polarizable solid/electrolyte
interfaces.

\subsection{ Scope and context of the article }

We first presented these ideas in a paper at the ELKIN International
Electrokinetics Symposium in Nancy, France in June 2006
~\cite{ELKIN06} and in a letter, which was archived
online in March 2007~\cite{large} and recently published ~\cite{large_new}.  
The present article is a review article with original material, 
built around the letter, where its basic arguments are further developed as follows:
\begin{enumerate}
\item We begin with a critical review of recent studies of induce-charge electrokinetic
  phenomena in section ~\ref{sec:expt}. By compiling results from the literature and performing our own simulations of other experiments, we systematically compare theory and experiment across a wide range of nonlinear electrokinetic phenomena. To motivate modified 
electrokinetic models, we also  
  review various concentrated-solution theories from electrochemistry and electrokinetics in sections 3 and 4.
\item In our original letter, the
  theoretical predictions of steric effects of finite ion
  sizes in electrokinetics were based on what we call the ``Bikerman's model" below~\cite{bikerman1942,kilic2007a}, a
  simple lattice-gas approach that allows analytical results. Here, we
  develop a general, mean-field local-density theory of volume constraints and
  illustrate it with hard-sphere liquid models~\cite{lue1999,biesheuvel2007}.
  In addition to the charge-induced thickening effect from the original letter, 
  we also consider the field-induced viscoelectric effect in the solvent proposed by Lyklema and
   Overbeek~\cite{lyklema1961,lyklema1994}, in conjunction with our models for steric effects. We also consider dielectric relaxation of the solution, which tends to enhance these effects.
\item We provide additional examples of new electrokinetic
  phenomena predicted by our models at large voltages. In the
  letter~\cite{large},  we predicted high-frequency flow reversal in
  ACEO (Fig.~\ref{fig:rev} below) and decay of ICEO flow at high
  concentration (Fig.~\ref{fig:B_MHS}). Here, we also predict two
  mechanisms for ion-specific, field-dependent mobility of polarizable
  colloids at large voltages. The first has to do with crowding
  effects on the redistribution of double-layer charge due to
  nonlinear capacitance, as noted by
  A. S. Dukhin~\cite{ASdukhin1993,ASdukhin2005}
  (Fig. ~\ref{fig:mobility}). The second results from a novel ion-specific
  viscosity increase at high charge density (Fig.~\ref{fig:sphere}).
\item We present a general theoretical framework of modified
  electrokinetic equations and boundary conditions for concentrated solutions and/or large voltages in  section~\ref{sec:disc}, 
  which could find many other applications in nonlinear electrochemical relaxation or electrokinetics. .
\end{enumerate}
In spite of these major changes, the goal of the paper remains the same: to
provide an overview of various physical aspects of electrokinetic
phenomena, not captured by classical theories, which become important
at large induced voltages. Here, we focus on general concepts,
mathematical models, and simple analytical predictions. Detailed
studies of some particular phenomena will appear elsewhere, e.g. 
Ref.~\cite{storey2008} on high-frequency flow reversal in AC
electro-osmosis.

There have been a few other attempts to go beyond dilute solution theory in electrokinetics, but in the rather different context of
linear ``fixed-charge'' flows in nanochannels at low surface
potentials. The first electrokinetic theories of this type may be those of Cervera et al.~\cite{cervera2001,cervera2003}, who used Bikerman's modified Poisson-Boltzmann (MPB) theory to account for the crowding of finite-sized ions during transport by conduction and electro-osmosis through a membrane nanopore. Independently, J. J. Horno et al.~\cite{lopez2007mpb,lopez2008mpb,aranda2009a,aranda2009b} also used Bikerman's model (albeit, attributed to others~\cite{wiegel1993,baker1999,borukhov1997} -- see below) to analyze linear electrophoresis of colloids in a concentrated electrolyte.  Recently, Liu et al. ~\cite{liu2008}  numerically implemented a more complicated 
MPB theory~\cite{outhwaite1980,outhwaite1983,bhuiyan2004} 
to predict effects of finite ion sizes,
electrostatic correlations, and dielectric image forces on
electro-osmotic flow and streaming potential in a
nanochannel. In these studies
of linear electrokinetic phenomena, effects beyond the dilute solution
approximation can arise due to nano-confinement, but, as we first noted in
Ref.~\cite{kilic2007a}, much stronger and possibly different crowding
effects also arise due to large induced voltages, regardless of
confinement or bulk salt concentration. Our goal here is to make a
crude first attempt to understand the implications of ion crowding for
nonlinear electrokinetic phenomena, using simple mean-field
approximations that permit analytical solutions in the thin
double-layer limit.

Similar models for double-layer charging dynamics are also starting to be developed for ionic liquids and molten salts~\cite{kornyshev2007,federov2008,federov2008b,oldham2008}, since describing  ion crowding is paramount in the absence of a solvent. Kornyshev  recently suggested using what we call the ``Bikerman-Freise" (BF) mean-field theory below to describe the differential capacitance of the double layer, with the bulk volume fraction of ions appearing as a fitting parameter to allow for a slightly different density of ions~\cite{kornyshev2007}. (An equivalent lattice-gas model was also developed long ago for the double layer in an ionic crystal by Grimley and Mott~\cite{grimley1947,grimley1950,macdonald1982}, and a complete history of related models is given below in Section~\ref{sec:history}.) The BF capacitance formula, extended to allow for a thin dielectric Stern layer, has managed to fit recent experiments and simulations of simple ionic liquids rather well, especially at large voltages~\cite{federov2008,federov2008b,lauw2009}. However, we are not aware of any work addressing electrokinetic phenomena in ionic liquids, so perhaps the mean-field electro-hydrodynamic models developed here for concentrated electrolytes at large voltages might provide a useful starting point, in the limit of nearly close packing of ions. 

As a by-product of this work, our attempts to model nanoscale phenomena in nonlinear electrokinetics may also have broader applicability in nanofluidics~\cite{eijkel2005,schoch2008}. Dilute-solution theory remains the state-of-the-art in mathematical modeling, and the main focus of the field so far has been on effects of  geometrical confinement, especially with overlapping double layers. The classical Poisson-Nernst-Planck and Navier-Stokes equations provide a useful first approximation to understand nanochannel transport phenomena, such as charge selectivity~\cite{zheng2003,pennathur2005a,pennathur2005b,baldessari2006}, mechanical-to-electrical power conversion efficiency~\cite{morrison1965,yao2003,daiguji2004,heyden2006,heyden2007,pennathur2007}, current-voltage characteristics~\cite{yossifon2009}, and nonlinear ion-profile dynamics ~\cite{mani2009,zangle2009}, but in some cases it may be essential to introduce new physics into the governing equations and boundary conditions to account for crowding effects and strong surface interactions. Molecular dynamics simulations of nanochannel electrokinetics provide crucial insights and can be used to test and guide the development of modified continuum models~\cite{lyklema1998,freund2002,qiao2003,thompson2003,joly2004,lorenz2008} . 

We stress that there are other important, developing areas of nonlinear electrokinetics, which are related, but outside the scope of this article. For example, we do not discuss the nonlinear electrophoresis of fixed-charge particles~\cite{mishchuk2002,shilov2003,ASdukhin2005}, which can result from surface conduction~\cite{dukhin1974,dukhin1993,chu2007,khair2008} ($Du > 0$) and advection-diffusion~\cite{mishchuk2002} ($Pe>0$), although we will analyze a different mechanism for field-dependent electrophoretic mobility for polarizable particles~\cite{ASdukhin1993}.   We also neglect nonlinear electrokinetic phenomena associated with strong salt concentration gradients, such as second-kind ``superfast'' electrophoresis~\cite{dukhin1991,mishchuk_review,mishchuk1995,barany1998,ben2002,barinova2008}, electro-osmotic fluid instability~\cite{rubinstein2000,rubinstein2001,zaltzman2007,kim2007,yossifon2008,yossifon2009}, and concentration-polarization shocks~\cite{mani2009,zangle2009}, which can now be directly observed in microfluidic systems~\cite{kim2007,yossifon2008,yossifon2009,zangle2009} and porous bead packings~\cite{leinweber2004,tallarek2005}. These effects result from ``super-limiting" normal current into a polarized surface, membrane, or nanochannel, which depletes the local salt concentration significantly and forces the diffuse charge out of equilibrium. In such cases, dilute solution theory is likely to remain valid up to large applied voltages in the regions of low ionic strength away from the surface, where the flow is mostly generated.  In contrast, our focus here is on metal structures and electrodes that do not sustain normal current, e.g. due to AC forcing and insufficient voltage to trigger Faradaic reactions. As a result, the applied voltage leads mostly to capacitive charging of the diffuse layer, and thus potentially to the crowding of counter-ions attracted to the surface. 
 
The article is organized as follows. In section~\ref{sec:expt}, we
review the standard low-voltage model for induced-charge electrokinetic
phenomena and its failure to explain certain key experimental
trends. We then review various attempts to go beyond dilute solution theory in electrochemistry and electrokinetics and analyze the effects of two types of new physics in nonlinear electrokinetic phenomena at large voltages: In section~\ref{sec:crowding}, we build on our recent work on
diffuse-charge dynamics at large applied
voltages~\cite{bazant2004,kilic2007a,kilic2007b} to argue that the
crowding of counterions plays a major role in induced-charge
electrokinetic phenomena by reducing the double-layer capacitance in
ways that are ion-specific and concentration-dependent; In
section~\ref{sec:viscosity}, we postulate that the local viscosity of
the solution grows with increasing charge density, which in turn
decreases the electro-osmotic mobility at high voltage and/or
concentration and introduces another source of ion specificity. Finally, in
section~\ref{sec:disc}, we present a theoretical framework of modified electrokinetic
equations, which underlies the results in
sections~\ref{sec:crowding}-\ref{sec:viscosity} and can be applied to general situations.

\section{ Background:  Theory versus Experiment } 
\label{sec:expt}

\subsection{ The Standard  Model }

We begin by summarizing  the "Standard Model" of induced-charge / AC electrokinetics, 
used in most theoretical studies, and bringing out some crucial
experimental trends it fails to capture.  
The general starting point for the Standard Model is the coupling of the 
Poisson-Nernst-Planck (PNP) equations of ion transport  to the 
Navier Stokes equations of viscous fluid flow. 
ICEO flows are rather
complex, so many simplifications from this  starting point 
have been made to arrive at an operational
model~\cite{murtsovkin1996,ramos1999,ajdari2000,iceo2004b,squires2006}.
For thin double layers (DL) compared to the geometrical length scales, 
the Standard Model is based on the assumption of "linear" or "weakly nonlinear"
charging dynamics~\cite{bazant2004}, which further requires that the applied voltage is small enough
not to  significantly perturb the bulk salt concentration, whether by
double-layer salt adsorption or Faradaic reaction currents. In this regime, the
problem is greatly simplified, and the 
electrokinetic problem decouples into one of electrochemical 
relaxation and another of viscous flow:
\begin{enumerate}
\item {\it Electrochemical relaxation. -- }
The first step is to solve Laplace's equation
for the electrostatic potential across the bulk resistance,
\begin{equation}
\del\cdot\Jb = \del\cdot (\sigma \Eb) = - \sigma \nabla^2 \phi = 0
\end{equation}
assuming Ohm's Law with a constant conductivity $\sigma$.  
A capacitance-like boundary condition for charging of the 
double layer at a blocking surface (which cannot pass normal current) then closes
the ``RC circuit"~\cite{bazant2004},
\begin{equation}
C_D \frac{d\Psi_D}{dt} = \sigma E_n,    \label{eq:RC}
\end{equation}
where the local diffuse-layer voltage drop $\Psi_D(\phi)$ (surface
minus bulk) responds to the normal electric field $E_n=-\hat{n}\cdot
\nabla \phi$. In the Standard Model, the bulk conductivity $\sigma$ and diffuse-layer
capacitance $C_D$ are usually taken to be constants, although these assumptions
can be relaxed~\cite{olesen2006,olesen_thesis,chu2006}. 
The diffuse layer capacitance is calculated from the PNP equations
by applying the justifiable assumption that 
the thin double layers are in thermal equilibrium; see section 3.1.
A compact
Stern layer or dielectric surface coating of constant capacitance
$C_S$ is often included~\cite{ajdari2000,bazant2004,olesen2006}, so
that only part of the total double-layer voltage $\Delta \phi$ is
dropped across the diffuse layer ``capacitor", 
\begin{equation}
\Psi_D = \frac{\Delta \phi}{1+ \delta} = \frac{C_S  \Delta \phi}{C_S + C_D} , \label{eq:PsiD}
\end{equation}
where $\delta = C_D/C_S$ is the diffuse-layer to compact-layer capacitance ratio.  
\item {\it Viscous flow. -- }
The second step is to solve for a (possibly unsteady) Stokes flow,
\begin{equation}
\rho_m \frac{\partial \ub}{\partial t} = - \del p + \eta_b \nabla^2 \ub, \ \ \del \cdot \ub = 0, 
\end{equation}
with the Helmholtz-Smoluchowski (HS) boundary condition for effective fluid slip outside the double layer,
\begin{equation}
  \ub_s = -b \, \Eb_t = -\frac{\varepsilon_b \Psi_D}{\eta_b}\, \Eb_t 
 \label{eq:HS}
\end{equation}
where $\Eb_t$ is the tangential field, $b = \varepsilon_b \zeta /
\eta_b$ is the electro-osmotic mobility, $\zeta$ is the zeta potential
at the shear plane ($= \Psi_D$ in the simplest models), and
$\varepsilon_b$, $\eta_b$, and $\rho_m$ are the permittivity, viscosity, and mass density of the
{\it bulk} solvent. Osmotic pressure gradients, which would modify the slip formula~\cite{rubinstein2001,zaltzman2007}, are neglected since the bulk salt concentration is assumed to be uniform.
\end{enumerate}
Although this model can be rigorously justified only for very small voltages, $\Psi_D
\ll kT/e$,  in a dilute
solution~\cite{gonzalez2000,iceo2004b,bazant2004}, 
it manages to describe many features of ICEO flows at much
larger voltages. 

There has been extensive theoretical work using the Standard Model, and it provides the basis for most of our understanding of induced-charge electrokinetics. 
In recent years, it has been widely used to model
nonlinear electrokinetic phenomena in microfluidic devices, such
as ACEO flows around electrode
pairs~\cite{ramos1999,green2000b,gonzalez2000,green2002,bown2006} and
arrays~\cite{ajdari2000,brown2000,bazant2006,urbanski2007,weiss2008,hilber2008,loucaides2007,burch2008},
traveling-wave electro-osmotic flows
(TWEO)~\cite{cahill2004,ramos2005,ramos2007}, ICEO interactions
between dielectric particles and
electrodes~\cite{nadal2002a,nadal2002b,ristenpart2003,ristenpart2007},
ICEO flow around metal
structures~\cite{iceo2004a,iceo2004b,levitan2005,zhao2007b,Wu:2008rc,Wu:2008qr,gregersen2009,gregersen2009b}
and dielectric corners~\cite{thamida2002,yossifon2006} and
particles~\cite{iceo2004b,yossifon2007,yariv2008a}, fixed-potential ICEO around
electrodes with a DC bias~\cite{iceo2004b,soni2007}, ICEP motion
of polarizable asymmetric  
particles~\cite{iceo2004a,iceo2004b,yariv2005,squires2006}, collections of interacting particles~\cite{rose2007,saintillan2006,saintillan2006b,saintillan2008,wu-li2009}, particles near walls~\cite{zhao2007,wall_preprint}, and particles in field gradients~\cite{squires2006}.
The Standard Model has had many successes in describing all of these phenomena, but it also has some fundamental shortcomings, when compared to experimental data. 

One possibility is that the underlying PNP/NS equations and boundary conditions are {\it physically} accurate, but the thin-DL approximation introduces large {\it mathematical} errors. A number of theoretical studies have allowed for arbitrary DL thickness in a dilute solution while solving the  linearized equations of ion transport and fluid flow in the regime of low voltages. This modeling approach has been applied to ACEO~\cite{gonzalez2000} and TWEO~\cite{gonzalez2008} flows over electrode arrays and ICEP particle motion in uniform~\cite{yariv2008b} or non-uniform~\cite{miloh2008} fields. For the model problem of ICEO flow around a thick metal stripe on a flat wall in a parallel electric field~\cite{bazant2004,soni2007}, Gregersen et al.~\cite{gregersen2009b} have recently compared full numerical solutions of the Poisson-Nernst-Planck (PNP) and Navier-Stokes (NS) equations with thin double-layer approximations in both the  linear regime (Standard Model) and the ``weakly nonlinear'' regime~\cite{bazant2004,chu2006} (including tangential surface conduction, but not diffuse-layer salt adsorption, surface reactions, and bulk concentration polarization, discussed below). For their model problem, they concluded that for micron-scale electrodes, the (outer) boundary-layer approximations can over-estimate ICEO velocities by  10\% for  thin double layers, and  by 100\% for double-layer thickness comparable to the electrode height. These errors could be reduced by constructing uniformly valid approximations for finite double layer thickness~\cite{bazant2005,bazant2006} (adding double layer contributions and subtracting the overlap), but the main point for us is that mathematical errors in thin double layer approximations (compared to full PNP/NS numerical solutions) vanish in the thin double layer limit and are small (of order $\lambda_D/L \ll 1$) for typical experimental situations in microfluidics. In particular, we cannot attribute the systematic and large (sometimes order of magnitude) discrepancies between theory and experiment discussed below to the thin double-layer approximation, especially if the asymptotic analysis is done carefully, going beyond the leading-order low-voltage approximation of the Standard Model. Instead, in this paper, we build the case that at least some of the discrepancies may be attributable to the breakdown of the underlying PBP/NS equations of dilute solution theory (and its boundary conditions), close to a highly charged surface.

\begin{table*}
\hspace{-1in}
\begin{tabular}[c]{|l|c|c|c|c|c|c|c|}
\hline 
Reference & Type of Flow & Solution & $c_{0}$ & $V_{max}^{induced}$ &
$\Lambda$ & $\zeta_{max}$ & $e\zeta_{max}/kT$\\
\hline
Green et al. 2000 ~\cite{green2000b} 
& ACEO electrode pair & KCl & 0.16mM & 1.0 V & 0.13 & 0.13 V & 5.2\\
&  &  & 0.67mM & 1.0 V & 0.055 & 0.055 V & 2.2\\
&  &  & 6.6mM & 2.5 V & 0.015 & 0.038 V & 1.52\\
\hline
Green et al. 2002 ~\cite{green2002} 
& ACEO electrode pair & KCl & 0.16mM & 0.5 V & 0.25$^{\ddag}$ & 0.125 V & 5\\
&  &  & 0.67mM & 0.5 V & 0.24$^\ddag$ & 0.12 V & 4.8\\
\hline
Studer et al. 2004 ~\cite{studer2004} 
& planar ACEO array & KCl & 0.1mM & 1.41 V & 0.18 & 0.25 V & 10\\
\hline
Ramos et al. 2005 ~\cite{ramos2005} 
& TWEO electrode array & KCl & 0.16mM & 0.5 V & 0.05 & 0.025V & 1\\
&  &  &  & 1.4V & 0.026 & 0.036 V & 1.44\\
\hline
Bown et al. 2006 ~\cite{bown2006} 
& Disk electrode ACEO & KCl & 0.43mM & 2.0 V & 0.0025 & 0.005 V & 0.2\\
\hline
Urbanski et al. 2007 ~\cite{urbanski2007} 
& 3D ACEO array & KCl & 3$\mu$M & 1.5 V & 0.2$^*$ & 0.3 V & 12\\
\hline
Bazant et al. 2007 ~\cite{microTAS2007} 
& planar ACEO array & KCl & 0.001mM & 0.75 V & 1$^*$ & 0.75 V & 30\\
&  &  & 0.003mM & & 0.88$^*$ & 0.66 V & 26.4\\
&  &  & 0.01mM & & 0.65$^*$ & 0.49 V & 19.6\\
&  &  & 0.03mM & & 0.47$^*$ & 0.35 V & 14\\
&  &  & 0.1mM &  & 0.41$^*$ & 0.31 V & 12.4\\
&  &  & 0.3mM &  & 0.24$^*$ & 0.18 V & 6.4\\
&  &  & 1mM & & 0.10$^*$ & 0.075 V & 3\\
\hline
Storey et al. 2008 ~\cite{storey2008} 
& planar ACEO array & KCl & 0.03mM & 0.75 V & 0.667 & 0.5 V & 20\\
\hline
Levitan et al. 2005 ~\cite{levitan2005} 
& metal cylinder ICEO & KCl & 1mM & 0.25 V & 0.4$^{\ddag}$ & 0.1 V & 4\\
\hline
Soni et al. 2007 ~\cite{soni2007} 
& fixed-potential ICEO & KCl & 1mM & 9.0 V & 0.005 & 0.045 & 1.8\\
\hline
Brown et al. 2000 ~\cite{brown2000} & ACEO array & NaNO$_{3}$
& 0.1mM & 1.7V & 0.068$^*$ & 0.115V & 4.6\\
&  &  &  & 1.41V & 0.062$^*$ & 0.087 V & 3.5\\
&  &  &  & 1.13V & 0.071$^*$ & 0.08 V & 3.2\\
&  &  &  & 0.85V & 0.079$^*$ & 0.067 V & 2.7\\
&  &  &  & 0.57V & 0.076$^*$ & 0.043 V & 1.7\\
&  &  &  & 0.28V & 0.081$^*$ & 0.023 V & 0.92\\
\hline
Urbanski et al. 2006 ~\cite{urbanski2006}
& ACEO array & water & $\approx \mu$M  & 1.5 V & 0.25$^*$ & 0.375 V & 15\\
&  &  &  & 1.0 V & 0.5$^*$ & 0.5 V & 20\\
\hline
Gangwal et al. 2008 ~\cite{gangwal2008} 
& Janus particle ICEP & water & $\approx \mu$M & 0.085 V & 0.14$^{\dagger}$ &
0.012 V & 0.48\\ 
Kilic \& Bazant 2008 ~\cite{wall_preprint} 
&  & NaCl & 0.1mM &  & 0.14$^{\dagger}$ & 0.012 V & 0.48\\
&  &  & 0.5mM &  & 0.105$^{\dagger}$ & 0.009 V & 0.36\\
&  &  & 1mM &  & 0.08$^{\dagger}$ & 0.007 V & 0.27\\
&  &  & 3mM &  & 0.048$^{\dagger}$ & 0.004 V & 0.16 \\
\hline
\end{tabular}
\caption{ \label{table:Lambda} 
Comparison of the standard low-voltage 
  model of induced-charge electrokinetic phenomena with experimental
  data (column 1) for a wide range of situations (column 2), although
  limited to a small set of aqueous electrolytes (column 3) at low bulk salt
  concentrations $c_0$ (column 4). In each case, the nominal maximum
  induced double-layer voltage $V_{max}$ is estimated (column 5). A crude
  comparison with the Standard Model is made by multiplying the predicted slip velocity  (\ref{eq:HS}) 
everywhere by a constant factor
  $\Lambda$ (column 6) for a  given $c_0$ and $V_{max}$. In
  addition to $\Lambda$ values from the cited papers, we have added entries to the table,
  indicated by $^*$, by fitting our own standard-model simulations to published experimental data.  Estimates 
indicated by
  $^\ddag$ assume a frequency-dependent constant-phase-angle impedance for the double layer, and those 
labeled by $^\dagger$ are affected by  
  particle-wall interactions, which are not fully understood. In each case, we also estimate the maximum 
induced
  zeta potential $\zeta_{max} = V_{max} \Lambda$  in Volts (column 7) and in units
  of thermal voltage $kT/e$ (column 8). }
\end{table*}

\subsection{ Open questions  }

\subsubsection{ The ``correction factor" }

Low-voltage, dilute-solution theories in nonlinear electrokinetics tend to over-predict fluid velocities, compared to experiments.  
A  crude way to quantify this effect in the Standard Model is to multiply the HS slip velocity
(\ref{eq:HS}) on all surfaces by a fitting parameter $\Lambda$, the
``correction factor'' introduced by Green et
al. ~\cite{green2000b,green2002}. This approach works best at low voltages and in very dilute solutions, but even in such a regime, we should stress that it is generally impossible to choose $\Lambda$ to fit complete flow profiles or multiple experimental trends, e.g. velocity vs. voltage and frequency, at the same time. Nevertheless, one can often make a meaningful fit of $\Lambda$ to reproduce a key quantity, such as the maximum flow rate or particle velocity.  Such a quantitative test of the model has 
been attempted for a number of data sets~
\cite{green2000b,green2002,levitan2005,bown2006,soni2007,storey2008,gangwal2008,wall_preprint}, 
but there has been no attempt to synthesize results from different types of experiments to seek general 
trends in the correction factor. 

As a background for our study, we provide a critical evaluation of
the Standard Model based on $\Lambda$ values for a wide range of
experimental situations. In Table~\ref{table:Lambda}, we have complied  
all available results from the literature. We have also added many entries
by fitting our own Standard-Model simulations to published
data, for which no comparison has previously been done. 

It is striking that $\Lambda$ is never larger than unity and can be orders of magnitude smaller. We managed to find only 
one published measurement where the Standard Model correctly predicts the maximum of the observed flow ($\Lambda=1$), from a recent 
experiment on ACEO pumping of  micromolar KCl by a planar, gold electrode array at relatively low voltage~
\cite{microTAS2007}, but even in that data set the model fails to predict weak flow reversal at high frequency and salt concentration dependence (see below). Remarkably, there has not yet been a single ICEO experiment where the model has been able to predict, or even to fit, how the velocity depends on the basic operating conditions -- voltage, AC frequency, and salt concentration -- let alone the dependence on surface and bulk chemistries. The greatest discrepancies come from ACEO pumping by 
a disk-annulus electrode pair~\cite{bown2006} ($\Lambda=0.0025$) and fixed-potential ICEO around a metal 
stripe~\cite{soni2007} ($\Lambda=0.005$), both in millimolar KCl and at high induced 
voltages $V_{max}$.  

In Table ~\ref{table:Lambda}, we have also used $\Lambda$ to convert the maximum 
nominal voltage $V_{max}$ induced across the double layer in each experiment to a maximum zeta potential 
$\zeta_{max} = \Lambda V_{max}$.
The range of $\zeta_{max}$ is much smaller for $\Lambda$, but still 
quite significant. It is clear that $\zeta_{max}$ rarely exceeds $10 kT/e$, regardless of the applied voltage. For very 
dilute solutions, the largest value in the Table, $\zeta_{max}=0.75 \mbox{V} = 30 kT/e$, comes from ACEO pumping of 
micromolar KCl~\cite{microTAS2007}, while the smallest values, $\zeta_{max} < 0.5 kT/e$, come from ICEP 
of gold-latex Janus particles in millimolar NaCl.  

\begin{figure}
\begin{center}
(a) \includegraphics[width=2.3in]{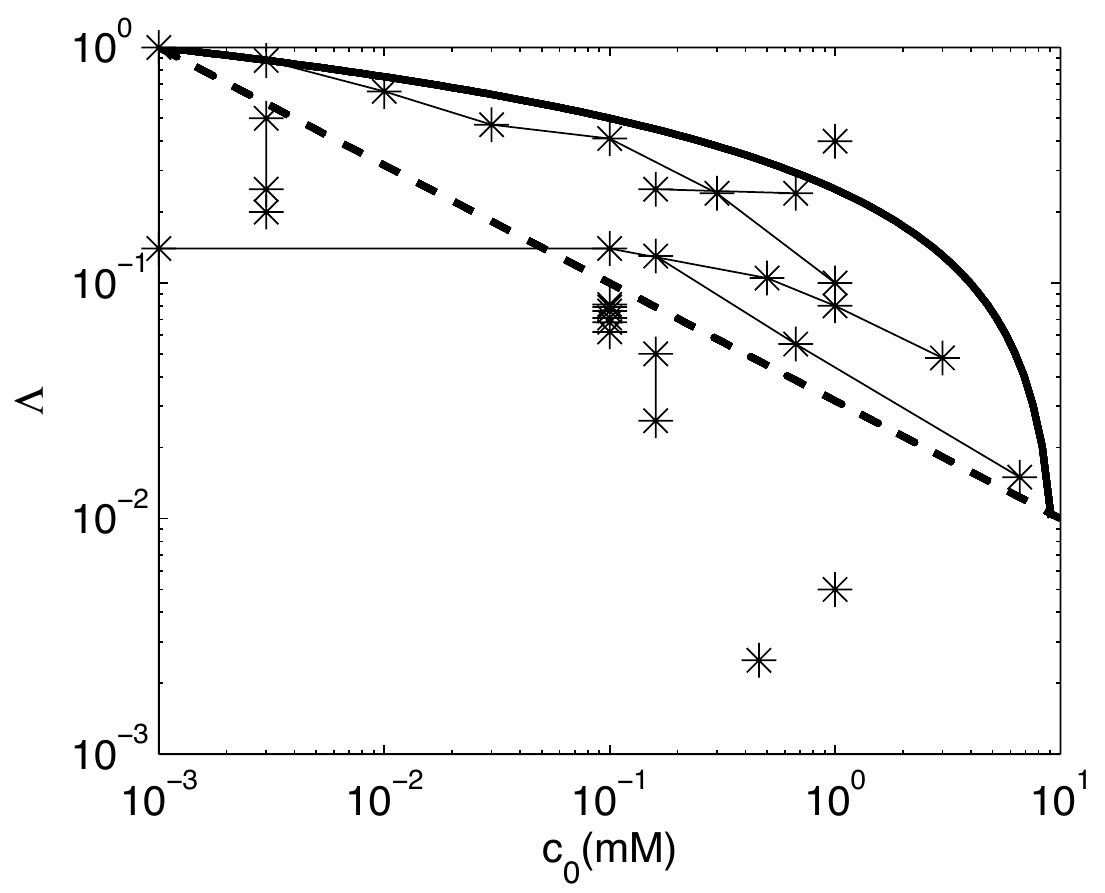}  (b) 
\includegraphics[width=2.3in]{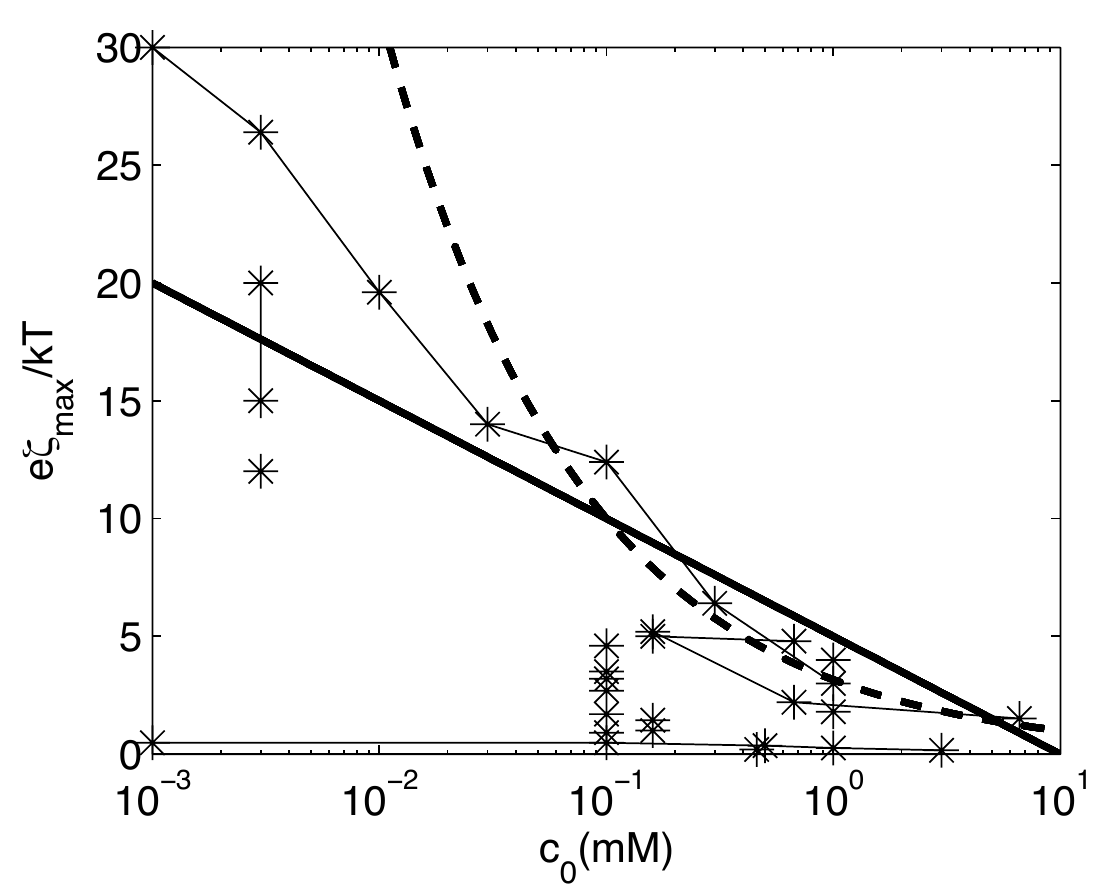} \\
(c) \includegraphics[width=2.3in]{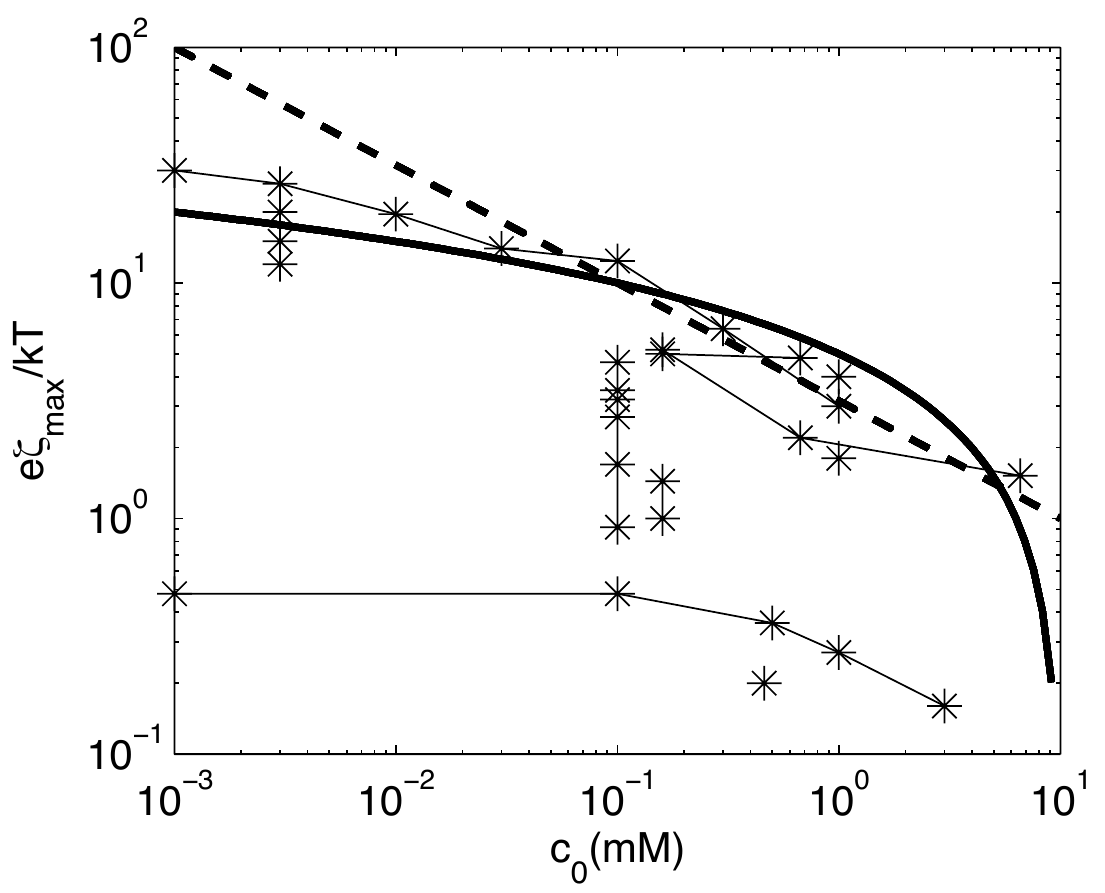}  (d) 
\includegraphics[width=2.3in]{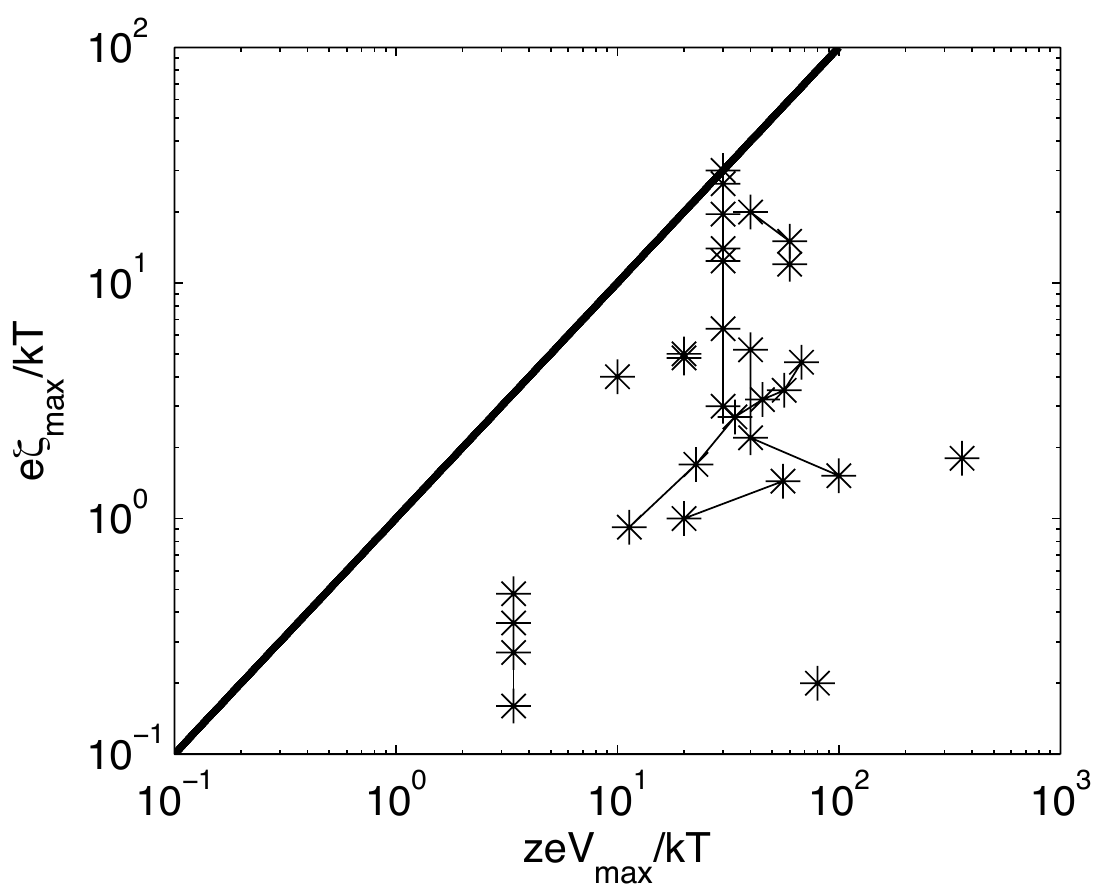} 
\caption{\label{fig:Lambda} General trends in the under-prediction of ICEO flow velocity by the Standard Model from Table~\ref{table:Lambda}, compared with (solid and dashed) scaling curves, simply to guide the eye. (a) Log-log plot of $\Lambda$ vs. $c_0$ compared with the curves $\Lambda=\sqrt{10^{-3}\mbox{mM}/c_0}$ (dashed) and $=\ln_{10}(10\mbox{mM}/c_0)/4$ (solid); (b) log-linear plot  and (c) log-log plot of $\zeta_{max}$ vs. $c_0$ compared with the curves $e\zeta_{max}/kT=\sqrt{10\mbox{mM}/c_0}$ (dashed) and $=5\ln_{10}(10\mbox{mM}/c_0)$ (solid); and (d) log-log plot of 
$\zeta_{max}$ vs. $V_{max}$ compared to $\zeta_{max} = V_{max}$ (solid).  Points from the same experiment (varying concentration or voltage) are connected by line segments. }
 \end{center}
\end{figure}

The values of $\Lambda$ and $\zeta_{max}$ from all the different experimental situations in Table~\ref{table:Lambda} are plotted versus 
$c_0$ and$V_{max}$ in Fig.~\ref{fig:Lambda}, and some general trends become evident. In Fig.~\ref{fig:Lambda}(b), we see that $\zeta_{max}$ decays strongly with increasing salt concentration and becomes negligible in most experiments above 10 mM. In Fig.~\ref{fig:Lambda}(c),  we see that $\zeta_{max}$ exhibits sub-linear growth with $V_{max}$ and appears to saturate below ten times times the thermal voltage ($10 kT/e = 0.25$ V at room temperature), or much lower values at high salt concentration. Even with applied voltages up to 10 Volts in dilute solutions, the effective maximum zeta potential tends to stay well below 1 Volt. From the perspective of classical electrokinetic theory, this implies that most of the voltage applied to the double layer is dropped across the immobile, inner "compact layer", rather than the mobile outer "diffuse layer", where electro-osmotic flow is generated.

This effect can be qualitatively, but not quantitatively, understood using the Standard Model. 
Many authors have assumed a 
uniform, uncharged Stern layer (or dielectric thin film) of permittivity $\varepsilon_S$ and thickness $h_S = \varepsilon_S/C_S$, acting as a capacitor in series with the diffuse layer. Via Eq.~(\ref{eq:PsiD}), this model implies 
\begin{equation}
\Lambda = \frac{1}{1+\delta}, \ \ \mbox{ with } \ \ 
\delta = \frac{C_D}{C_S} = \frac{\varepsilon_b}{\varepsilon_S} \frac{h_S}{\lambda_D}
= \frac{\lambda_S}{\lambda_D},
 \label{eq:Laminv}
\end{equation}
where $\lambda_D$ is the Debye-H\"uckel screening length (diffuse-layer thickness), which takes the form 
\begin{equation}
\lambda_D = \sqrt{ \frac{\varepsilon_b kT}{2(ze)^2 c_0} }\label{eq:lambda}
\end{equation}
for a $z:z$ electrolyte, 
and $\lambda_S$ is an effective width for the Stern layer, if it were a capacitor with the same dielectric constant as the bulk. 
Inclusion of the Stern layer only transfers the large, unexplained variation in the correction factor $\Lambda$ to the 
parameter $\lambda_S$ (or $C_S=\varepsilon_b/\lambda_S$) without any theoretical prediction of why it should vary so much with voltage, concentration, and  geometry. Using these kinds of equivalent circuit models applied to differential capacitance measurements~\cite{bockris_book}, electrochemists sometimes infer a tenfold reduction in permittivity in the Stern layer versus bulk water, $\varepsilon_b/\varepsilon_S\approx 10$, but, even if this were always true, it would still be hard to explain the data. For many experimental situations in Table~\ref{table:Lambda}, the screening length $\lambda_D$ is tens of nanometers, or hundreds of molecular widths, and the effective Stern-layer width $\lambda_S$ would need to be much larger -- up to several microns -- to predict the observed values of $\Lambda \ll 1$. In contrast, if we take the physical picture of a Stern monolayer literally, then $h_S$ should be only a few Angstroms, and $\lambda_S$ at most a few nanometers, so there is no way to justify the model. As noted in early papers by Brown et al.~\cite{brown2000} and Green et al.~\cite{green2000b} , it is 
clear that the effective diffuse-layer voltage (or induced zeta potential) is not properly
described by the Standard Model under typical experimental conditions. 

\subsubsection{ Electrolyte dependence }

\begin{figure}
\begin{center}
(a)\includegraphics[width=2.6in]{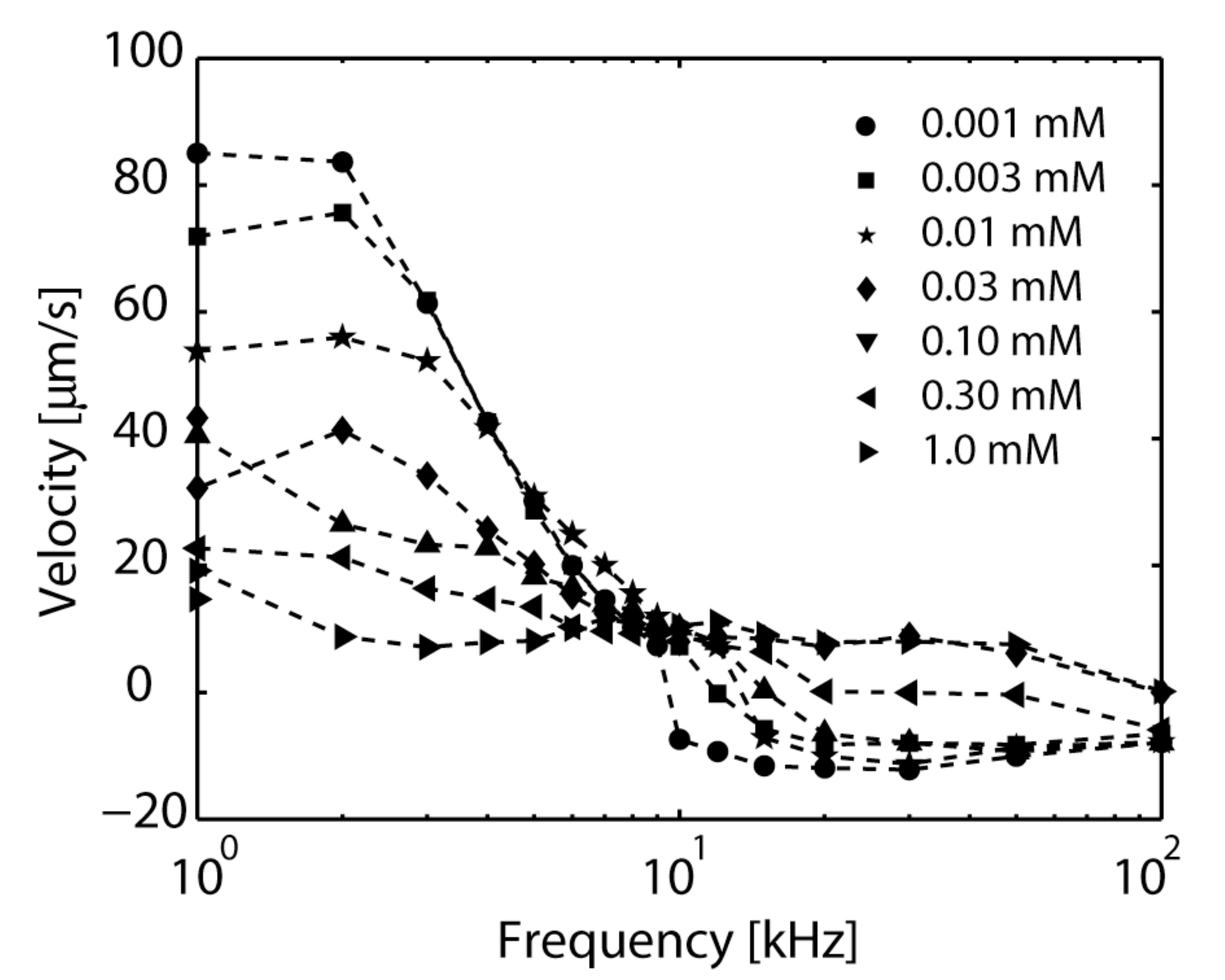} \nolinebreak
(b)\includegraphics[width=2.4in]{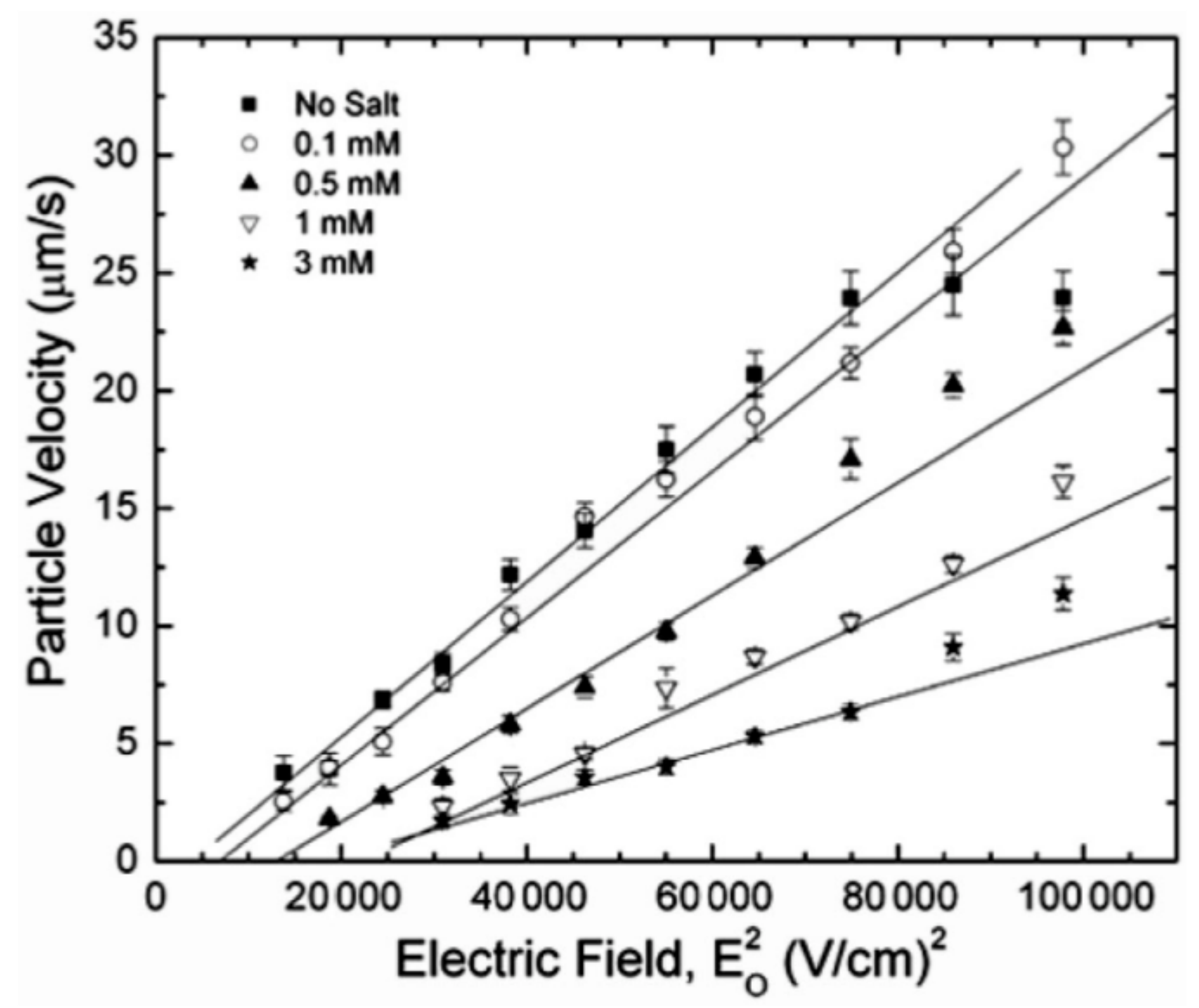}
\caption{ \label{fig:expt} Typical experimental data (included in the estimates of Table 1) for two different types of nonlinear, induced-charge electrokinetic phenomena showing qualitative features not captured by the Standard Model, or the underlying electrokinetic equations of dilute solution theory. (a) Velocity of ACEO pumping of dilute aqueous solutions of KCl around a microfluidic loop by an asymmetric planar Au electrode array with the geometry of Refs~\cite{brown2000,studer2004} versus AC frequency at constant voltage, 3 Volts peak to peak ($V_{max} = 1.5$ V), reproduced from Ref.~\cite{microTAS2007}. The data exhibits the unexplained flow reversal at high frequency ($10-100$ kHz) and strong concentration dependence first reported in Ref.~\cite{studer2004}. (b) Velocity of ICEP motion of 5.7 $\mu$m metallo-dielectric Janus particles versus field-squared at different concentrations of NaCl in water at constant 1kHz AC frequency, reproduced from Ref.~\cite{gangwal2008}. The data shows a similar decay of the velocity with increasing bulk salt concentration, which becomes difficult to observe experimentally above 10mM, in both experiments.
}
\label{fig:Exp_data}
\end{center}
\end{figure}

In addition to overestimating experimental velocities, the standard
model fails to predict some important phenomena, even
qualitatively. For example, ICEO flows have a strong sensitivity to
solution composition, which is under-reported and unexplained. Most
experimental work has focused on dilute
electrolytes~\cite{brown2000,green2000b,green2000,levitan2005}. (See Table~\ref{table:Lambda}.) 
Some recent experiments suggest a logarithmic decay of the induced
electro-osmotic mobility, $b \propto \ln(c_c/c_0)$, with bulk
concentration $c_0$ seen in KCl for ACEO
micropumps~\cite{studer2004,microTAS2007}, in KCl and CaCO$_3$ for
ICEO flows around metal posts~\cite{levitan_thesis}, and in NaCl for
ICEP motion of metallo-dielectric Janus
particles~\cite{gangwal2008}. This trend is visible to some extent at moderate concentrations in Fig.~\ref{fig:Lambda}(b) over a wide range of experimental conditions, although a power-law decay also gives a reasonable fit  at high salt concentrations.  Two examples of different nonlinear electrokinetic phenomena (ACEO fluid pumping and ICEP particle motion) showing this unexplained decay with concentration are shown in Fig.~\ref{fig:expt}. 

In all experiments, such as those in Fig.~\ref{sec:expt}, the flow practically
vanishes for $c_0 > 10$ mM, which is two orders of magnitude below the
salinity of most biological fluids and buffer solutions ($c_0 > 
0.1$ M). Experiments with DC~\cite{schasfoort1999,wouden2005} and
AC~\cite{wouden2006,soni2007} field-effect flow control, where a gate voltage
controls the zeta potential of a dielectric channel surface, have
likewise been limited to low salt concentrations below 10 mM in a
variety of aqueous solutions. Sodium carbonate/bicarbonate buffer solutions have also been used in ICEO mixers with platinum structures~\cite{Wu:2008qr}, but experimental data was only reported at a low ionic strength of 2.5 mM after dilution by water, and not for the original 1 M solutions.  Remarkably, out of all the experimental work reviewed in this section, we could not find any observations of induced-charge electrokinetic phenomena at salt concentrations above 30 mM in water.  

The Standard Model seems
unable to explain the decay of flow with increasing salt concentration quantitatively, although it does aid in qualitative understanding.  Substituting the Debye-H\"uckel screening length for a binary $z:z$ electrolyte in (\ref{eq:Laminv}) we obtain
\begin{equation}
\Lambda = \frac{1}{1 + \sqrt{c_0/c_c}} \sim \sqrt{\frac{c_c}{c_0}} \ \ \mbox{ for } \ c_0 \gg c_c
\end{equation}
where
\begin{equation}
c_c = \frac{kT}{2\varepsilon_b} \left( \frac{\varepsilon_S}{h_S z e} \right)^2
= \frac{\varepsilon_b kT}{2 (ze)^2 \lambda_S^2} 
\end{equation}
is a crossover concentration, above which the flow decays like the inverse square-root of concentration. As noted above, it is common to attribute
the theoretical over-prediction of ICEO flows, even in very dilute
solutions to a large voltage drop across the compact layer ($\delta\gg
1$), but this would imply a strong concentration dependence ($c_0 \gg c_c$) that is
not observed. Alternatively, fitting the compact-layer capacitance to
reproduce the transition from dilute to concentrated solution behavior
($c_0 \approx c_c$, $\delta\approx 1$) would eliminate the correction factor in dilute
solutions ($\delta \ll 1$), making the theory again over-predict the
observed velocities. For example, such difficulties are apparent in 
Ref.~\cite{wall_preprint} where this argument applied to the data of
Gangwal et al~\cite{gangwal2008} for ICEP motion of metallo-dielectric
Janus particles (Fig.~\ref{fig:expt}(a)).

Beyond the dependence on salt concentration, another failing of
dilute-solution theory is the inability to explain the experimentally
observed ion-specificity of ICEO phenomena. At the same bulk
concentration, it has been reported that ICEO flow around metal
posts~\cite{levitan_thesis}, ACEO pumping by electrode
arrays~\cite{microTAS2007} and AC-field induced interactions in
colloids~\cite{sides2001} depend on the ions. Comparing experiments
under similar conditions with different electrolytes or different
metal surfaces further suggests a strong sensitivity to the chemical
composition of the double layer, although more systematic study is
needed. In any case, none of these effects can be captured by the
Standard Model, which posits that the ions are simply mathematical
points in a dielectric continuum and that the surface is a homogeneous
conductor or dielectric; all specific physical or chemical properties
of the ions, solvent molecules, and the surface are neglected.

\subsubsection{ Flow reversal }

In many situations of large induced voltages, the Standard Model does
not even correctly predict the direction of the flow, let alone its
magnitude. Flow reversal was first reported around tin particles
in water~\cite{gamayunov1992}, where the velocity agreed with the
theory~\cite{murtsovkin1996,gamayunov1986} sketched in Fig. 1a only for micron-sized
particles and reversed for larger ones ($90-400 \mu$m). The transition
occured when several volts was applied across the particle and
reversal was conjectured to be due to Faradaic
reactions~\cite{gamayunov1992}. In this regime, reverse flows have recently been observed around large (millimeter scale) copper washers and steel beads with flow patterns resembling second-kind electro-osmosis~\cite{barinova2008} (see below); although the field was kept below the level causing gas bubbles at the anodic side of the metal, copper dendrites (resulting from electrodeposition) were observed on the cathodic side in dilute CuCl$_2$ solutions, implying normal currents and concentration gradients.

In microfluidic systems, flow reversal has also been observed
at high voltage ($> 2$ V) and high frequency (10-100 kHz) in ACEO
pumping by $10 \mu$m-scale planar electrode arrays for dilute
KCl~\cite{studer2004,urbanski2006,microTAS2007,garcia2006}, as shown in Fig.~\ref{fig:expt}(b), 
although not for water in the same pump geometry~\cite{brown2000,urbanski2006}. 
Non-planar 3D stepped
electrodes~\cite{bazant2006} can be designed that do not exhibit flow
reversal, as demonstrated for KCl~\cite{urbanski2007} and water~\cite{chien-chih}, but certain
non-optimal 3D geometries can still reverse, as shown in 
water~\cite{urbanski2006}.  In the latter case the frequency spectrum
also develops a double peak with the onset of flow reversal around 3
Volts peak to peak. In travelling-wave electro-osmosis (TWEO) in
aqueous electrolytes~\cite{cahill2004,ramos2005}, strong flow
reversal at high voltage has also been observed, spanning all
frequencies~\cite{ramos2005,garcia2006,garcia2008}, and not yet fully understood.

Flow reversal of ACEO was first attributed to Faradaic reactions under
different conditions of larger voltages (8-14 V) and frequencies (1-14
MHz) in concentrated NaCl solutions ($0.001-0.1$ S/m) with a $100
\mu$m-scale T-shaped electode pair composed of different metals (Pt,
Al, Chromel)~\cite{chang2004}. Indeed, clear signs of Faradaic
reactions (e.g. gas bubbles from electrolysis of water) can always be observed at sufficiently large
voltage, low frequency and high
concentration~\cite{studer2004,chang2004}. In recent TWEO experiments~\cite{garcia2008,garcia2009}, signatures of Faradaic reactions (including pH gradients from electrolysis) have been correlated with low-frequency flow reversal at high voltage and bulk electroconvection has been implicated~\cite{garcia2009} (see below). Under similar conditions
another possible source of flow reversal is AC electrothermal
flow driven by bulk Joule heating~\cite{gonzalez2006}, which has been
implicated in reverse pumping over planar electrode arrays at high
salt concentrations~\cite{wu2007}. Closer to standard ACEO conditions,
e.g. at 1-2 V and 50-100 Hz in water with Au electrode arrays, flow
reversal can also be induced by applying a DC bias voltage of the same
magnitude as the AC voltage~\cite{wu2005,wu2006,lian2009}. Reverse ACEO flow
due to ``Faradaic charging'' (as opposed to the standard case of
``capacitive charging'') is hypothesized to grow exponentially with
voltage above a threshold for a given the electrolyte/metal
interface~\cite{chang2004,wu2006}, but no quantitative theory has been
developed.

Simulations of the standard low-voltage model with Butler-Volmer
kinetics for Faradaic reactions have only managed to predict weak flow
reversal at low frequency in
ACEO~\cite{ajdari2000,olesen2006,olesen_thesis} and
TWEO~\cite{ramos2007,gonzalez2008}. In the case of ACEO with a planar, asymmetric
electrode array, this effect has recently been observed using
sensitive ($\mu$m/s) velocity measurements in dilute KCl with Pt
electrodes at low voltage ($< 1.5$ V) and low frequency ($ < 20$
kHz)~\cite{gregersen2007}.  Faradaic reactions can also produce an oscillating quasi-neutral diffusion layer between the charged diffuse layer and the uniform bulk, due to the normal flux of ions involved in reactions, and  it has been shown via a low-voltage, linearized analysis of TWEO that  flow reversal can arise  in the case of ions of unequal diffusivities due to enhanced diffusion-layer forces on the fluid~\cite{gonzalez2008}.  Recently, flow reversal has been successfully predicted in TWEO by such a model allowing for bulk electroconvection in regions of pH gradients from electrolysis reactions and compared to experimental data~\cite{garcia2009}. However, the theory is still incomplete, and it seems that current models cannot predict the
strong ($> 100 \mu$m/s), high-frequency ($> 10$ kHz) flow reversal
seen in many ACEO and TWEO
experiments~\cite{studer2004,urbanski2006,garcia2006,microTAS2007}. Faradaic
reactions generally reduce the flow at low frequency by acting as a
resistive pathway to ``short circuit'' the capacitive charging of the
double layer~\cite{olesen2006,ramos2007}, and diffusion-layer phenomena are also mostly limited to low frequency.
Resolving the apparent paradox of high-frequency flow reversal is a major motivation for our study.

\subsection{ Nonlinear dynamics in a dilute solution }

Dilute-solution theories generally predict that nonlinear effects 
dominate at low frequency. One reason is that the differential
capacitance $C_D$ of the diffuse layer, and thus the ``RC'' time for
capacitive charging of a metal surface, grows exponentially with
voltage in nonlinear Poisson-Boltzmann (PB) theory. The familiar PB
formula for the diffuse-layer differential capacitance of a symmetric
binary electrolyte~\cite{lyklema_book_vol2,bazant2004},
\begin{equation}
  C^{PB}_{D}(\Psi_D)=\frac{\varepsilon_b}{\lambda_{D}}\cosh\left(\frac{ze\Psi_{D}}{2kT}
  \right) \label{eq:cdpb}
\end{equation}
was first derived by Chapman~\cite{chapman1913}, based on Gouy's
solution of the PB model for a flat diffuse layer~\cite{gouy1910}. It
has been shown that this nonlinearity shifts the dominant flow to
lower frequencies at high voltage in ACEO~\cite{olesen2006} or
TWEO~\cite{gonzalez2005} pumping. It also tends to suppress the flow
with increasing voltage at fixed frequency, since there is not
adequate time for complete capacitive charging in a single AC period.

At the same voltage where nonlinear capacitance becomes important,
dilute-solution theory also predicts that salt
adsorption~\cite{bazant2004,marescaux2009,chu2006,suh2008,olesen2009} and tangential
conduction~\cite{chu2006,chu2007} by the diffuse layer also occur and
are coupled to (much slower) bulk diffusion of neutral salt, which
would enter again at low frequency in cases of AC forcing. If
concentration gradients have time to develop, then they generally
alter the electric field (``concentration polarization'') and can
drive bulk electroconvection~\cite{garcia2009},  diffusio-osmotic
slip~\cite{prieve1984,anderson1989,rubinstein2001} and in some cases, non-equilibrium space charge and second-kind
electro-osmotic flow~\cite{dukhin1991,mishchuk1995,mishchuk_review,zaltzman2007,olesen2009} (if the bulk
concentration goes to zero, at a limiting current).

Bulk gradients in salt concentration may play a crucial role in induced-charge electrokinetics, especially at voltages large enough to drive Faradaic reactions. 
Concentration polarization has been demonstrated around electrically floating and (presumably) blocking metal posts in DC fields and applied to microfluidic demixing of electrolytes~\cite{leinweber2006}. In nonlinear electrokinetic theory, diffusion-layer phenomena have begun to be considered in low-voltage, linearized analysis of TWEO with Faradaic reactions~\cite{gonzalez2008}, and flow reversal of TWEO has been attributed to bulk electro-convection in regions of concentration polarization, for asymmetric electrolytes with unequal ionic mobilities ~\cite{garcia2009}. Such effects could be greatly enhanced in the strongly nonlinear regime and as yet unexplored. Including all of
these effects in models of induced-charge electrokinetic phenomena
presents a formidable mathematical challenge. 

To our knowledge, such complete nonlinear modeling within the framework of
dilute solution theory has only begun to be accomplished in the case
of ACEO pumping (albeit without Faradaic reactions) in the Ph.D. thesis of Olesen~\cite{olesen_thesis} and the papers of Suh and Kang~\cite{suh2008,suh2009}
by applying
asymptotic boundary-layer methods to the classical electrokinetic
equations in the thin-double-layer limit. At  
least in this representative case, all of the nonlinear large-voltage
effects in dilute solution theory in the solution phase tend to make the agreement with
experiment worse than in the Standard Model~\cite{olesen_thesis}. The flow is greatly
reduced and shifts to low frequency, while the effects of salt
concentration and ion-specificity are not captured. Similar
conclusions have been reached by a recent numerical and experimental
study of fixed-potential ICEO for DC bias of 9 Volts~\cite{soni2007},
where the correction factor is found to be $\Lambda=0.005$ for the
linear theory, but only $\Lambda=0.05$ if nonlinear capacitance
(\ref{eq:cdpb}) and surface conduction from PB theory are included in
the model (albeit without accounting for bulk concentration
gradients). 

On the other hand, Suh and Kang~\cite{suh2009} have recently shown that the experiments of Green et al. on ACEO over a symmetric electrode pair~\cite{green2002} can be better described by nonlinear dilute solution theory (without surface conduction, but with oscillating diffusion layers), if the no-flux boundary condition for ideally polarizable electrodes is replaced by a model of surface adsorption of ions~\cite{suh2008}, using a Langmuir isotherm applied at the Stern plane~\cite{mangelsdorf1998a,mangelsdorf1998b}. This approach relies on the classical, but somewhat arbitrary, partitioning of the double layer into an outer diffuse part (where ``free" ions can drive fluid flow, governed by continuum equations of dilute solution theory) and an inner compact part (where ``adsorbed" ions cannot, described by boundary conditions). It makes sense to describe ions that make contact with the electrode surface (breaking free of their own solvation shells and penetrating that of the metal, i.e. jumping from the ``outer Helmholtz plane" to the ``inner Helmholtz plane"~\cite{bockris_book}) via an adsorption boundary condition, but the transition to dilute solution behavior is surely not so abrubt. Indeed, we will see that concentrated solution theories of the liquid phase can predict an effectively immobile compact layer forming at high voltage and advancing into the solution.

Certainly, more theoretical work is needed on nonlinear
dynamics  of electrolytes in response to large voltages, especially in the presence of Faradaic reactions, but we believe the time has come
to also question the validity of the underlying electrokinetic
equations themselves. Based on experimental and theoretical results for
induced-charge electrokinetic phenomena, we conclude that dilute
solution theories may not properly describe the dynamics of
electrolytes at large voltages. In the following sections, we consider
some fundamental changes to the Standard Model and the underlying electrokinetic equations, while preserving the mean-field and local-density approximations, which permit a simple mathematical description in terms of partial differential equations. We review relevant aspects of concentrated-solution theories  and develop some new ideas as well. Through a variety of model problems in nonlinear electrokinetics, we make theoretical predictions using modified electrokinetic equations, which illustrate qualitatively new phenomena, not predicted by the Standard Model and begin to resolve some of the experimental puzzles highlighted above.

\begin{figure}
\begin{center}
\includegraphics[width=4in]{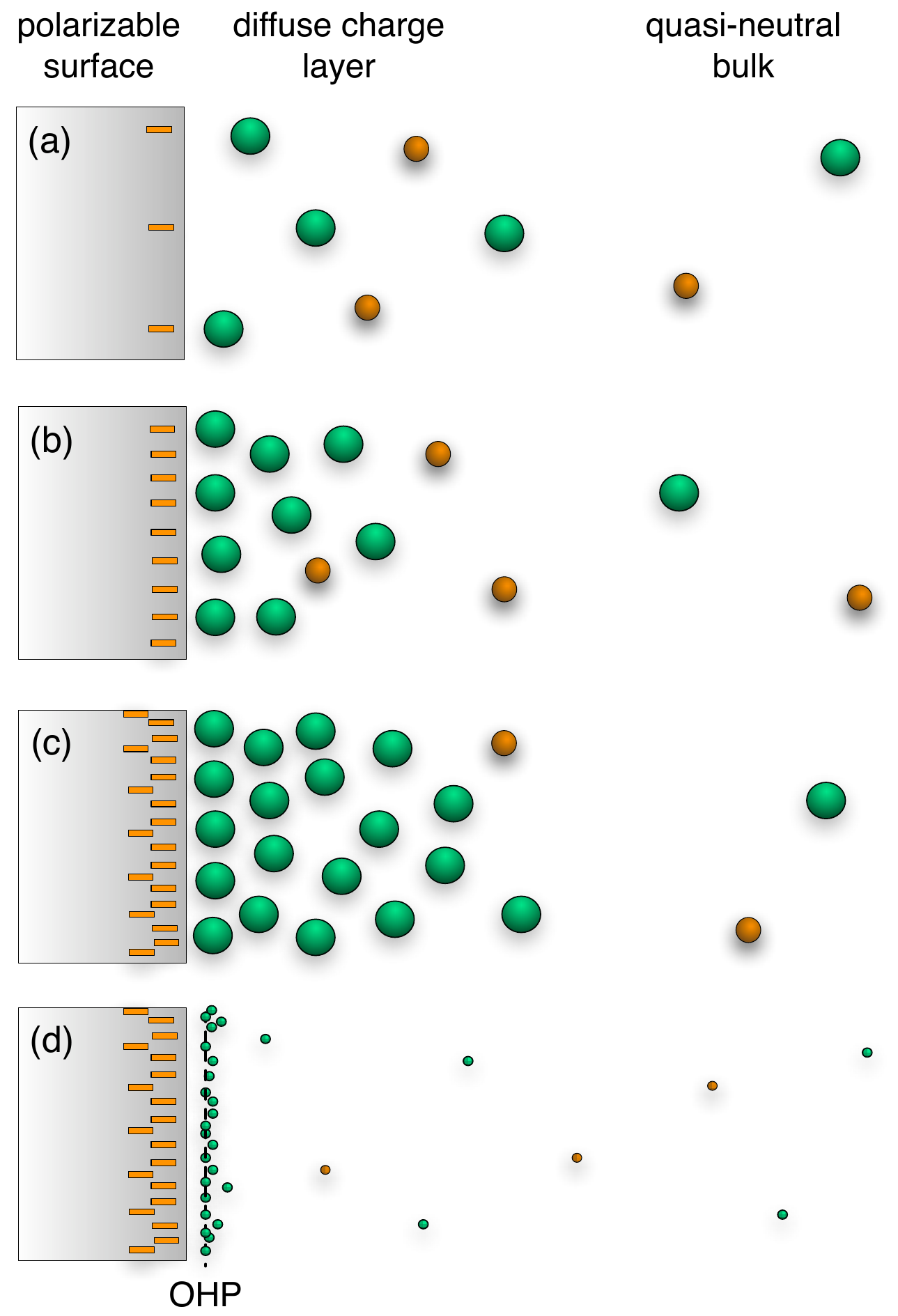}
\caption{\label{fig:cartoons} Sketch of solvated counterions (larger
  green spheres) and co-ions (smaller orange spheres) near a
  polarizable surface. (a) At small induced voltages, $\Psi_D \ll
  \Psi_c$, the neutral bulk is only slightly perturbed with a
  diffuse-charge layer of excess counterions at the scale of
  $\lambda_D$. (b) At moderate voltages, $\Psi_D \approx \Psi_c$, the
  diffuse layer contracts, as described by Poisson-Boltzmann (PB)
  theory. (c) At large voltages, $\Psi_D \gg \Psi_c$, the counterions
  inevitably become crowded, causing  expansion of the diffuse
  layer compared to the predictions of the classical
  Gouy-Chapman-Stern model, sketched in (d), which is based PB theory
  for point-like ions with a minimum distance of approach, the ``outer
  Helmholtz plane'' (OHP), to model solvation of the surface.  }
  \end{center}
\end{figure}

\section{Crowding effects in a concentrated solution}
\label{sec:crowding}

\subsection{ Mean-field local-density approximations}

\subsubsection{ Modified Poisson-Boltzmann theories }

All dilute solution theories, which describe point-like ions in a
mean-field approximation, break down when the crowding of ions becomes
significant, and steric repulsion and correlations potentially become
important.  If this can be translated into a characteristic length
scale $a$ for the distance between ions, then the validity of Poisson-Boltzmann
theory is limited by a cutoff concentration $c_{max}=a^{-3}$, which is
reached at a fairly small diffuse-layer voltage,
\begin{equation}
 \Psi_c = -\frac{kT}{ze}
  \ln\left(\frac{c_{max}}{c_0} \right)  = \frac{kT}{ze} \ln(a^3c_0).  \label{eq:psic}
\end{equation}
where $z$ is the valence (including its sign) and $c_0$ the bulk
concentration of the counterions. In a dilute solution of small ions,
this leads to cutoffs well below typical voltages for ICEO flows. For
example, even if only steric effects are taken into account, with e.g.
$a=3$ \AA (for solvated bulk K$^+$ - Cl$^-$ interactions
~\cite{mancinelli2007}), then $\Psi_c \approx 0.33 V$ for
$c_0=10^{-5}$ M and $z=1$.

To account for the obvious excess ions in PB theory,
Stern~\cite{stern1924} long ago postulated a static compact monolayer
of solvated ions~\cite{bockris_book}.  A similar cutoff is also
invoked in models of ICEO flows, where a constant capacitance is added
to model the Stern layer and/or a dielectric coating, which carries
most of the voltage when the diffuse-layer capacitance (\ref{eq:cdpb})
diverges.  However, it seems unrealistic that a monolayer could
withstand most of the voltage drop in induced-charge electrokinetic
phenomena at a blocking surface (e.g. without dielectric breakdown~\cite{jones1995}). In any case, a {\it dynamical} model is
required for a ``condensed layer'' that is built and destroyed as the
applied field alternates. As sketched in Fig.~\ref{fig:cartoons}, the
condensed layer forms in the diffuse part of the double layer and thus
should be described by the same ion transport equations. For a non-ideally blocking surface, it may also be necessary to account for surface adsorption of ions (breaking free of solvation shells) and Faradaic reactions (electron transfer) via compact-layer boundary conditions (see below), but these should be applied to a model of the diffuse layer that allows for the crowding of ions near the surface, which must occur to some degree with increasing voltage.

A plethora of ``modified Poisson-Boltzmann'' (MPB) theories have been
proposed to describe equilibrium ion profiles near a charged wall. As described in recent reviews~\cite{kilic2007a,biesheuvel2007,grochowski2008,vlachy1999,attard1996,macdonald1987,levin2002}), there are many possible modifications to describe different physical effects, such as dielectric relaxation, electrostatic correlations and volume constraints. In this paper, we focus on the simplest continuum models, which are based on the local-density and mean-field approximations. In spite of various limitations discussed below, this class of models is very convenient (if not required) for  mathematical analysis and numerical simulation of time-dependent nonlinear problems (based on modified Poisson-Nernst-Planck equations~\cite{kilic2007b,baker1999,olesen2009}).

The starting point for continuum modeling is a theory for the excess electrochemical potential of an ion
\begin{equation}
\mu_i^{ex}=\mu_i - \mu_i^{ideal} = kT \ln f_i, 
\end{equation}
relative to its ideal value in a dilute solution,
\begin{equation}
\mu_i^{ideal}=kT \ln c_i + z_i e \phi,  \label{eq:mu_ideal}
\end{equation}
where $c_i$ is the mean concentration and $f_i$ is the chemical activity coefficient. (Equivalently, one can write $\mu_i = kT\ln(\lambda_i) + z_ie \phi$, where $\lambda_i = f_i c_i$ is the absolute chemical activity~\cite{newman_book}.)  
In the {\it mean-field approximation} (MF), the electrostatic potential $\phi$ in (\ref{eq:mu_ideal}) self-consistently
solves the MPB equation, 
\begin{equation}
-\nabla\cdot(\varepsilon \nabla\phi) = \rho = \sum_i z_i e c_i,   \label{eq:PoissonMPB}
\end{equation}
where the source of the electric field acting on an individual ion is the mean charge density $\rho$, rather than the sum of fluctuating discrete charges. Time-dependent
modified PNP equations then express mass
conservation with gradient-driven fluxes~\cite{kilic2007b}, as described below.

In the asymptotic limit of thin double layers, it is often justified to assume that the ions are in thermal equilibrium, if the normal current is not too large and the nearby bulk salt concentration is not too low~\cite{bazant2005,chu2005}, even in the presence of electro-osmotic flow~\cite{rubinstein2001,zaltzman2007}. In terms of electrochemical potentials, the algebraic system
$\{\mu_i=\mbox{constant}\}$ then determines the ion profiles $c_i$ in the
diffuse layer, which lead to effective surface conservation
laws~\cite{chu2007}.  In dilute-solution theory ($\mu^{ex} = 0$), this procedure yields
the Boltzmann distribution, 
\begin{equation}
c_i(\psi) = c_i^0 \exp\left(\frac{-z_i e \psi }{ kT}\right),
\end{equation}
where $\psi=\phi - \phi_b$ is the potential
relative to its bulk value $\phi_b$ just outside the double layer. For a symmetric binary electrolyte ($z_\pm = \pm z$), substituting into (\ref{eq:PoissonMPB}) yields the standard form of the PB equation
\begin{equation}
(\varepsilon \psi^\prime)^\prime = -\rho(\psi) = 2 c_0 z e 
\sinh\left(\frac{ze\psi}{kT}\right),  \label{eq:rhoGC}
\end{equation}
from the Gouy-Chapman model of the double layer~\cite{gouy1910,chapman1913}, where $\varepsilon$ is typically set to a constant bulk value of $\varepsilon_b$ (which we relax below). In that case, by scaling length to $\lambda_D$ and the potential to $kT/ze$, the PB equation takes the simple dimensionless form, $\tilde{\psi}^{\prime\prime} = \sinh \tilde{\psi}$.

In concentrated-solution
theories, the simplest and most common approach is based on the {\it local density approximation} (LDA), where $\mu^{ex}$ depends only on the local ion densities, in the same way as in a homogeneous bulk system~\cite{kilic2007a,biesheuvel2007,grochowski2008,dicaprio2003,dicaprio2004}. The 
choice of a model for $\mu_i^{ex}$ then yields a modified charge density profile $\rho$, differing from (\ref{eq:rhoGC}) with increasing voltage. In this section, we focus on entropic effects in bulk lattice-gas and hard-sphere models, where $\mu_i^{ex}$ depends only on the local ion concentrations, but not, e.g., explicitly on the distance to a wall~\cite{qiao2003,qiao2004} or non-local integrals of the ion concentrations~\cite{zhang1992,zhang1993,tang1994}, as discussed below. 

For any MF-LDA model, the equilibrium charge density can be expressed as a function of the potential, $\rho(\psi)$, but the Boltzmann distribution (\ref{eq:rhoGC}) is modified for non-ideal behavior, as we now demonstrate. Here, we set the permittivity to its bulk value, $\varepsilon=\varepsilon_b=$constant, but below we will extend the derivation for field-dependent dielectric response $\varepsilon(E)$  in section \ref{sec:dielectric}. Given $\rho(\psi)$, we integrate the MPB equation (multiplied by $\psi'$) to obtain the electrostatic pressure, $p_e$ and normal electric field $E_D$ at the inner edge of the diffuse
layer,
\begin{equation}
  p_e(\Psi_D)  =\frac{1}{2}\varepsilon E_D^2 = \int_{\Psi_D}^0  \rho(\psi)d\psi.   \label{eq:E}
\end{equation}
From the total diffuse charge per unit area, 
\begin{equation}
q = -\mbox{sign}(\Psi_D)\sqrt{2\varepsilon p_e(\Psi_D)}, \label{eq:q}
\end{equation}
we then arrive at a general formula for the differential capacitance,
\begin{equation}
  C_D(\Psi_D) = -\frac{dq}{d\Psi_D} =  -\rho(\Psi_D)
  \sqrt{\frac{2\varepsilon}{p_e(\Psi_D)}},  \label{eq:cdgen}
\end{equation}
which reduces to (\ref{eq:cdpb}) in a dilute solution. 

\subsubsection{ The Bikerman-Freise formula }
\label{sec:history}

Since most MPB models are not analytically tractable, we first illustrate
the generic consequences of steric effects using the oldest and simplest mean-field theory~\cite{stern1924,bikerman1942,freise1952,borukhov1997,kilic2007a,kornyshev2007}. This model has a long and colorful history of rediscovery in different communities and countries (pieced together here with the help of P. M. Biesheuvel, Wageningen).   It is widely recognized that O. Stern in 1924~\cite{stern1924} was the first to cutoff the unphysical divergences of the Gouy-Chapman model of the double layer~\cite{gouy1910,chapman1913}  by introducing the concept of a ``compact layer" or ``inner layer" of solvent molecules (and possibly adsorbed ions) forming a thin mono-molecular coating separating an electrode from the ``diffuse layer" in the electrolyte phase. The resulting two-part model of the double layer has since become ingrained in electrochemistry~\cite{bockris_book}. Over the years, however, it has somehow been overlooked that in the same ground-breaking paper~\cite{stern1924}, Stern also considered volume constraints on ions in the electrolyte phase and in his last paragraph, remarkably, wrote down a modified charge-voltage relation [his Eq. (2')] very similar to Eq.~(\ref{eq:qnu}) below, decades ahead of its time.  We have managed to find only one reference to Stern's formula, in a footnote by Freise~\cite{freise1952}.

Although Stern had clearly introduced the key concepts, it seems the first complete MPB model with steric effects in the electrolyte phase was proposed by J. J. Bikerman in 1942, in a brilliant, but poorly known paper~\cite{bikerman1942}. (Bikerman also considered polarization forces on hydrated-ion dipoles in non-uniform fields, discussed below.)  Over the past sixty years, Bikerman's MPB equation has
been independently reformulated by many authors around the
world, including Grimley and Mott (1947) in England~\cite{grimley1947,grimley1950},  Dutta and Bagchi (1950)  in
India~\cite{dutta1950,bagchi1950a,bagchi1950b,dutta1954}, Wicke and Eigen (1951) in Germany
~\cite{eigen1951,wicke1952,eigen1954}, Strating and Wiegel (1993) in the Netherlands~\cite{strating1993a,strating1993b,wiegel1993,wiegel1991}, 
Iglic and Kralj-Iglic (1994) in
Slovenia ~\cite{iglic1994,kralj-iglic1996,bohinc2001,bohinc2002}, and 
Borukhov, Andelman and Orland (1997) in Israel and
France~\cite{borukhov1997,borukhov2000,borukhov2004}.  For an early review of electrolyte theory, which cites papers of Dutta, Bagchi, Wicke and Eigen up to 1954 (but not Bikerman or Freise, discussed below), see Redlich and Jones~\cite{redlich1955}.   

Unlike Bikerman, who applied continuum volume constraints to PB theory, most of these subsequent 
authors derived the same MPB model starting from
the bulk statistical mechanics of ions and solvent molecules on a cubic
lattice of spacing $a$ in the continuum limit (MF-LDA) where the concentration
profiles vary slowly over the lattice. (More recently, ``semi-discrete" lattice-gas models have also been developed, which consist of discrete layers described via mean-field approximations, without taking the continuum limit in the normal direction~\cite{macdonald1987,macdonald1982,macdonald1984,kenkel1984}.)  While early authors were concerned with departures from PB theory in concentrated electrolytes~\cite{bikerman1942,dutta1950,eigen1954} or ionic crystals~\cite{grimley1947,grimley1950}, recent interest in the very same mean-field model proposed by Bikerman has been motivated by a wide range of modern applications involving electrolytes with large ions or biological molecules~\cite{wiegel1993,borukhov1997,borukhov2000,borukhov2004}, polyelectolytes~\cite{biesheuvel2004,israels1994,gonzalez2004},  polymeric electrolytes~\cite{soestbergen2008}, electromagnetic waves in electrolytes~\cite{baker1999}, electro-osmosis in nanopores~\cite{cervera2001,cervera2003}, electrophoresis of colloids~\cite{lopez2007mpb,lopez2008mpb,aranda2009a,aranda2009b}, and solvent-free ionic liquids ~\cite{kornyshev2007,federov2008,oldham2008}, in addition to our own work on dilute electrolytes in large applied voltages ~\cite{kilic2007a,kilic2007b,large_new}.

\begin{figure}
(a)\includegraphics[width=2.4in]{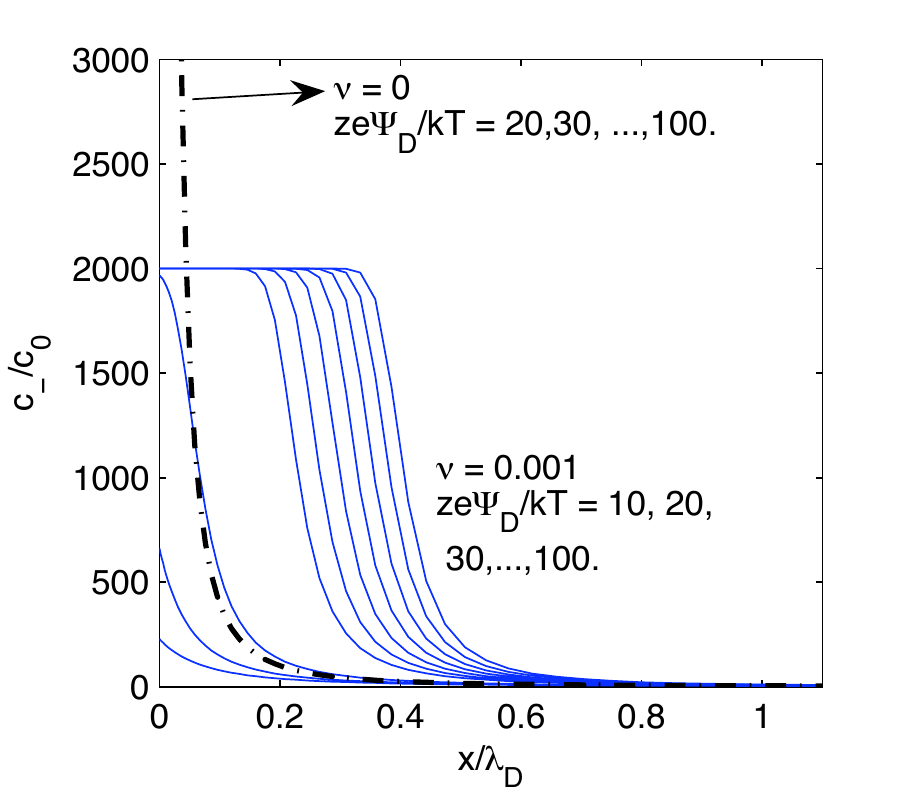}
(b)\includegraphics[width=2.7in]{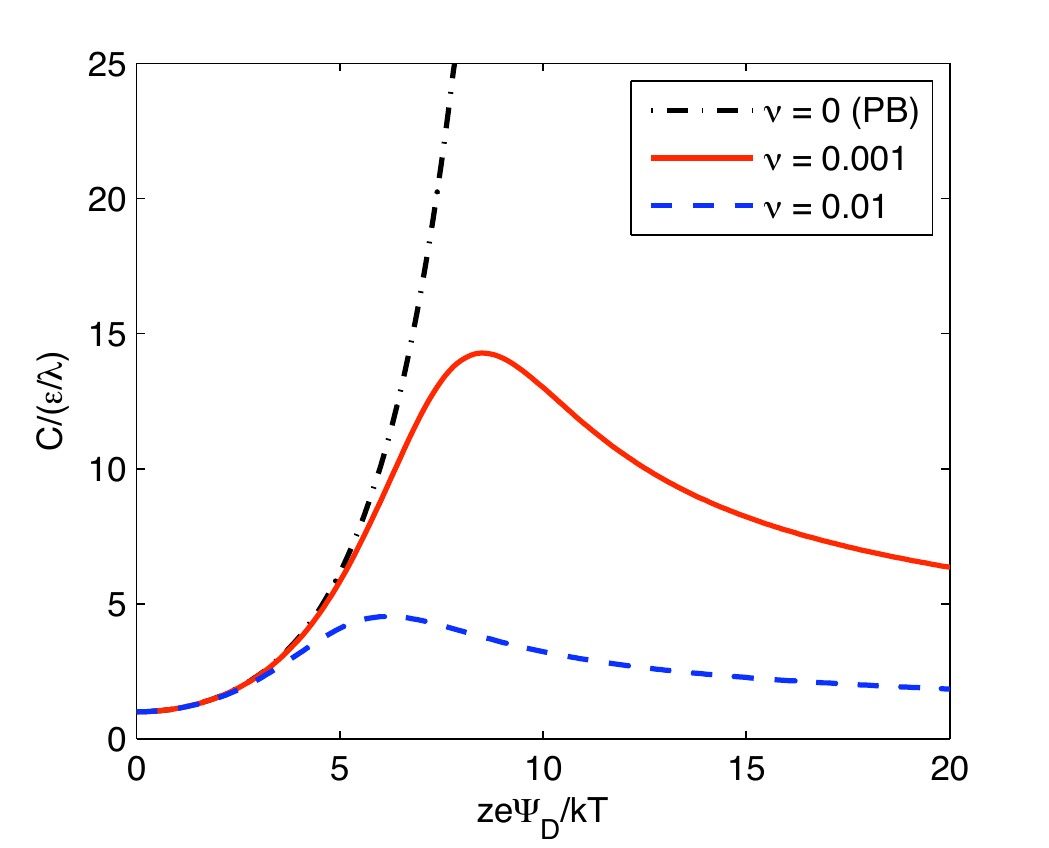}
\caption{\label{fig:C} (a) The equilibrium distribution of counterions
  in a flat diffuse layer for large applied voltages
  $ze\Psi_D/kT=10,20,\ldots, 100$ predicted by Poisson-Boltzmann
  theory (PB) and Bikerman's modified theory (MPB) taking into account
  an effective (solvated) ion size $a$, where $\nu = 2a^3 c_0 = 0.001$ is the bulk
  volume fraction of solvated ions.  (b) The diffuse-layer differential
  capacitance $C_D$ vs. voltage predicted by PB (\ref{eq:cdpb}) ($\nu=0$) and
  MPB (\ref{eq:cdnu}) ($\nu > 0$), scaled to the low-voltage
  Debye-H\"uckel limit $\varepsilon/\lambda_D(c_0)$.   }
\end{figure}

In the present terminology, Bikerman's model corresponds to an excess
chemical potential 
\begin{equation}
\mu_i^{ex} = - kT\, \ln(1- \Phi) \ \ \mbox{ (Bikerman) },  \label{eq:bike}
\end{equation}
associated with the entropy of the solvent, where $\Phi = a^3 \sum_i
c_i$ is the local volume fraction of solvated ions on the
lattice~\cite{biesheuvel2007}. For now, we also assume a symmetric
binary electrolyte, $c_+^0 = c_-^0 = c_0$, $z_\pm = \pm z$, to obtain
an analytically tractable model. As shown in Fig.~\ref{fig:C}(a), when
a large voltage is applied, the counterion concentration exhibits a
smooth transition from an outer PB profile to a condensed layer at
$c=c_{max}=a^{-3}$ near the surface. Due to the underlying lattice-gas model for excluded volume, the ion profiles 
effectively obey Fermi-Dirac statistics,
\begin{equation}
c_\pm = \frac{c_0 e^{\mp ze\psi/kT}}{1 + 2\nu \sinh^2(ze\psi/2kT)}, \label{eq:cMPB}
\end{equation}
where $\nu = 2a^3c_0 = \Phi_{bulk}$ is the bulk volume fraction of
solvated ions.  Classical Boltzmann statistics and the Gouy-Chapman PB 
model are recovered in the limit of point-like ions, $\nu=0$. 

For a flat double layer, similar results can be obtained with the even
simpler Composite Diffuse Layer model of Kilic et al.~\cite{kilic2007a} (also termed the ``cutoff 
model'' in
Ref.~\cite{grochowski2008}), where an outer PB diffuse layer is
abrubtly patched with an inner condensed layer of only counterions at the uniform, maximal
charge density. This appealingly simple construction requires assumptions about the shape of the condensed layer (e.g. a plane), so its position can be determined only from its thickness or total charge. Even if it can be uniquely defined, the cutoff model introduces discontinuities in the co-ion concentration (which drops to zero in the condensed layer) and in the gradient of the counter-ion concentration, although the same is also true of Stern's original model of the compact layer.  In this work we focus on Bikerman's model since it is the simplest general model of
steric effects that remains analytically tractable; unlike the cutoff
model, it predicts smooth ionic concentration profiles in any
geometry and can be naturally extended to time-dependent
problems~\cite{kilic2007b,baker1999}.

Substituting the equilibrium ion distributions  (\ref{eq:cMPB}) into Poisson's equation (\ref{eq:PoissonMPB}), we obtain Bikerman's MPB equation, 
\begin{equation}
(\varepsilon \psi^\prime)^\prime = \frac{ 2 c_0 z e 
\sinh(ze\psi/kT) } { 1 + 2\nu \sinh^2(ze\psi/2kT) },  \label{eq:BikeMPB}
\end{equation}
which has the simple dimensionless form  (scaling length to $\lambda_D$, potential to $kT/ze$), 
\begin{equation}
\tilde{\psi}^{\prime\prime} = \frac{ \sinh \tilde{\psi} }{ 1+2\nu \sinh^2(\tilde{\psi}/2)}, 
\end{equation}
extending the PB equation for a nonzero volume fraction $\nu$ of ions in the bulk solution.
Integrating across the diffuse layer, the charge-voltage relation (\ref{eq:q}) takes the form,
\begin{equation}
q_\nu = \mbox{sgn}(\Psi_D) 2ze c_0\lambda_D \sqrt{\frac{2}{\nu} \ln \left[ 1 + 2\nu \sinh^2\left(\frac{ze\Psi_D}{2kT}\right)\right]}  \label{eq:qnu}
\end{equation}
which was probably first derived by Grimley~\cite{grimley1950} in a lattice-gas theory of diffuse charge in ionic crystals~\cite{grimley1947}, independent of Bikerman. Grimley's formula has the same form as Stern's surprising Eq. (2') noted above~\cite{stern1924}, but it has all the constants correct and clearly derived. Recently, Soestbergen et al.\cite{soestbergen2008} have given a slightly different formula approximating (\ref{eq:qnu}) that is easier to evaluate numerically for large voltages, and applied it to ion transport in epoxy resins encapsulating integrated circuits.

Although Stern and Grimley derived the modified form of the charge-voltage relation with volume constraints, they did not point out its striking qualitative differences with Gouy-Chapman dilute-solution theory, noticed by several recent authors~\cite{kilic2007a,kornyshev2007}. This important aspect was apparently first clarified by Freise~\cite{freise1952}, who took the derivative of (\ref{eq:qnu}) and derived the differential capacitance (\ref{eq:cdgen}) in the form
\begin{equation}
C_{D}^{\nu }=
\frac{\frac{\varepsilon}{\lambda_{D}}\sinh\left(\frac{ze|\Psi_{D}|}{kT}\right)}
{[1+2\nu\sinh^2\left(\frac{ze\Psi_{D}}{2kT}\right)]
\sqrt{\frac{2}{\nu}\ln \left[1+2\nu\sinh^2\left(\frac{ze\Psi_{D}}{2kT}\right)\right]}}.
\label{eq:cdnu}%
\end{equation}
and pointed out that $C_D^{\nu}$ decays at large voltages. This elegant formula, also derived by Kilic et al.~\cite{kilic2007a} and Kornyshev~\cite{kornyshev2007} (and Oldham~\cite{oldham2008} for the ionic-liquid limit $\nu=1$), predicts the opposite dependence of Chapman's formula (\ref{eq:cdpb}) from dilute-solution theory, which  diverges exponentially with $|\Psi_D|$. Since Chapman~\cite{chapman1913} is given credit in ``Gouy-Chapman theory"  for first deriving the capacitance formula (\ref{eq:cdpb}) for Gouy's original PB model~\cite{gouy1910}, we suggest calling Eq. ~(\ref{eq:cdnu})
the ``Bikerman-Freise formula" (BF), in honor of Bikerman, who first postulated the underlying MPB theory, and Freise, who first derived and interpreted the modified differential capacitance. By this argument,  it would be reasonable to also refer to the general MPB model as "Bikerman-Freise theory", but we will simply call it "Bikerman's model" below, following Refs.~\cite{cervera2001,cervera2003,biesheuvel2007}.

According to the BF formula (\ref{eq:cdnu}), as shown in Fig.~\ref{fig:C}(b), {\it the 
differential capacitance increases following PB theory
up to a maximum near the critical voltage $\Psi_D \approx \Psi_c$, and
then decreases as the square-root of the voltage},
\begin{equation}
  C_D^\nu \sim \sqrt{ \frac{ze\varepsilon_b}{2a^3 |\Psi_D|}},  \label{eq:Clarge}
\end{equation}
because the effective capacitor width grows due to steric effects, as
seen in Fig.~\ref{fig:C}(a). 
In stark contrast, the PB
diffuse-layer capacitance diverges exponentially according to
Eq. (\ref{eq:cdpb}), since point-like ions pile up at the surface.
Although other effects,
such as compact layer compression and specific adsorption of ions~\cite{macdonald1987,bockris_book} (discussed
below) can cause the differential capacitance to increase at intermediate voltages,
this effect is quite general. As long as the surface continues to
block Faradaic current, then the existence of steric volume
constraints for ions implies the growth of an extended condensed layer at
sufficiently large voltages, and a concomitant, universal decay of the
differential capacitance. Indeed, this effect can be observed for
interfaces with little specific adsorption, such as NaF and KPF$_6$ on
Ag \cite{valette1981,valette1982} or Au \cite{hamelin1982,hamelin1987,tymosiak2000}. 
(See Fig.~\ref{fig:valette} below for fits to simple mean-field models.) The same square-root 
dependence at large voltage can 
also be observed in experiments~\cite{kornyshev2007} and simulations~\cite{federov2008,federov2008b} of ionic liquids at 
blocking electrodes, with remarkable accuracy. We conclude that the decay of the double-layer 
differential capacitance at large voltage is a universal 
consequence of the crowding of finite-sized, mobile charge carriers near a highly charged, blocking surface.

\begin{figure}
\begin{center}
\includegraphics[width=3.5in]{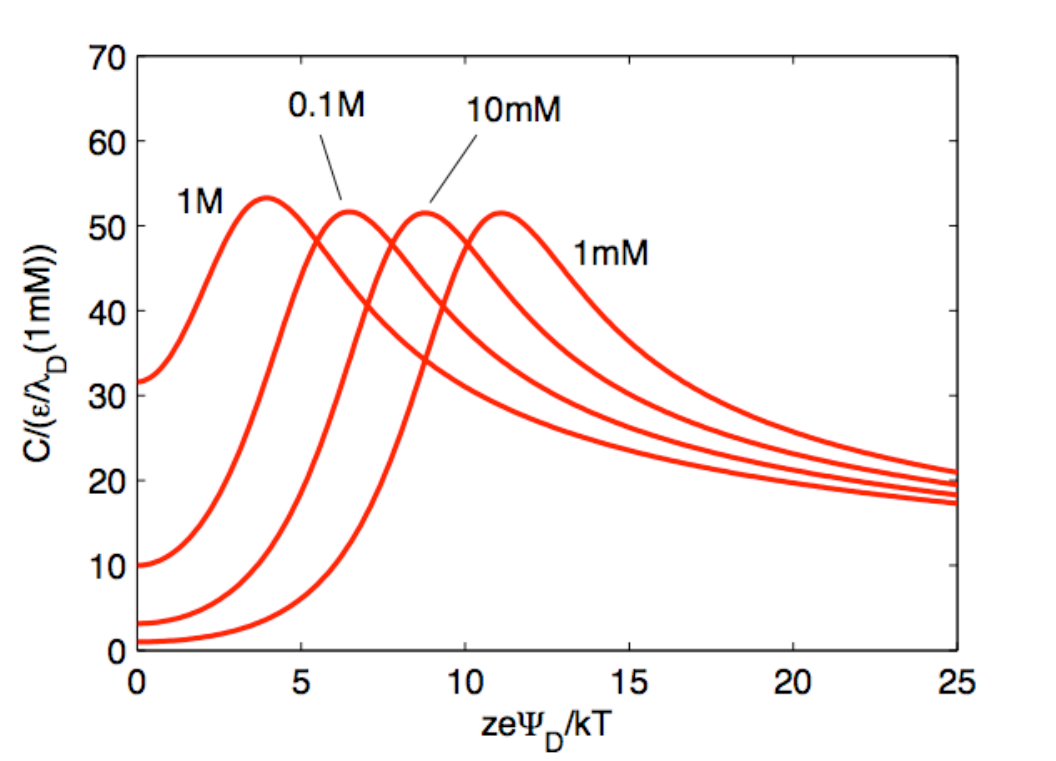}
\caption{ \label{fig:Cmolar} Differential capacitance $C_D$ vs. voltage in Bikerman's MPB model (\ref{eq:cdnu}) with $a=4$ \AA for $\nu$ values corresponding to $c_0 = 1, 10 ,100$ mM. In contrast to Fig.~\ref{fig:C}(b), here $C_D$ is scaled to a single constant,  $\varepsilon/\lambda_D(1 \mbox{mM})$, for all concentrations. }
\end{center}
\end{figure}

The BF formula (\ref{eq:Clarge}) also illustrates
another general feature of double-layer models with steric
constraints, shown in Fig.~\ref{fig:Cmolar}: {\it The differential
capacitance at large voltages is independent of bulk salt
  concentration, but ion specific} through $z$ and $a$. This
prediction is reminiscent of Stern's picture
picture~\cite{stern1924} of an inner, compact layer of solvent molecules 
which carries the majority of a large double layer voltage, 
compared the outer, diffuse layer described by
dilute PB theory.  This picture is ubiquitous in electrochemistry~\cite{bockris_book}, and first gained acceptance based on Grahame's famous experiments on
mercury drop electrodes~\cite{grahame1947,macdonald1962,macdonald1987}. The significant difference,
however, is that the condensed layer forms continuously in
the solution near the inner edge of the diffuse layer due to ion
crowding effects in a general model of the electrolyte phase, which is
not restricted to flat quasi-equilibrium double layers. 


\subsubsection{ Hard-sphere liquid models }

Although Bikerman's model describes steric effects in a convenient and
robust analytical form, the bulk ionic volume fraction $\nu$ is best
viewed as an empirical fitting parameter.  For crystalline solid
electrolytes, its microscopic basis in a lattice model is realistic,
but even then, the thinness of the condensed layer, comparable to the
lattice spacing at normal voltages, calls the continuum limit into
question. For liquid electrolytes involved in electrokinetic
phenomena, it would seem more realistic to start with the ``restricted
primitive model'' of charged hard spheres in a uniform dielectric
continuum~\cite{blum1990} in developing better MPB
models~\cite{lue1999,dicaprio2003,biesheuvel2007}. From this
theoretical perspective, Bikerman's lattice-based model has the
problem that it grossly under-estimates steric effects in hard-sphere
liquids; for example, in the case of a monodisperse hard-sphere
liquid, the volume excluded by a particle is eight times its own
volume ~\cite{porkess1977,sparnaay1958,biesheuvel2007}. Although we
focus on electrolytes at large voltages, it is also interesting to
consider the MF-LDA dynamics of charged hard spheres to model
other systems, such as dense 
colloids~\cite{dufreche1999,biesheuvel2005}, polyelectrolytes~\cite{biesheuvel2008poly,devos2008}, and ionic
liquids~\cite{kornyshev2007,federov2008}.

\begin{figure*}
(a) \includegraphics[width=2.5in]{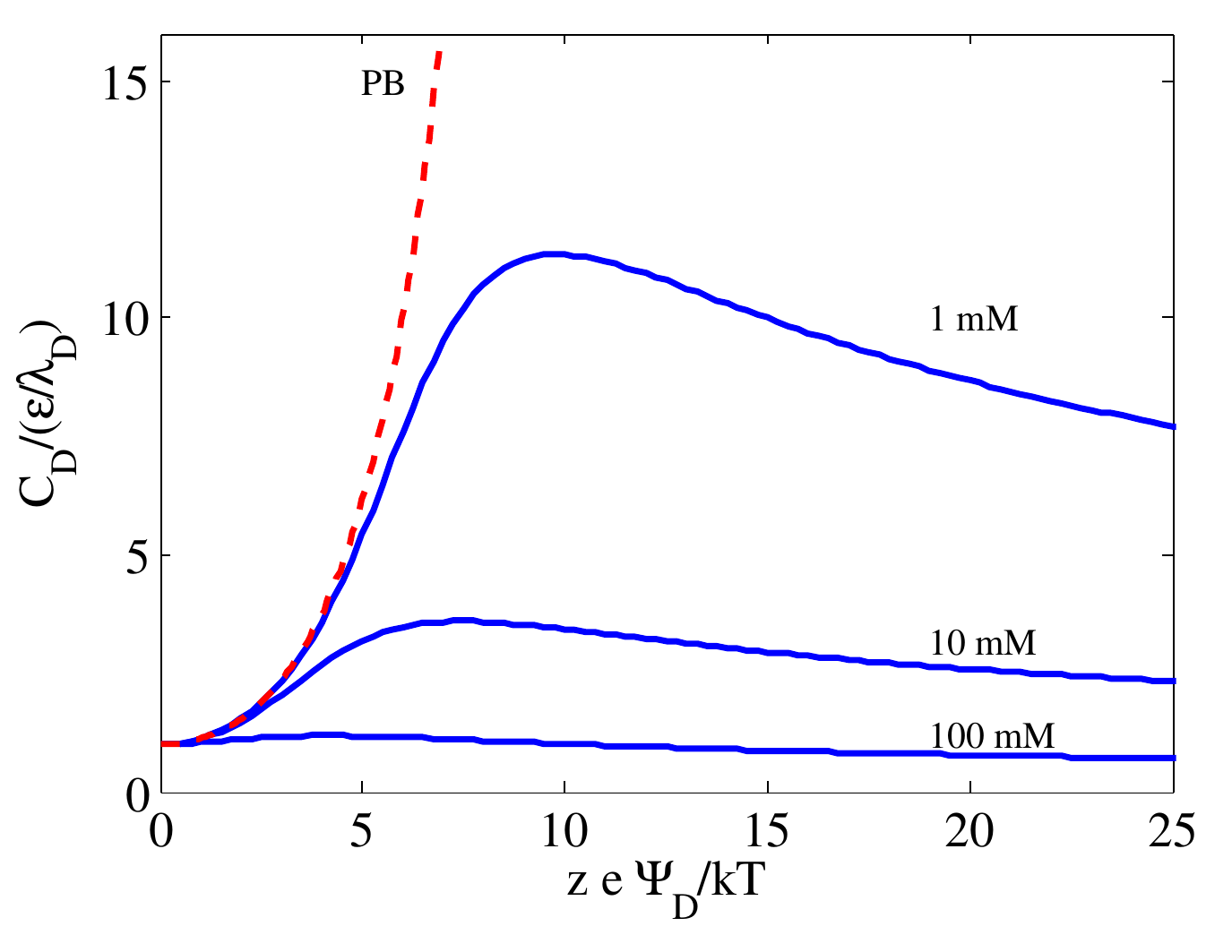}
(b) \includegraphics[width=2.5in]{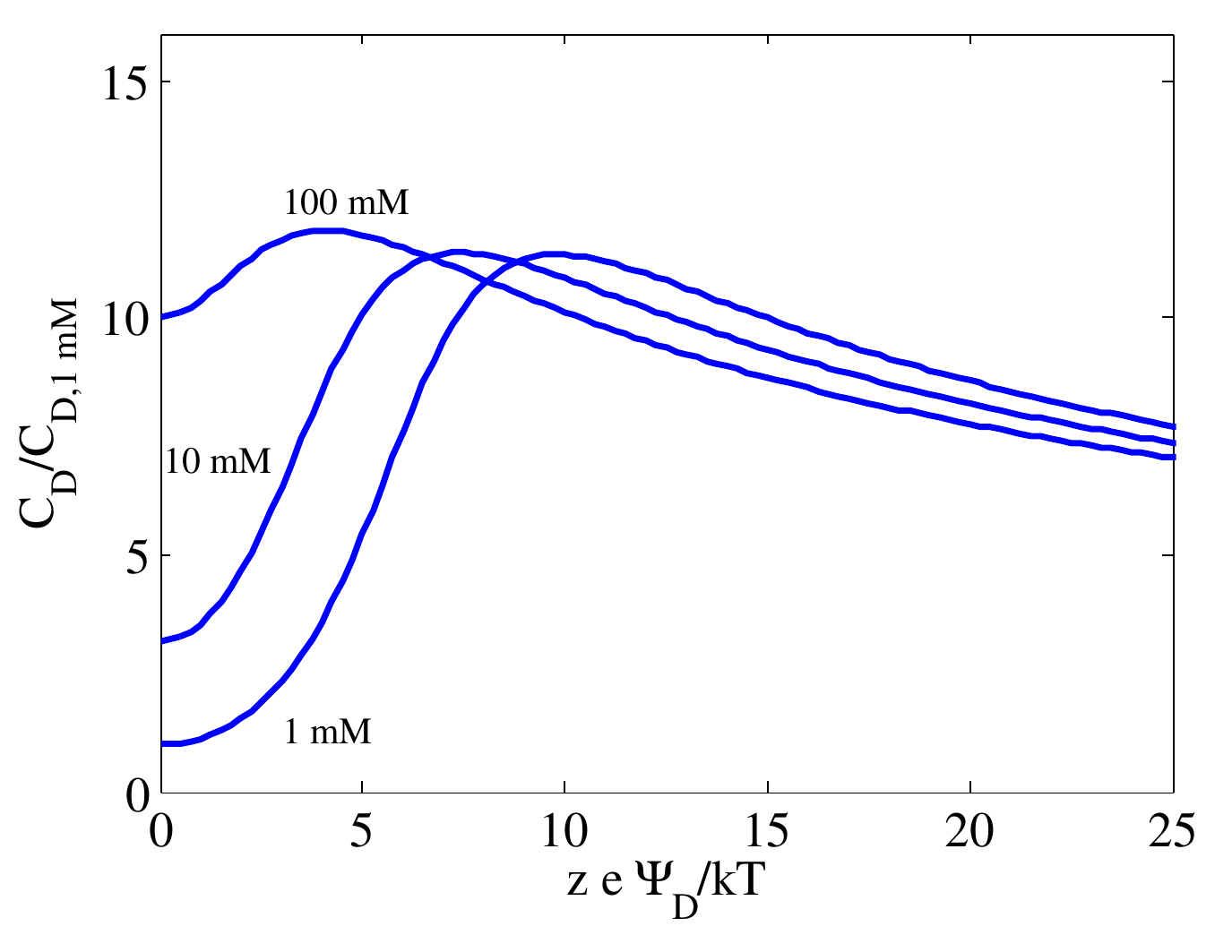}\\
(c) \includegraphics[width=2.5in]{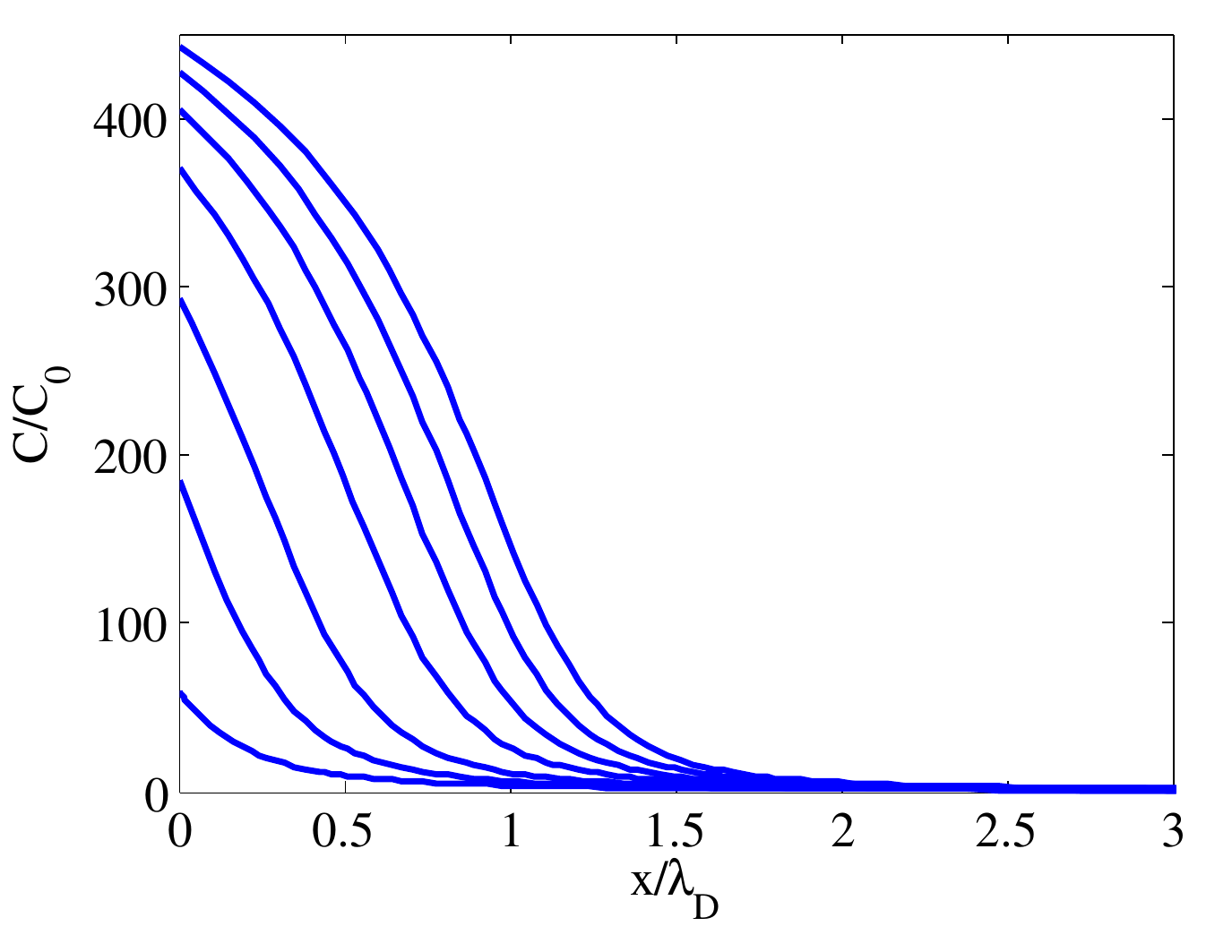}
(d) \includegraphics[width=2.5in]{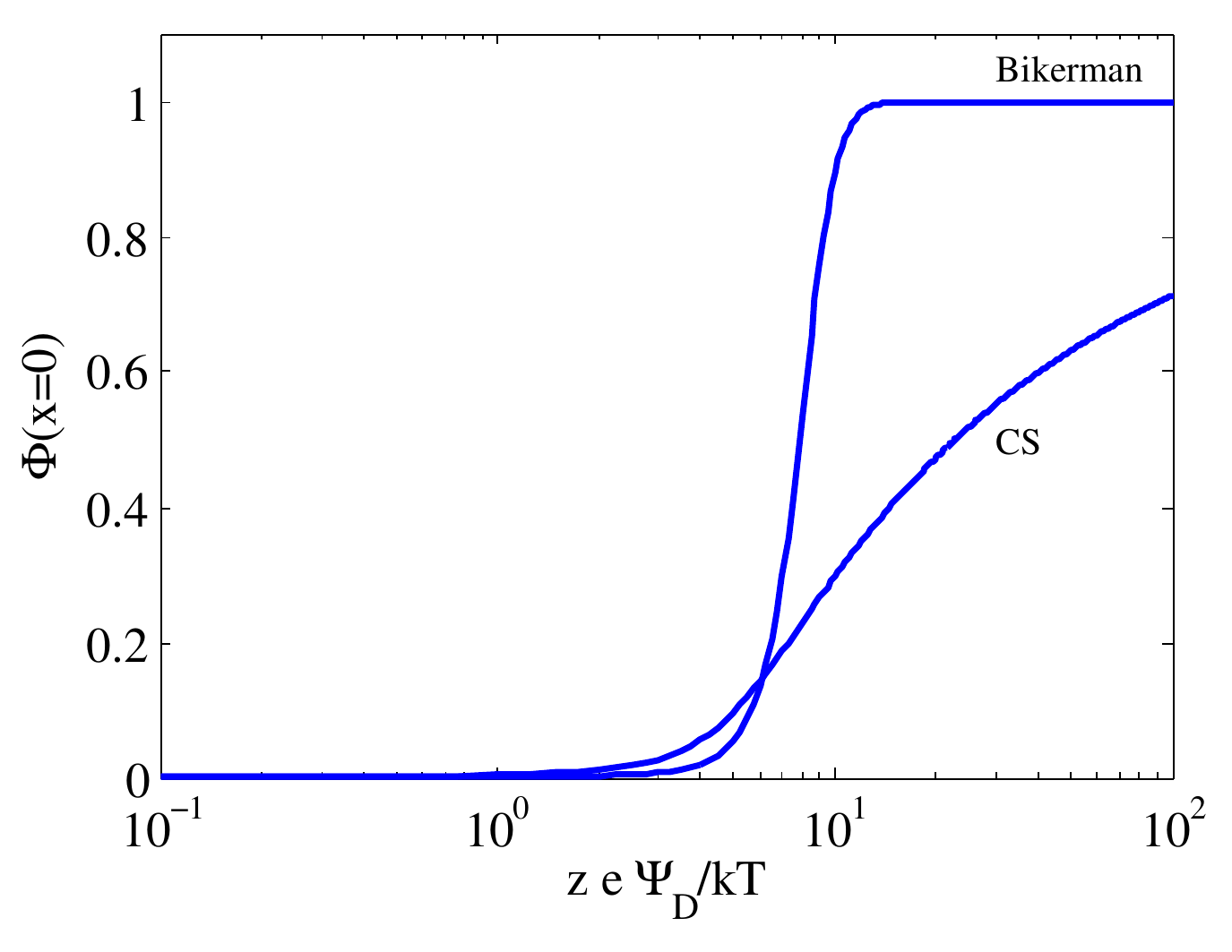}
\caption{\label{fig:CS} Modified Poisson-Boltzmann theory for a binary
  solution of charged hard spheres of diameter $a=4$ \AA using the
  Carnahan-Starling (CS) equation of state (\ref{eq:CS}). (a,b) The
  diffuse-layer differential capacitance vs. voltage, analagous to
  Fig.~\ref{fig:C} b and Fig. ~\ref{fig:Cmolar}, respectively. (c) The counterion density
  profile in the diffuse layer at voltages
  $ze\Psi_D/kT=5,10,20,40,60,80,100$ and concentration of $c_o = 10$ mM, analogous to
  Fig.~\ref{fig:C}(a). (d) The surface counterion density vs. voltage
  at $c_o = 10$ mM in the CS and Bikerman MPB models.  }
\end{figure*}

Various MF-LDA approximations of $\mu_i^{ex}$ for bulk hard-sphere liquids can be
used to develop more sophisticated MPB models, which yield
similar qualitative behavior of the diffuse-layer differential
capacitance~\cite{dicaprio2003,biesheuvel2007}, due to the generic
arguments given above. For example, the Carnahan-Starling (CS)
equation of state for a bulk monodisperse hard-sphere liquid
corresponds to the following excess chemical
potential~\cite{carnahan1969,russel_book},
\begin{equation}
  \frac{\mu_i^{ex}}{kT}=  \frac{\Phi(8-9\Phi+3\Phi^2)}{(1-\Phi)^3} \ \ \mbox{
    (Carnahan-Starling)}  \label{eq:CS}
\end{equation}
Although this algebraic form precludes analytical results, it is much
simpler to evaluate numerically and incorporate into continuum models
of electrokinetic phenomena than are more sophisticated MPB
approximations which go beyond the LDA, e.g.  based on self-consistent correlation
functions~\cite{outhwaite1980,outhwaite1983,bhuiyan2004,bhuiyan2005}
or density functional
theory~\cite{henderson_book,gillespie2003,reszko2005}, which require
solving nonlinear integro-differential equations, even for a flat
double layer in equilibrium. As shown in Fig.~\ref{fig:CS}(b), the
simple CS MPB model predicts capacitance curves similar to
Fig.~\ref{fig:Cmolar} with Bikerman's model, respectively, only with
more realistic salt concentrations~\cite{biesheuvel2007}. In
particular, the differential capacitance in Bikerman's model
ressembles that of CS MPB if an unrealistically large hydrated ion
size $a$ (or large bulk volume fraction $\nu$) is used, due to the
under-estimation of liquid steric effects noted above.

In spite of similar-looking capacitance curves, however, there are
important differences in the ionic profiles predicted by the two
models. As shown in Fig~\ref{fig:C}(a), in Bikerman's model steric
effects are very weak until the voltage becomes large enough to form a
thin condensed layer at maximum packing.
 As such, the width of the diffuse layer at
typical large voltages is still an order of magnitude smaller than the
Debye length $\lambda_D$ relevant for small voltages. In contrast,
steric effects in a hard-sphere liquid are  stronger and cause the
diffuse layer to expand with voltage as shown in
Fig.~\ref{fig:CS}(c). The widening of the diffuse layer reduces its
differential capacitance, but without forming the clearly separated
condensed layer predicted by Bikerman's model. As shown in
Fig.~\ref{fig:CS}(d), the counterion density at the surface in the CS
MPB model increases more slowly with voltage as compared to the Bikerman
model. These differences will be important when we discuss the viscosity 
effects in Section 4. 

An advantage of the hard-sphere approach to volume constraints is that it has a simple extension to mixtures of unequal particle  sizes~\cite{hansen_book} which can be applied
to general multicomponent electrolytes~\cite{lue1999,biesheuvel2007,dicaprio2003,devos2008,biesheuvel2008poly}.   According to the Boublik-Mansoori-Carnahan-Starling-Leland (BMCSL) equation of state~\cite{boublik1970,mansoori1971}, the excess chemical potential of species $i$ in a mixture of $N$ species of hard spheres with different diameters $\{a_i\}$ is given by
\begin{eqnarray}
\frac{\mu_i^{ex}}{kT} &=& -\left(1 + \frac{2 \xi_2^3a_i^3}{\Phi^3} - \frac{3 \xi_2^2 a_i^2}{\Phi^2}\right)\ln(1-\Phi)
+ \frac{3\xi_2 a_i + 3 \xi_1 a_i^2 \xi_0 + a_i^3}{1-\Phi} \nonumber \\
&+& \frac{3 \xi_2 a_i^2}{(1-\Phi)^2}\left(\frac{\xi_2}{\Phi} + \xi_1a_i\right)
- \xi_2^3 a_i^3 \frac{\Phi^2 - 5 \Phi + 2}{\Phi^2(1-\Phi)^3} \ \ \ \ \  \mbox{(BMCSL)}   \label{eq:BMCSL}
\end{eqnarray}
where $\xi_n = \sum_{j=1}^N \Phi_j a_j^{n-3}$,  $\Phi_j$ is the volume fraction of species $j$, and $\Phi = \sum_{j=1}^N\Phi_j$ is the total volume fraction of ions. Although this formula may seem complicated, it is an algebraic expression that can be easily expanded or evaluated numerically and thus is much simpler than statistical theories based on integral equations~\cite{hansen_book,vlachy1999,attard1996}. The first BMCSL correction to dilute solution theory is simply,
\begin{equation}
\frac{\mu_i^{ex}}{kT} \sim \sum_{j=1}^N \left(1 + \frac{a_i}{a_j}\right)^3 \Phi_j
 \end{equation}
The BMCSL MPB model for asymmetric electrolytes predicts the segregation of ions of different size and/or charge in the diffuse layer~\cite{biesheuvel2007} and has been applied to adsorption phenomena in polyelectrolyte layers~\cite{devos2008,biesheuvel2008poly}. The broken
symmetry between ions of different sizes is an important qualitative effect, which we will show 
implies new electrokinetic phenomena at large voltages,
regardless of the model.

{\it A word of caution: } In spite of its mathematical convenience, the local-density approximation is known to provide a poor description of confined hard-sphere liquids, even in equilibrium. For example, it cannot capture density oscillations near a wall or two-point correlation functions of hard spheres. These features can be approximated by various {\it weighted density approximations} (WDA) in statistical mechanics, which redefine the  local reference densities $\bar{c}_i$ as averages over the inhomogeneous densities $c_i$ with a suitable weight function~\cite{tarazona1985,tarazona1987,curtin1985,meister1985,groot1987,levin2002}. The non-local weighted densities are used in place of the local densities in chemical potential expressions for homogeneous systems, such as (\ref{eq:CS}) and (\ref{eq:BMCSL}) for hard spheres, and several prescriptions for the weight function are available. (See section ~\ref{sec:general}.) Non-local MF-WDA models have been applied to electrolytes to quantify excluded volume effects involving ions and solvent molecules~\cite{tang1992a,zhang1992,zhang1993,tang1994}. Recent Monte-Carlo simulations of hard-sphere counterion profiles around a hard-sphere macro-ion have shown that LDA can perform even worse than dilute-solution PB theory, while WDA theories are able to fit the simulations well~\cite{antypov2005}. Even the simplest WDA, however, is a non-local continuum theory and thus requires solving nonlinear integral equations for the equilibrium densities, and integro-partial differential equations for time evolution. Clearly, LDA-based partial differential equations are much better suited for mathematical modeling, if they can capture enough of the essential physics in a given situation.

\subsubsection{ Interpretation of the effective ion size } 

In order to apply our modified electrokinetic equations, 
we stress that the effective diameter of a 
solvated ion in various MF-LDA theories is not simply related to its bare atomic size and can exhibit very  
different trends. Smaller bare ions tend to be more heavily solvated
and therefore have larger effective diameters \cite{stokes_book}. 
Effective solvated ion sizes  depend on the size and charge  of the ions, the nature of the solvent, 
the ion concentration, and temperature -- as well as the mathematical models used in their definitions. 
Table \ref{table:ionsize} compares bare ionic diameters in crystalline solids   
to effective solvated-ion diameters inferred from bulk properties \cite{nightingale1959}
and ``hard-sphere" diameters inferred from viscosity measurements \cite{gering2006}, both in aqueous solutions.
In these models used to interpret experimental data, the hard sphere radius is essentially a collision size, whereas the 
the effective solvated radius is an effective size for transport properties, similar to a Stokes 
radius.  The effective solvated radius is generally larger than the hard sphere value. Both are greatly exceed the bare diameter and exhibit roughly opposite trends with the chemical identity of the ion. 

\begin{table*}
\begin{center}
\begin{tabular}{|c|ccc|}
	\hline
Ion &  $d_x$  (\AA)  &   $d_s$ (\AA)  & $d_v$  (\AA)  \\
	\hline
Li$^+$   &  1.20 &  7.64 &  4.2   \\
Na$^+$   &  1.90  &  7.16 &  4.0  \\
K$^+$    &  2.66  & 6.62 & 3.8  \\
Cl$^-$   &  3.62  & 6.64 & 3.6   \\
	\hline
\end{tabular}
\hspace{.5in}  \includegraphics[height=1in]{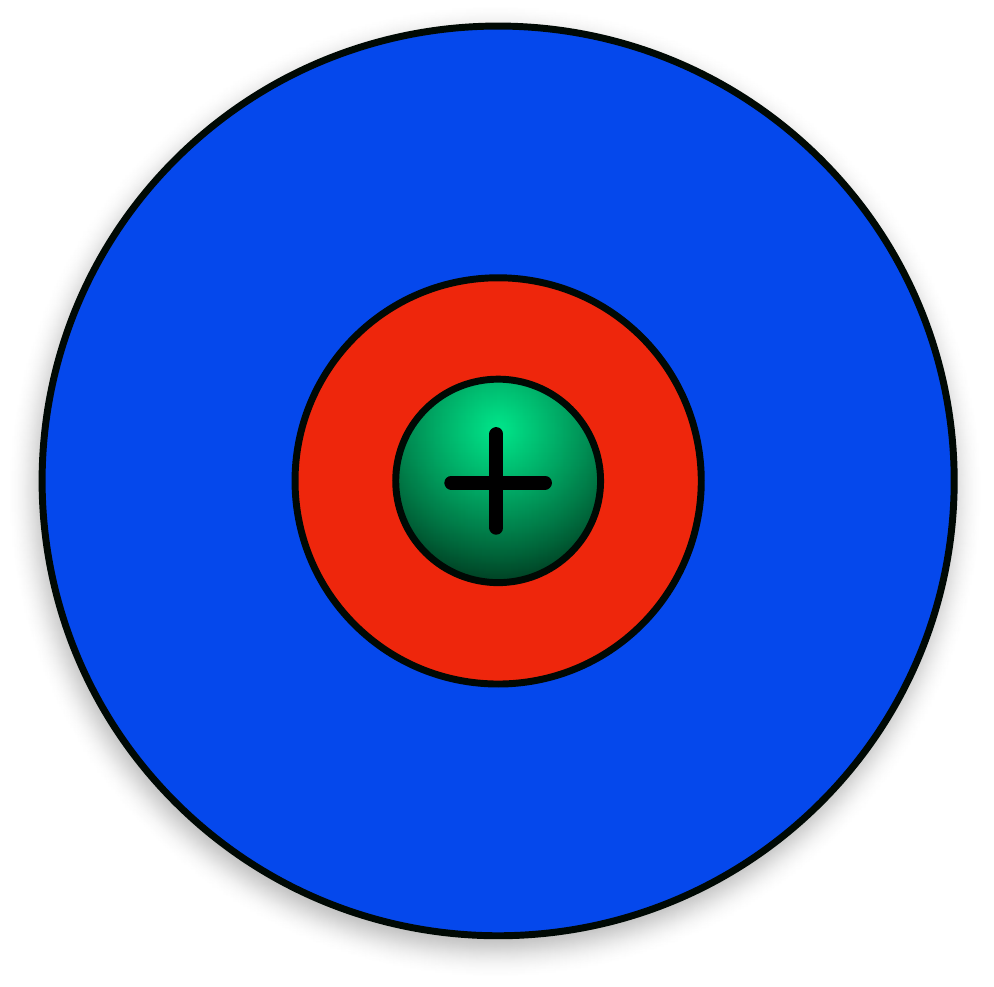}
\caption{ \label{table:ionsize} Comparison of the bare ion diameter in a crystalline solid,  
$d_x$, with the effective solvated diameter $d_s$ in water from bulk transport measurements (akin to a Stokes diameter)  \cite{nightingale1959} and the ``hard-sphere" diameter (akin to a collision cross section) inferred from viscosity data $d_v$ in dilute aqueous solution  \cite{gering2006} for some common ions used in nonlinear electrokinetic experiments.  The figure depicts an ion with its effective hard-sphere and solvation shells, in red and blue respectively. Note that the effective sizes $d_s$ and $d_v$ in solution are much larger than the bare ion size $d_x$ and exhibit different trends. In the text, we argue that the appropriate effective ion size  $a$ in our models of highly charged double layers may be approximated by $d_s$, and possibly larger. 
}
\end{center}
\end{table*}

What is the appropriate effective ion size $a$? Unlike the models used to infer the various ion diameters in Table ~\ref{table:ionsize}, our models seek to capture crowding effects in a highly charged double layer, rather than in a neutral bulk solution (albeit still using MF-LDA theories). As such, it is important to think of crowded counterions of the same sign and not a neutral mixture of oppositely charged ions (where our models reduce to the Standard Model in typical situations with dilute electrolytes). Below, we will argue that the crowding of like-charged counterions in large electric fields leads to some different physical effects, not evident in the neutral bulk liquid. Among them, we can already begin to discuss solvation effects. In the bulk, ions cannot reach very high concentrations due to solubility limits, but a condensed layer of counterions cannot recombine and is unaffected by solubility (except for the possibility of electron transfer reactions near the surface).  Moreover, like-charged ions cannot easily ``share" a solvation shell and become compressed to the hard-sphere limit, since the outer surfaces of the polarized solvation shells have the same sign and yield electrostatic repulsion. Therefore, we propose that the ``ion size" in our models is an effective solvated ion size at high charge density, which is much larger than the bare crystalline and hard-sphere ion sizes in Table~\ref{table:ionsize} and may also exceed the solvated ion size inferred from bulk transport models. This physical intuition is borne out by the comparisons between theory and experiment below for nonlinear electrokinetics, although we will not claim to reach any quantitive molecular-level conclusions.

\subsubsection{ Comparison with experiments on blocking surfaces without ion adsorption }

\begin{figure*}
\begin{center}
\includegraphics[height=2in]{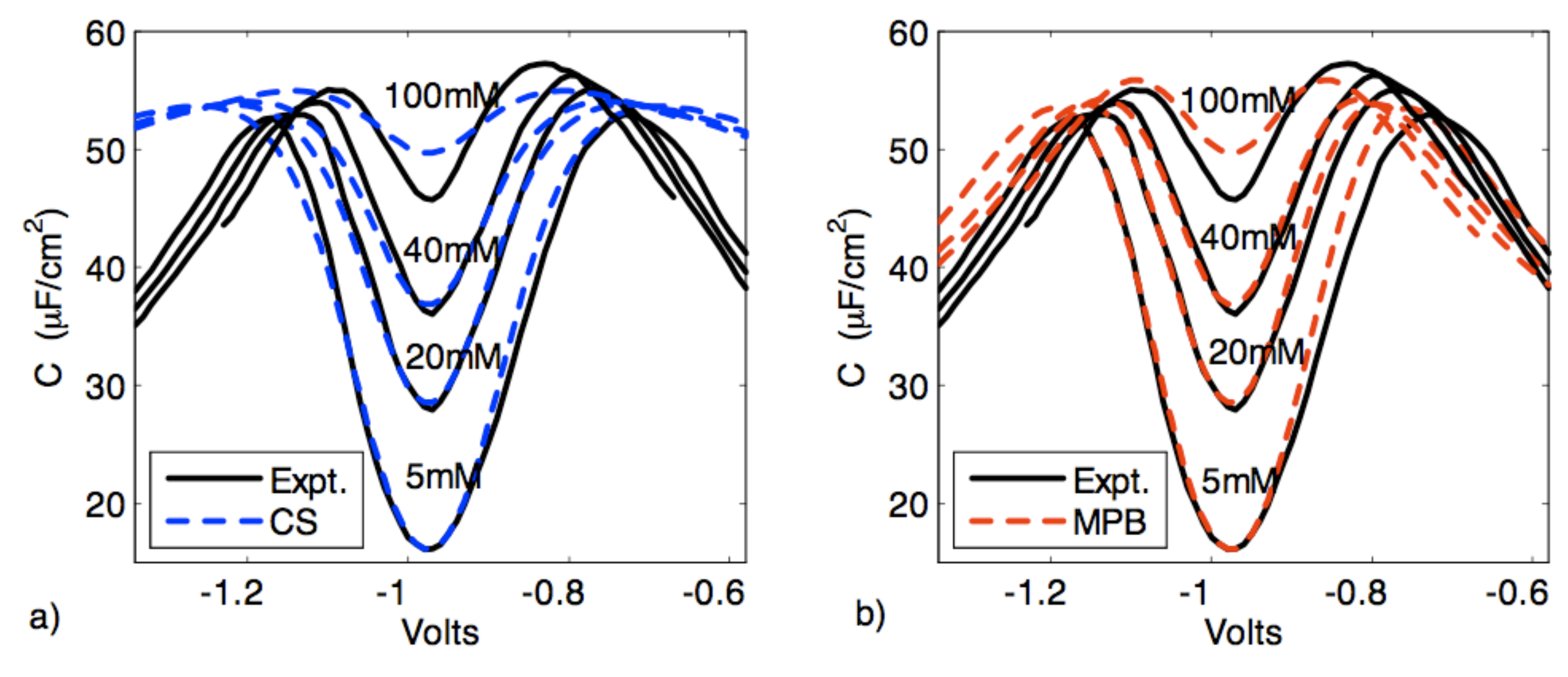}
\caption{\label{fig:valette}  
Fitting of simple MF-LDA models of finite-sized ions to the experimental data of Valette (Fig. 3 of Ref.~\cite{valette1981}) for the differential capacitance of the aqueous KPF$_6\ |$ Ag double layer. For all curves, the electrode area is scaled by a surface roughness factor $R = 1.1$, and the theoretical diffuse-layer capacitance $C_D$ from (\ref{eq:cdgen}) is taken in series with a constant Stern-layer capacitance $C_S = 125 \mu$F/cm$^2$. The only other adjustable parameter in each panel is the effective solvated ion size $a$.  In (a), we use a hard-sphere radius of $a=4$ \AA in the Carnahan-Starling model (\ref{eq:CS}), while in (b) we use a lattice spacing of $a=11$ \AA in Bikerman's MPB model (\ref{eq:bike})-(\ref{eq:cdnu}).
 }
 \end{center}
\end{figure*}

To illustrate these points, in Fig.~\ref{fig:valette} we fit our simple mean-field models to experimental data of Valette~\cite{valette1981} for the differential capacitance of a standard electrochemical interface with negligible surface adsorption of ions: aqueous KPF$_6$ solution with single-crystal Ag electrodes. These experiments and others were cited by Di Caprio et al.~\cite{dicaprio2003,dicaprio2004} as motivation to develop mean-field hard-sphere models for double-layer capacitance, but they concluded that ``a quantitative comparison with the experimental results is not possible at this stage".  Here, we proceed naively to directly test the predictions of two simple theories. To model the liquid electrolyte, we use the simplest CS and Bikerman MPB models, each of which has only one adjustable parameter, the effective ion size $a$. To model the surface (and admittedly, to aid in the fitting), we introduce two standard fitting parameters:  (i) a  "surface roughness factor" $R_S$ multiplying the nominal electrode area, and  (ii) a Stern-layer capacitance $C_S$ in series with the diffuse-layer capacitance, ostensibly to model the dipolar electrode solvation layer, since we already take into account the crowding of finite-sized solvated ions. Unlike Valette~\cite{valette1981}, who adjusts $R_S$ for each measurement, we choose a fixed value of $R_S=1.1$ for all the data, since this parameter should reflect fixed surface profile variations at scales larger than the inner part of the double layer (where ion crowding occurs) and thus should not depend on concentration or voltage. Perhaps more seriously, we assume a constant Stern capacitance $C_S$ for simplicity, in order to highlight what nonlinear trends can be captured by the MPB liquid model alone. This is rather different from the standard fitting procedure in electrochemistry dating back to Grahame~\cite{grahame1947,bockris_book}, which essentially {\it defines} the compact-layer capacitance empirically to account for all effects not predicted by the dilute-solution PB model of the diffuse layer. 

First, we consider the CS excess chemical potential (\ref{eq:CS}) for a hard-sphere liquid in the general MPB formula for the differential capacitance (\ref{eq:cdgen}). As shown in Fig.~\ref{fig:valette}, a reasonably good fit is obtained using $C_S = 125 \mu$F/cm$^2$ and an effective ion diameter of $a= 4$ \AA, comparable to the various ion sizes (in particular for $K^+$) inferred from completely different measurements. Near the point of zero charge and in dilute solutions ($< 100$ mM), the modified theory reduces to the Gouy-Chapman formula for a dilute solution (\ref{eq:cdpb}), and we obtain an excellent fit of the data, consistent with volumes of prior work in electrochemistry ~\cite{grahame1947,macdonald1962,macdonald1987,bockris_book}. At the highest salt concentration ($c_0=$100 mM), the theory underestimates the capacitance, although this is could be improved by adjusting $C_S$ for each salt concentration, as is commonly done. Of course, it is more interesting for us to focus on the regime of large voltage, where ion crowding occurs in the model. From Fig.~\ref{fig:valette}(a) it is clear that the CS model cannot fit the fairly sudden decay of $C_D$ beyond the maximum, although it predicts the qualitative loss of bulk salt concentration dependence (due to the dominance of near-surface crowding effects, as described above). This is due to the gradual onset of crowding effects for hard spheres in large fields, illustrated earlier in Fig.~\ref{fig:CS}.

Next, we fit Bikerman's MPB model to the same experimental data by fixing the parameters of the surface, $R_S=1.1$ and $C_S= 125 \mu$F/cm$^2$, and varying only the ion-size parameter $a$. We might expect a worse fit because the model (based on a lattice gas) has less statistical justification in the liquid state, but the results in Fig~\ref{fig:valette}(b) show a significant improvement in fitting the voltage dependence, due to the more sudden onset of steric effects with increasing voltage versus the CS model (Fig.~\ref{fig:CS}). Overall, Bikerman's model provides an impressive fit to the concentration and voltage dependence of $C_D$ across the entire experimental range, considering that it has {\it only one adjustable parameter} $a$ to describe the liquid (and we added two more, $R_S$ and $C_S$, for the surface). 

On the other hand, the excellent fit by Bikerman's model in Fig.~\ref{fig:valette}(b) is achieved with a surprisingly large value for the effective ion size, $a = 1.1$ nm, at least twice the typical solvated ion sizes in the bulk liquid from Table \ref{table:ionsize}.  Interestingly, very similar large values of $a$ result from fitting Bikernan's model to completely different experiments on ACEO pumping by electrode arrays~\cite{storey2008}, as discussed below. These results may point to correlation effects at high charge density or other neglected effects (see below), which are fortuitously fitted well by Bikerman's model with an enlarged ion size.

 It would be interesting to extend our capacitance analysis to electrolytes with little ion adsorption on Au surfaces~\cite{hamelin1982,hamelin1987,tymosiak2000}, since gold surfaces are widely used in experimental studies of induced-charge electrokinetic phenomena (including most of the entries Table 1). Given the successful fit  in Fig.~\ref{fig:valette}(b), it would also be interesting to perform ICEO experiments using silver surfaces and KPF$_6$ solutions or other electrolytes with similar capacitance behavior, showing little surface adsorption of ions. By taking the differential capacitance from fitting independent measurements, one can directly test other assumptions in the model, such as the HS slip formula (or its modifications, derived below).

\subsubsection{ Dielectric response in a concentrated solution }
\label{sec:dielectric}

We close our discussion of mean-field models of ion crowding by briefly considering nonlinear dielectric response. We have already mentioned that Bikerman, in his pioneering paper on effects of finite ion size~\cite{bikerman1942},  also  modeled the polarization of solvated ions by the local electric field, in terms of induced dipoles. A related effect is the polarization of a Stern monolayer of solvent dipoles on a metal electrode, which plays a major role in classical models of the compact inner layer~\cite{macdonald1962,macdonald1987}. This approach was extended by Macdonald and Kenkel~\cite{macdonald1982,macdonald1984,kenkel1984} to also describe the diffuse part of the double layer by postulating discrete layers of finite-sized dipoles separating layers of ions, where each layer is treated by a mean-field approximation.   

Closer in spirit to the continuous MF-LDA theories of this section, Abrashkin, Andelman and Orland~\cite{abrashkin2007} recently derived a ``Modified Dipolar PB" equation for a symmetric binary electrolyte by considering an equilibrium lattice gas of both ions and dipoles of (the same) finite size, thus generalizing Bikerman's model (or its many reincarnations cited above, including Borukhov, Andelman, Orland~\cite{borukhov1997}) in a natural way to allow for a variable dipole density. In our notation, their dipolar MPB equation takes the form
\begin{equation}
\varepsilon_0 \psi^{\prime\prime} + \frac{c_d p_0}{a^3} \left( \frac{ {\cal G}(p_0 \psi^\prime/kT) }{{\cal D}} \right)^\prime 
 =  \frac{2 c_0 ze}{a^3} \frac{\sinh(ze\psi/kT)}{{\cal D}}  \label{eq:DMPB}
\end{equation}
where $\varepsilon_0$ is the vacuum permittivity, $c_d$ is the bulk density of dipoles of moment $p_0$ on a lattice of spacing $a$ given by $c_d + 2c_0 = a^{-3}$, where $c_0$ is the bulk salt concentration; the second term describes  solvent polarization as a generalization of the classical expression for a dilute solution of dipoles (below), where the function 
\begin{equation}
{\cal G}(\tilde{E}) = \frac{\sinh\tilde{E}}{\tilde{E}}\, {\cal L}(\tilde{E}) = \frac{\cosh \tilde{E}}{\tilde{E}} - \frac{\sinh\tilde{E}}{\tilde{E}^2}
\end{equation}
relates the nonlinear polarization to the Langevin function ${\cal L}(x) = \coth(x)-1/x$, which arises in the limiting probability distribution of a sum of independent, randomly oriented unit vectors (also known as Rayleigh's random walk~\cite{hughes_book}); the third term is the free charge density, where the familiar PB expression (\ref{eq:rhoGC}) appears in the numerator; the effect of volume constraints enters through a generalized ``Bikerman factor'',
\begin{equation}
{\cal D} = 2c_0 \cosh\left( \frac{ze\psi}{kT}\right) + c_d \frac{ \sinh(p_0\psi^\prime/kT)}{p_0\psi^\prime/kT}
\end{equation}
which rescales both the solvent polarization and the free charge density.  In the limit of linear dielectric response, $p_0\psi^\prime/kT \to 0$, the dipolar MPB equation Eq.~(\ref{eq:DMPB}) reduces to Bikerman's MPB equation (\ref{eq:BikeMPB}), where the permittivity has the effective bulk value 
\begin{equation}
\varepsilon_b = \varepsilon_0 + \frac{ c_d p_0^2 }{ 3kT} . 
\end{equation}
In the limit of point-like ions and dipoles $a \to 0$, Eq. (\ref{eq:DMPB}) reduces to the PB equation with the classical field-dependent permittivity,
\begin{equation}
\varepsilon(E) = \varepsilon_0 + \frac{c_d p_0}{E} {\cal G}\left( \frac{p_0 E}{kT} \right)   \label{eq:sat}
\end{equation} 
It would be interesting to develop this model further, test it against experiments, and eventually apply it to electrokinetics. In its present form,  however, the dipolar MPB theory defies the common wisdom in electrochemistry by predicting a large increase in permittivity in the inner part of the double layer, due to a rise in the dipole density near the surface~\cite{abrashkin2007}.  Perhaps this discrepancy is due to the breakdown of the LDA~\cite{antypov2005} already noted above, and different predictions might result from non-local WDA theories with dielectric relaxation~\cite{zhang1992,tang1992a,zhang1993,tang1994}.

Based on half a century of fitting experimental capacitance data (with admittedly simple models), electrochemists have come to believe that that the local dielectric constant of an electrolyte is generally {\it reduced} in the inner part of the double layer,  due to the alignment of solvent dipoles in the large local field (``dielectric saturation") as in (\ref{eq:sat}); more precisely, in aqueous solutions,  the compact Stern layer (defined as the inner part of the double layer not described by PB theory) is inferred to have an effective permittivity $\varepsilon_S$ smaller than that of bulk water $\varepsilon_b$ by roughly an order of magnitude, e.g. reduced from $\varepsilon_b=78\varepsilon_0$ to $\varepsilon_S=6\varepsilon_0$ ~\cite{bockris_book}. This conclusion comes mostly from fitting compact layer models with nonlinear dielectric response~\cite{grahame1947,macdonald1962,macdonald1987}, but also from some models assuming similar  nonlinear dielectric properties in the diffuse layer.  

Grahame~\cite{grahame1950} was perhaps the first to analyze the structure of the diffuse layer with a field-dependent permittivity, using PB theory and the empirical form
\begin{equation}
\varepsilon(E) = \varepsilon_S + \frac{\varepsilon_b - \varepsilon_S}{\left(1 + (E/E_s)^2\right)^m}
\label{eq:epsE}
\end{equation}
where $m$ is an empirical  exponent, conjectured to be in the range $0< m < 2$ to avoid overly sudden onset of dielectric saturation above the characteristic field strength $E_s$. 
For any permittivity model such as (\ref{eq:epsE}), where dielectric saturation sets in above a characteristic field strength $E_s$,  the importance of variable permittivity in regions of  diffuse charge is measured by the dimensionless parameter
\begin{equation}
\tilde{E}_s = \frac{E_s}{(kT/ze  \lambda_D)} = E_s \sqrt{ \frac{\varepsilon_b}{2 c_0 kT }}
= \sqrt{ \frac{p_s}{c_0 kT} }
\end{equation}
which compares the critical electric field $E_s$ to the characteristic diffuse-layer field (at low voltage), $E_0 =  kT/ze  \lambda_D$, where the screening length $\lambda_D$ is defined using the bulk permittivity $\varepsilon_b$. Equivalently, the dimensionless parameter $\tilde{E}_s^2$ compares the critical electrostatic pressure for dielectric saturation, $p_s = \frac{1}{2}\varepsilon_b E_s^2$, to the bulk osmotic pressure, $c_0 kT$. For $\tilde{E}_s\gg 1$, the diffuse layer maintains a constant permittivity for typical voltages of order $kT/ze$.

To study effects of nonlinear dielectric response in the present context, we extend Grahame's analysis by deriving the differential capacitance of the double layer for any MPB theory with an arbtirary field-dependent permittivity $\varepsilon(E)$. Given the (non-Boltzmann) equilibrium charge density profile $\rho(\psi)$, we use Poisson's equation (in the normal coordinate), $-\left(\varepsilon \psi^\prime\right)^\prime= \rho(\psi)$, to write $\rho(\psi) \psi^\prime =  h(E)E^\prime$, where $E=-\psi^\prime$ and $h(E)=\varepsilon^\prime(E)E^2 + \varepsilon(E)E$. Next, we integrate from the bulk ($\psi=0$, $E=0$) to the inner edge of the diffuse layer ($\psi=\Psi_D$, $E=E_D$) to obtain 
\begin{equation}
H(E_D) \equiv \int_0^{E_D} h(E) dE = \frac{1}{2}\left( \varepsilon(E_D)E_D^2 + \int_0^{E_D} \varepsilon^\prime(E) E^2 dE \right) = \int_{\Psi_D}^0 \rho(\psi) d\psi = p_e(\Psi_D)  \label{eq:H}
\end{equation}
The electrostatic pressure at the surface $p_e(\Psi_D)$ must be non-negative, as is the function $H(E_D)$ from its definition since $\varepsilon>0$. Therefore, there are an even number of nonzero solutions to the algebraic equation $H(E_D)=p_e(\Psi_D)$, typically two for a monotonic $\varepsilon(E)$, and we select the physical solution by requiring that $E_D$ and $\Psi_D$ have the same sign. (It can be shown that unique solution exists if $h(E)/E=\varepsilon + \varepsilon^\prime E>0$ for all $E$.) To derive the differential capacitance $C_D(\Psi_D)$, we 
integrate Poisson's equation  across the diffuse layer (i.e. apply Gauss' law) to obtain the total charge density, $q=-\varepsilon(E_D)E_D$. Inserting $E_D(\Psi_D)$ from (\ref{eq:H}), we finally obtain a general formula for the differential capacitance of the diffuse layer,
\begin{equation}
C_D(\Psi_D) = -\frac{\rho(\Psi_D)}{H^{-1}(p_e(\Psi_D))} \label{eq:CDepsE}
\end{equation}
valid for any $\rho(\psi)$ and $\varepsilon(E)$, which reduces to (\ref{eq:cdgen}) in the case of a constant permittivity, $\varepsilon=\varepsilon_b$.

It is instructive to consider Grahame's model (\ref{eq:epsE}) with $m=\frac{1}{2}$, since it permits an analytical solution for $C_D$. In this case, we obtain the function 
\begin{equation}
\frac{H(E)}{E_s^2} = \frac{1}{2}\varepsilon_S \left( \frac{E}{E_s} \right)^2  + (\varepsilon_b-\varepsilon_S)  
\left[ 1 - \sqrt{ 1  + \left( \frac{E}{E_s} \right)^2 } \right]
\end{equation}
which can be inverted analytically by solving a quadratic equation for $E^2$, taking a square root, and choosing the physical solution as described above. Substitution into (\ref{eq:CDepsE}) then yields an explicit formula for $C_D(\Psi_D)$ in terms of the charge density $\rho(\Psi_D)$ and  electrostatic pressure $p_e(\Psi_D)$ at the surface, valid for any MPB model with field-dependent permittivity $\varepsilon(E)$.  For example, in Bikerman's model, these functions are given by 
\begin{equation}
\rho(\psi) = - \frac{2zec_0 \sinh(ze\psi/kT) }{1 + 2 \nu \sinh^2 (ze\psi/2kT) } 
\ \ \ \mbox{ and } \ \ \ p_e(\psi) = \frac{c_0 kT }{\nu} \ln \left[ 1 + 2 \nu \sinh^2 \left(\frac{ze\psi}{2kT}\right) \right]  \label{eq:rhope}
\end{equation}
where PB theory corresponds to the limit $\nu\to 0$.

\begin{figure*}
\begin{center}
\includegraphics[width=5in]{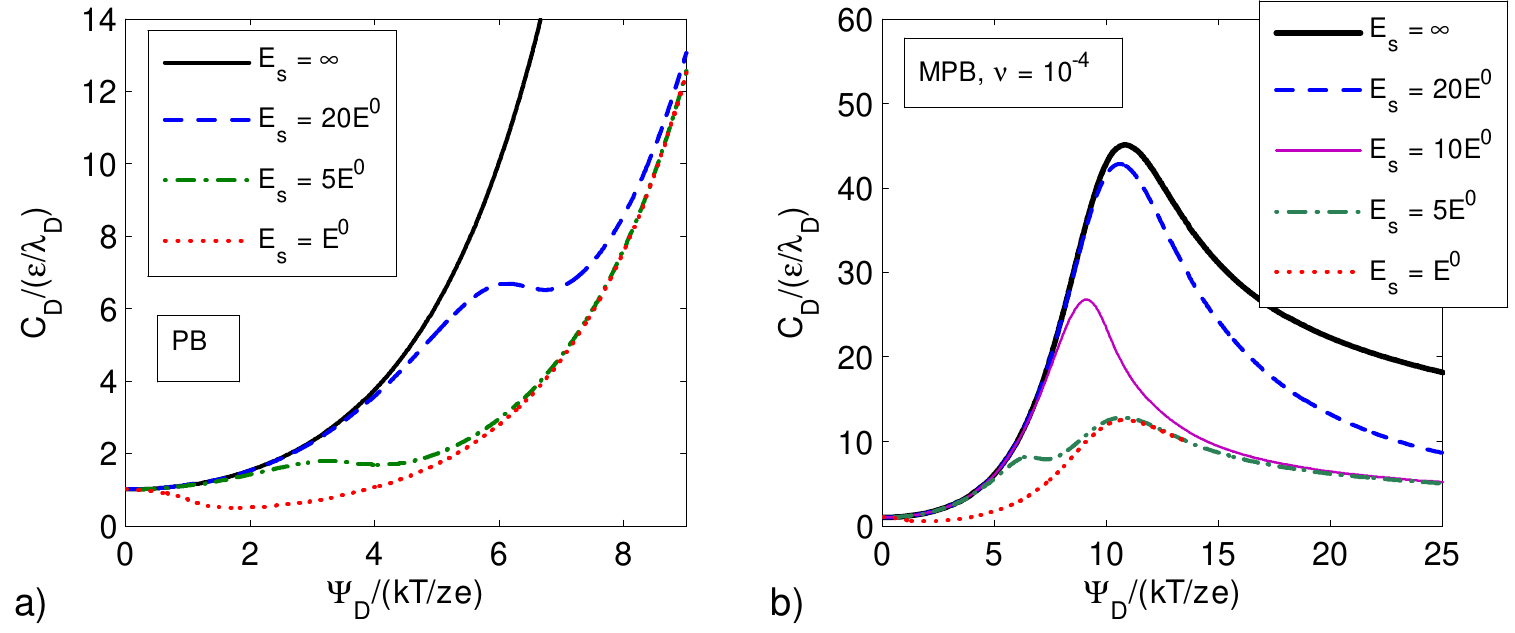}
\caption{\label{fig:Cd_epsE}  
The effect of dielectric saturation on the differential capacitance of the diffuse layer, given by (\ref{eq:CDepsE})-(\ref{eq:rhope}), for (a) PB theory ($\nu=0$) and (b) Bikerman's MPB theory with $\nu = 10^{-4}$. The field-dependent permittivity $\varepsilon(E)$ given by Grahame's model (\ref{eq:epsE}) with $m=\frac{1}{2}$. The characteristic field strength for dielectric saturation $E_s$ is scaled to $E_0 = kT/ze\lambda_D$, where $\lambda_D = \sqrt{\varepsilon_b kT/2(ze)^2c_0}$ is the screening length based on the bulk permittivity.
 }
 \end{center}
\end{figure*}

In Fig. ~\ref{fig:Cd_epsE}, we show the effects of dielectric saturation on the  diffuse-layer differential capacitance, using our analytical formula for Bikerman's MPB theory. In all cases, the capacitance shows a transition from low-field dependence with $\varepsilon \approx \varepsilon_b$  to a very similar high-field dependence with $\varepsilon \approx \varepsilon_S$, which differs only by a multiplicative factor $\varepsilon_S/\varepsilon_b$. The transition occurs when the electric field  strength at the surface reaches $E_s$. Applying PB theory for low voltages, this happens 
 at a transition voltage of roughly $\tilde{\Psi}_D = ze\Psi_D/kT = 2 \sinh^{-1}(\tilde{E}_s/2)$, consistent with the plots in Fig.~\ref{fig:Cd_epsE}. For PB theory, shown in (a), upon increasing the voltage, dielectric saturation leads to a fairly sudden decrease in $C_D$, followed by another exponential increase, corresponding to the PB formula (\ref{eq:cdpb}) with $\varepsilon_b$  replaced by $\varepsilon_S$. This physical interpretation of dielectric saturation followed by ``electro-constriction" is also contained in classical models of the Stern layer, which can lead to excellent fits of capacitance data for interfaces with little surface adsorption, such as NaF at a mercury drop,  although only for positive polarization ~\cite{macdonald1954_b,macdonald1962,macdonald1987}. Perhaps the asymmetry with negative polarization not captured by PB theory with dielectric saturation, which is also seen in NaF on silver surfaces~\cite{valette1981,valette1982}, and may be better captured by asymmetric MPB models with different ion sizes. Here, we do not attempt such fitting, but simply discuss our exact result for dielectric saturation in Bikerman's MPB theory with one ion size. 

As shown in Fig.~\ref{fig:Cd_epsE}(b), the shape of the theoretical capacitance curves depends on whether dielectric saturation occurs at lower or higher voltage than ion crowding. If dielectric saturation sets in first, then the transition $\varepsilon_b \to \varepsilon_S$ leads to  second small peak at low voltage, separate from the main peak corresponding to ion crowding; if crowding sets in first, then dielectric saturation only causes the capacitance to drop more quickly at higher voltage. Overall, the effect of dielectric saturation is similar to that of increasing the ion size. As noted below and in Ref.~\cite{storey2008}, this effect may contribute to the artificially large ions sizes inferred from Bikerman's model without dielectric saturation, and allow more realistic, smaller ion sizes to be used.
 
 It is interesting that so much structure in the differential capacitance can be predicted by simple mean-field models of the diffuse layer, without introducing a compact layer with additional parameters to fit large voltage behavior. Indeed, various modified models of the diffuse layer already mimic the formation of a compact layer with increasing voltage, as ions become crowded in an inner condensed layer where the local permittivity drops. The ``Stern plane" or ``outer Helmholtz plane" effectively expands and contracts in response to the applied voltage, in ways that may be captured by the same continuum equations describing the diffuse layer. Below in section ~\ref{sec:viscosity}, we will argue that the viscosity increases in the condensed region, and thus the ``shear plane" also effectively moves in response to the voltage.

\subsection{ Implications for nonlinear electrokinetics }

The cutoff and eventual decrease of diffuse-layer capacitance at large voltages for blocking surfaces (without Faradaic reactions or adsorption of ions) is robust to variations in the model and has important
consequences for nonlinear electrokinetics. Here, we provide two
examples of induced-charge electrokinetic phenomena, where any MPB theory
with volume constraints is able to correct obvious failures of PB
theory. As shown above, dielectric saturation only enhances these effects, but as a first approximation we assume constant permittivity $\varepsilon=\varepsilon_b$ in our calculations. This is also consistent with theoretical estimates~\cite{lyklema1961,lyklema1994} and atomistic simulations~\cite{freund2002} (at low voltages), which suggest that dielectric saturation in the diffuse layer is weak compared to other effects. Our results suggest that incorporating crowding effects into the electrokinetic equations may be essential in many other situations in electrolytes or ionic liquids, whenever the voltage or salt concentration is large.

\begin{figure}
\begin{center}
(a) \includegraphics[width=2in]{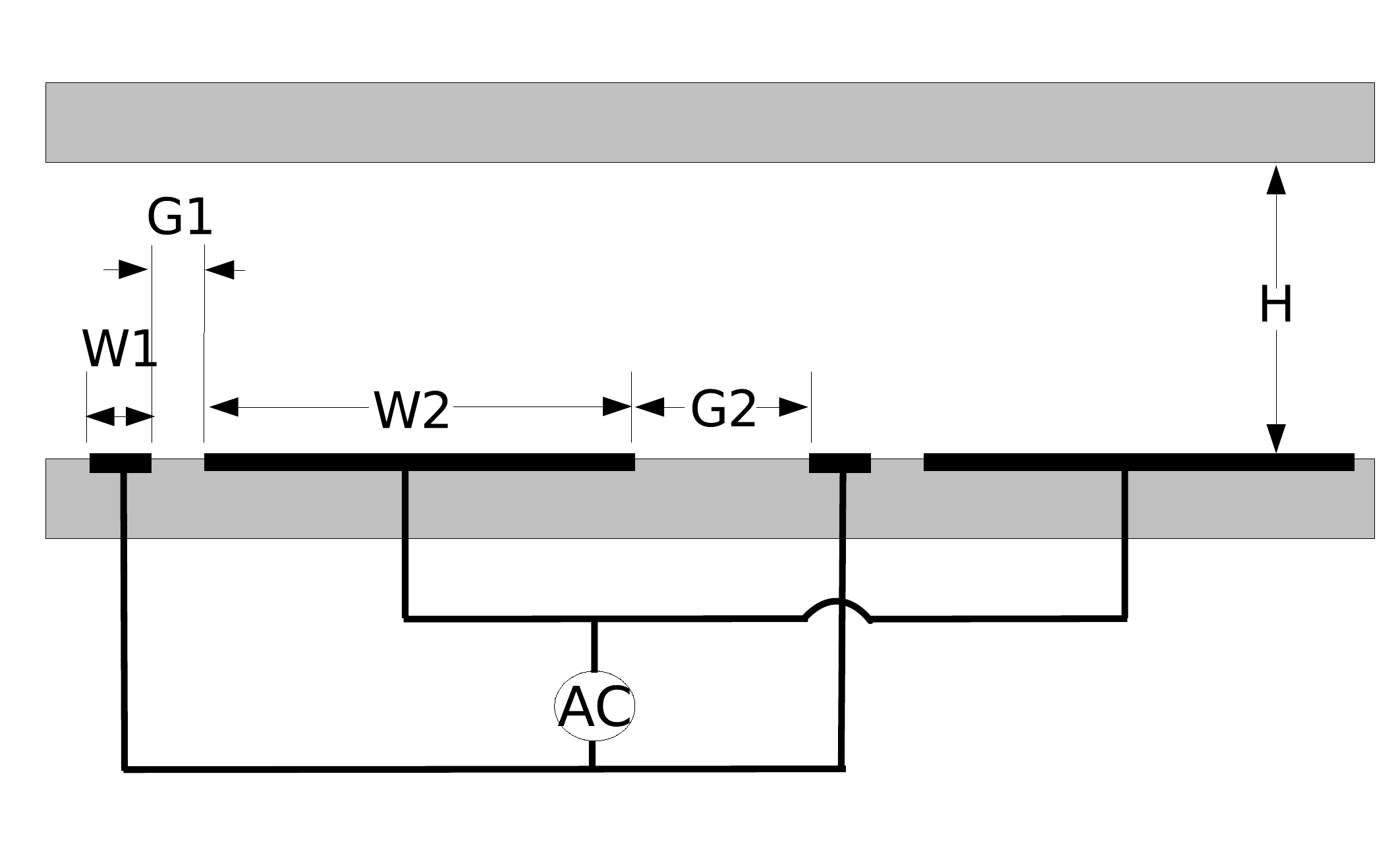} 
(b) \includegraphics[width=2.8in]{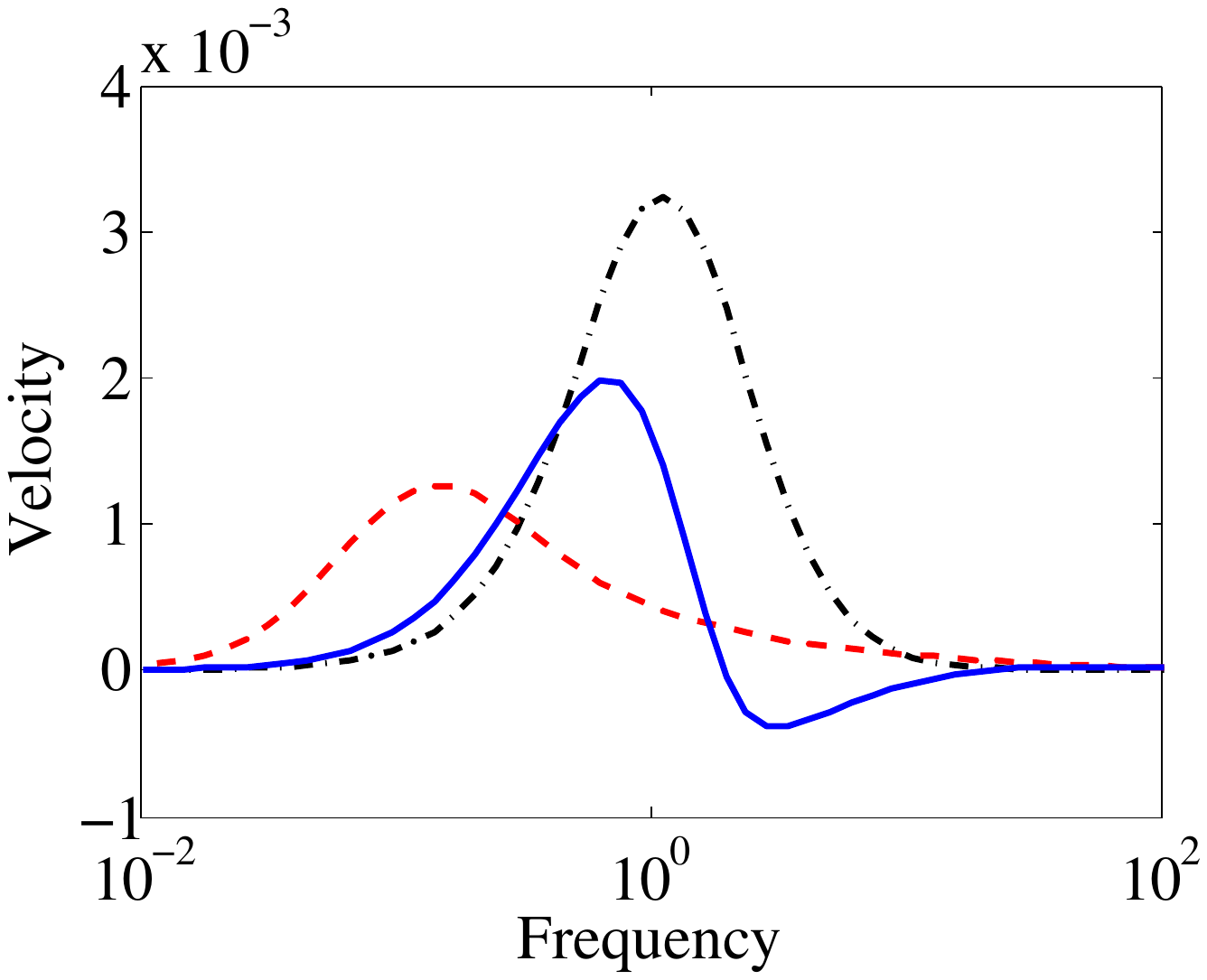}
\caption{\label{fig:rev} (a) One period of an asymmetric array of
  planar microelectrodes in an ACEO pump studied in
  experiments~\cite{brown2000,studer2004,urbanski2006,microTAS2007}
  and simulations with the low-voltage
  model~\cite{brown2000,ramos2003,olesen2006,bazant2006} with $W1=4.2~
  \mathrm{\mu m}$ , $W2=25.7~\mathrm{\mu m}$, $G1=4.5~ \mathrm{\mu m}$
  , and $G2=15.6 ~\mathrm{\mu m}$. (b) The dimensionless flow rate
  versus frequency for different models. In the low-voltage limit
  $V\ll kT/e=25$ mV, low-voltage models predict a single peak (black dash-dot
  line). For a typical experimental voltage, $V=100 kT/e = 2.5$ V, PB
  theory breaks downs and its capacitance (\ref{eq:cdpb}) shifts the
  flow to low frequency (red dashed line) and Stern capacitance is
  needed to prevent the capacitance from diverging.  Accounting for steric
  effects (\ref{eq:cdnu}) with $\nu = 0.01$ (solid blue line) reduces the
  shift and predicts high frequency flow reversal, similar to
  experiments~\cite{studer2004,microTAS2007}.
}
\end{center}
\end{figure}

\subsubsection{\label{sec:highfreqrev}  High-frequency flow reversal of AC electro-osmosis } 

Steric effects on the double-layer capacitance alone suffice to
predict high-frequency flow reversal of ACEO pumps, without invoking
Faradaic reactions. Representative results are shown in
Fig.~\ref{fig:rev}, and the reader is referred to
Ref.~\cite{storey2008} for a more detailed study. Numerical
simulations of a well studied planar pump
geometry~\cite{brown2000,studer2004,urbanski2006,microTAS2007} with
the Standard Model in the linearized low-voltage regime~\cite{ramos1999,ajdari2000,ramos2003,bazant2006} 
predicts a single peak in flow rate versus frequency at all voltages. If the
nonlinear PB capacitance (\ref{eq:cdpb}) is
included~\cite{olesen2006,olesen_thesis}, then the peak is reduced and
shifts to much lower frequency (contrary to experiments), due to
slower charging dynamics at large
voltage~\cite{bazant2004,chu2006}.  As shown in Fig.~\ref{fig:rev}, the
BF capacitance for Bikerman's MPB model of steric effects (\ref{eq:cdnu}) reduces the peak
shift and introduces flow reversal similar to experiments. 

This result is the
first, and to our knowledge the only, theoretical prediction of
high-frequency flow reversal in ACEO. The physical mechanism for flow reversal in our model can be easily 
understood as follows: At low voltages, the pumping direction is set by the
larger electrode, which overcomes a weaker reverse flow driven by the smaller electrode. 
At large voltages, however, the more highly charged, smaller
electrode has its $RC$ charging time reduced by steric effects, so at
high frequency it is able to charge more fully in a single AC period
and thus pump harder than the larger electrode.

\begin{figure}
\begin{center}

\begin{tabular}{cc}
(a) \includegraphics[width=2.2in]{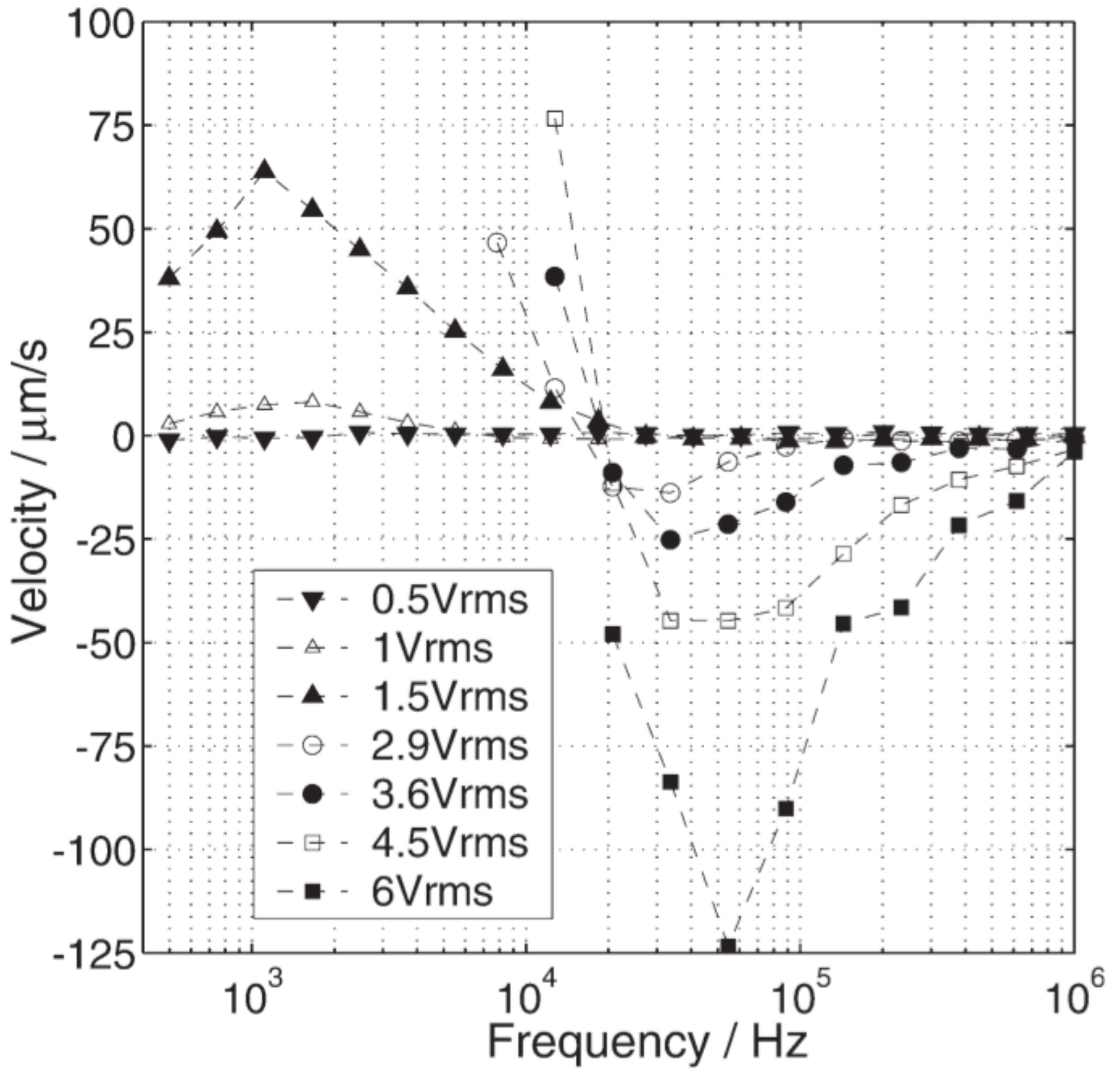} &
(b) \includegraphics[width=2.5in]{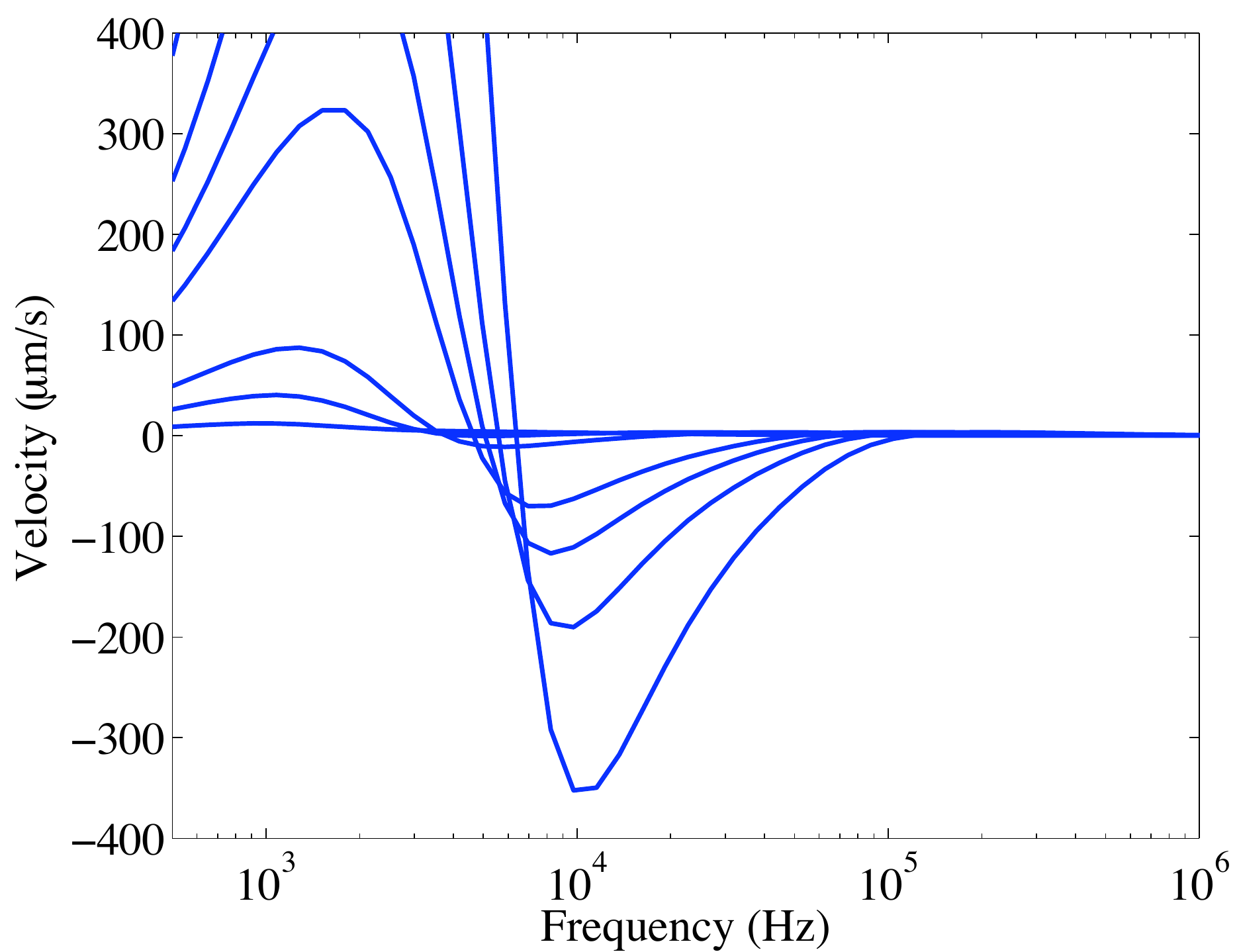} \\
(c) \includegraphics[width=2.5in]{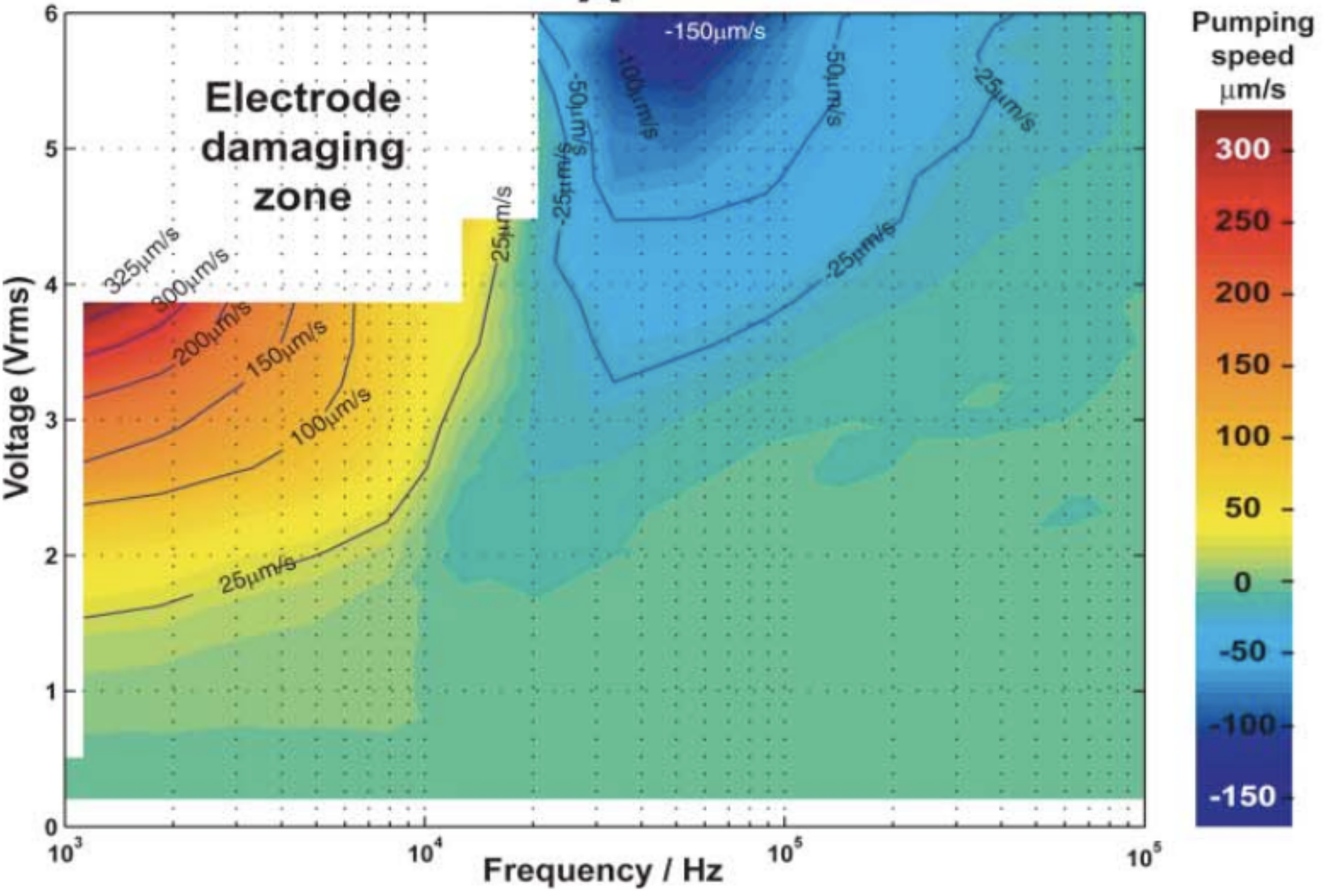} &
(d) \includegraphics[width=2.5in]{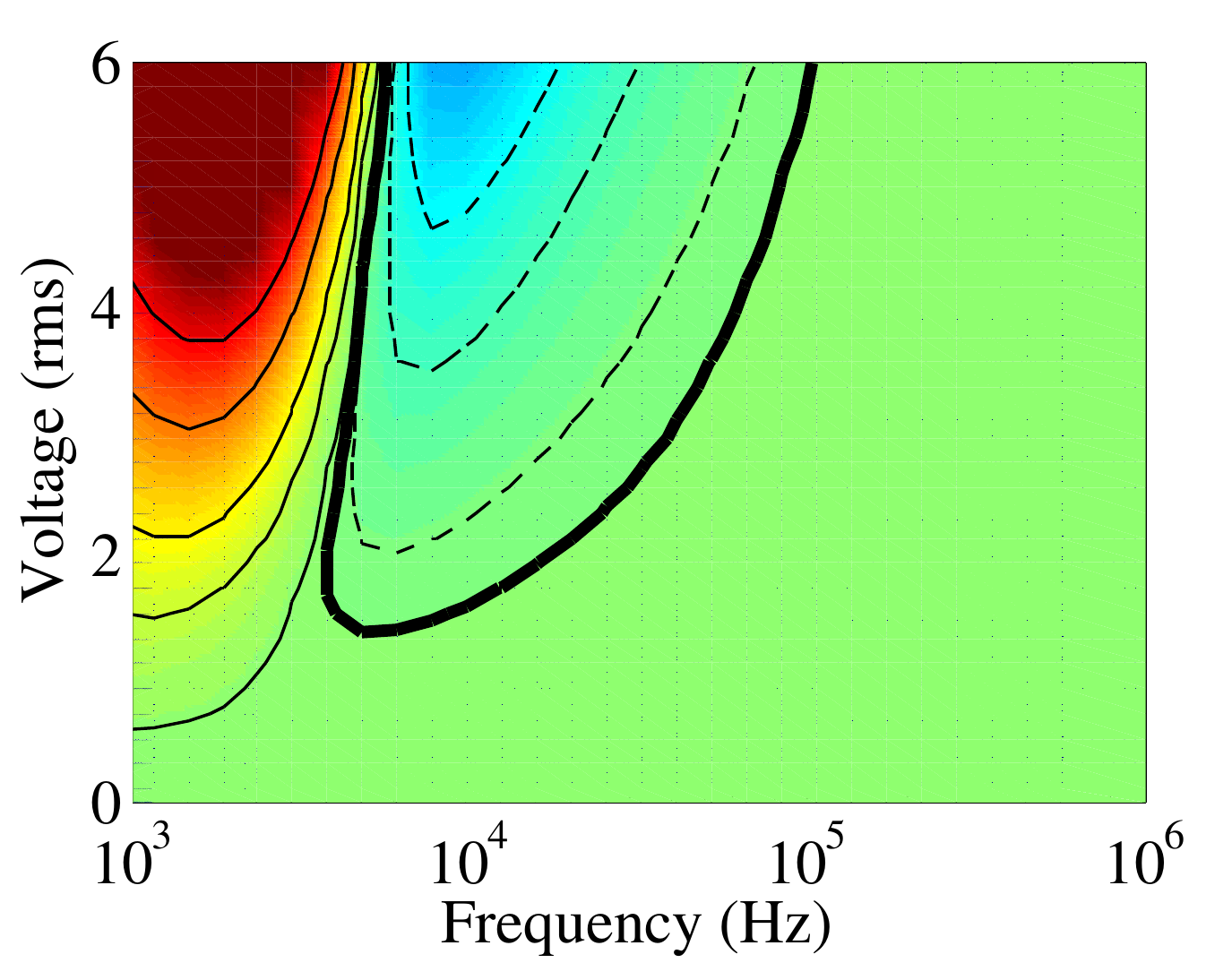} \\
(e) \includegraphics[width=2.5in]{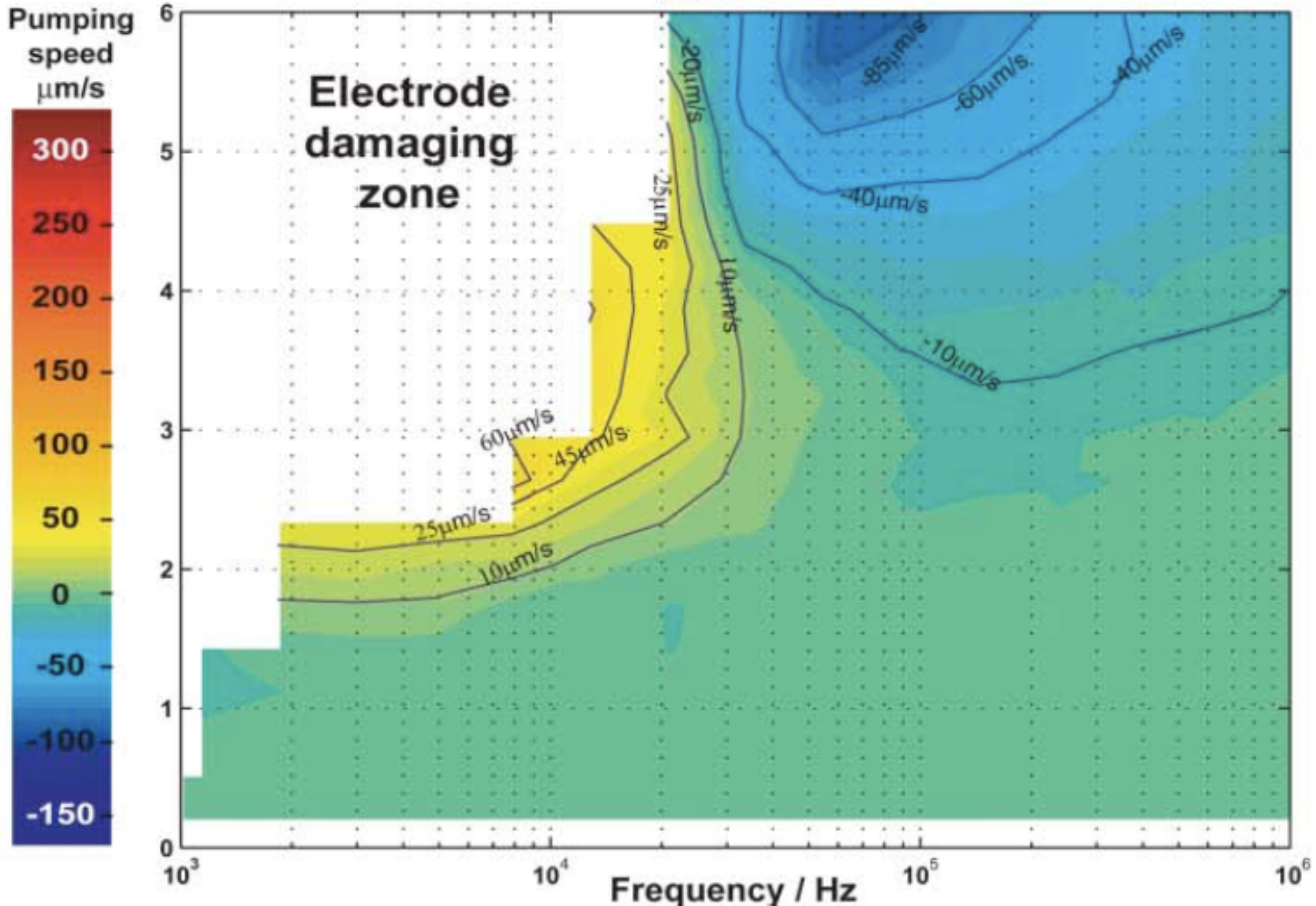} &
(f) \includegraphics[width=2.5in]{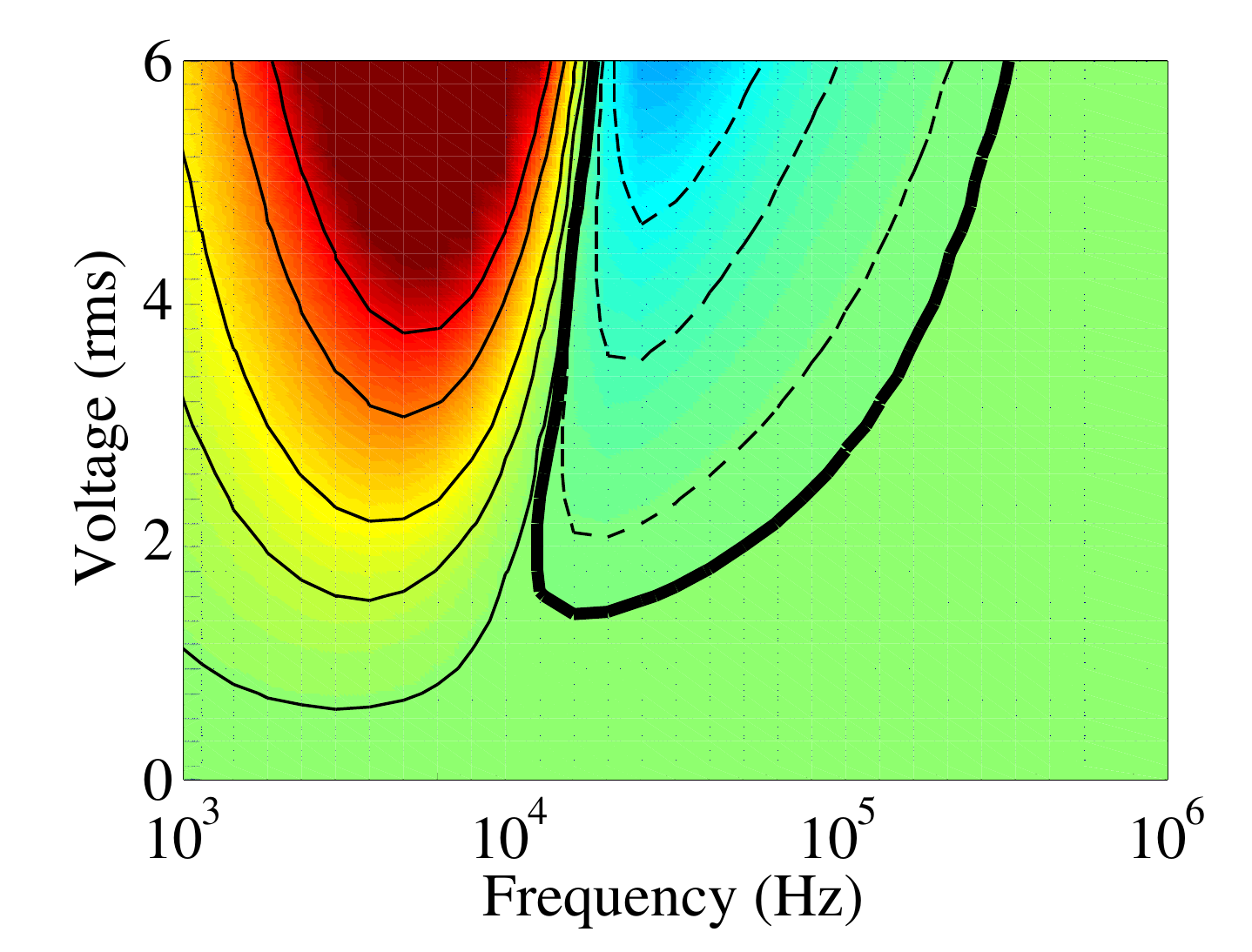} \\
\end{tabular}
\caption{\label{fig:ACEOexpt} (a) Velocity of ACEO  pumping in 0.1mM KCl  by a planar electrode array around a microfluidic loop versus frequency at different voltages from the experiments of Studer et al.~\cite{studer2004}. (b) Simulations by Storey et al. ~\cite{storey2008} of the same flow using the Standard Model with Bikerman's MPB theory (\ref{eq:cdnu}) for the double-layer differential capacitance with only one fitting parameter, $a=4.4$ nm or $\nu=0.01$. 
Countour plots of ACEO pumping velocity contours in frequency-voltage space for (c) 0.1 mM and (e) 1.0 mM KCl from experiments of Studer et al.~\cite{studer2004}, compared to 
simulations under the same conditions using Bikerman's MPB theory with $\nu=0.01$ in (d) and (f), respectively.  
Red indicates forward flow and blue 
reverse flow. The solid contour lines show positive velocity contour and 
the dashed show reverse flow. The heavy solid contour is the zero velocity contour in the simulations.  
}
\end{center}
\end{figure}

As shown in Fig.~\ref{fig:ACEOexpt}, the MPB model is able to reproduce experimental data for ACEO pumping of 
dilute KCl rather well, including the dependence on both voltage and frequency. 
Through  Fig.~\ref{fig:ACEOexpt}, we compare simulations to experimental data at two different concentrations.
In the left column we show experimental data 
and on the right we show the corresponding simulations using Bikerman's MPB theory for the double 
later capacitance.  As in experiments, the flow 
reversal arises at 10-100 kHz frequency and high voltage, without shifting appreciably the main peak below 10 kHz
 frequency (which is hard to see in experiments at high voltage due to electrolysis). This is all the more remarkable, 
since the model has only one fitting parameter, the effective ion size $a$, and does not include any additional Stern-layer
 capacitance. As seen in ~\ref{fig:ACEOexpt} (a) and (b),  
the magnitude of the flow is over-estimated by a roughly a factor of two ($\Lambda \approx 0.4$), but this 
is much better than in most predictions of the Standard Model (in Table 1),
 which fail to predict flow reversal under any circumstances.

In spite of this success, we are do not claim a complete understanding of flow reversal in ACEO. 
One difficulty with these results is 
that the effective ion size in Bikerman's model needed to fit the data 
is unrealistically large. For the simulations to reproduce the experiments 
we seem  to typically require $\nu=0.001-0.01$, which implies an overly
small bulk ion spacing $l_0=(2c_0)^{-1/3}=\nu^{-1/3}a$, or overly large ion size $a$.
For example, for the  $c_0 = 0.1$ mM KCl data shown in  Fig.~\ref{fig:ACEOexpt} (a) and (b), 
we use  $a = 4.4$ nm in the model, which is clearly unphysical. As noted above and
shown in Fig.~\ref{fig:CS}, this can be attributed at least
in part to the signficant under-estimation of steric effects in a
liquid by the simple lattice approximation behind Bikerman's model.

Indeed, hard-sphere liquid models tend to improve the agreement
betweeen simulation and experiment, and this increases our confidence in 
the physical mechanism of ion crowding at large voltage. 
Using the CS MPB model for
monodisperse charged hard spheres in the same simulations of ACEO
pumping allows a smaller value of the ion size. For example, the 0.1 mM KCl shown in 
Fig. \ref{fig:ACEOexpt} (a)  can be fit by using  $a = 2.2$ nm (instead of 4.4 nm for Bikerman), 
and the magnitude of the velocity also gets closer to the experimental data 
($\Lambda \approx 0.7$).  Assuming a reduced permittivity
in the condensed layer   could further yield $a \approx 1$ nm ($\approx$
10 atomic diameters)~\cite{storey2008}. This value is more realistic but
still considerably larger than the bulk hydrated ion sizes in KCl and
NaCl. For the commonly used electrolytes in
ICEO experiments (see Table ~\ref{table:Lambda}), the cation-anion radial
distribution function from neutron scattering exhibits a sharp
hard-sphere-like first peak at 3\AA, although the water structure is
strongly perturbed out to the second neighbor shell (up to 1nm), as if
under electrostriction. Anion-anion correlations are longer ranged and
softer, with peaks at 5 \AA and 7 \AA, but unfortunately such data is
not available for crowded like charges within the double layer at high
voltage. Perhaps under such conditions the effective hard-sphere radius 
grows due to strong correlations, beyond the mean-field approximation.

We have already noted that similar over-sized ionic sizes are needed to fit electrochemical capacitance data with MPB models. (See Fig.~\ref{fig:valette}, where $a=1.1$ nm in Bikerman's model yields an excellent fit of capacitance data for KPF$_6 \ | \ $Ag.) In the case of hard-sphere MPB models,  
DiCaprio et al. have likewise concluded that, in spite of 
good qualitative predictions, effective sphere radii over 1.2 nm 
are required to fit capacitance data~\cite{dicaprio2003}.
The fact we reach similar conclusions in the completely different
context of ACEO pumping suggests that various neglected effects, in addition to dielectric saturation (section ~\ref{sec:dielectric}), may be
extending the apparent scale for crowding effects in MPB models, as
discussed below.

Along with overly large ion sizes, another 
difficulty is that the simulated results in 
Fig.~\ref{fig:ACEOexpt} (d) and (f), 
are reproducing  data from two different concentrations 
using the same  value of $\nu=0.01$ in Bikerman's MPB theory. 
Since the concentration 
varies between the two experiments by a factor of 10, the simulations
are using different values of the ion size,  $a$: 4.4 nm at 0.1 mM and 2 nm at 1 mM. 
Physically, it would make sense to use the 
same ion size in the model regardless of concentration.
If we use the same ion size to fit data from both concentrations, the 
comparison worsens significantly, 
representing one of the key difficulties of 
fitting model to experimental data. 
 
The simple steric models (without dielectric saturation) can thus provide good qualitative agreement between the measured 
and predicted frequency responses. However, they can not 
predict the proper frequency response as we change concentration and hold the ion size fixed. 
In the model calculations, the frequency response at high voltage always shows
regions of  forward and reverse flow. As concentration 
changes,  features in the modeled frequency response (e.g. forward and reverse peaks,  crossover frequency) 
shift along  the frequency axis as the RC charging time  changes with concentration.
Since the capacitance of the double layer is insensitive to bulk concentration
at high voltage (see Fig. \ref{fig:Cmolar}), the RC charging time 
depends on concentration through the bulk resistance.
Thus the location of peaks on the frequency response should 
vary approximately  linearly with concentration; 
experiments at high voltage indicate a much weaker dependence.
The difficulty in matching the frequency response of experiments to simulations 
may be due to the neglect of Faradaic reactions and other effects discussed below. 

At least at a qualitative level, changes in the double-layer charging
time due to crowding effects likely also play a role in the
sensitivity of ACEO flow to the solution chemistry. For example, flow
reversal in our models is related to an ion-specific scale for
crowding effects. In cases of asymmetric electrolytes, there may also
be two different frequency responses at large voltages, one for
positive and another for negative charging of the double
layer. Perhaps this effect is responsible for the double-peaked
frequency spectrum of ACEO pumping in water with non-planar electrodes
at high voltage~\cite{urbanski2006}. In multicomponent electrolytes,
the situation is even more complicated, since large voltages can
induce segregation of different counterions, opposite to PB
predictions, e.g. with smaller ions condensing closest to the surface,
even if the larger ions have high bulk concentration or carry more
charge. These effects can be predicted by multi-component hard-sphere MPB
models~\cite{biesheuvel2007,dicaprio2003,dicaprio2004} consistent with x-ray
reflectivity measurements on mixed double
layers~\cite{shapovalov2006,shapovalov2007}, so there is hope that
applying such models in our weakly nonlinear formalism for ACEO may
also be fruitful.

\subsubsection{ Field-dependent electrophoretic mobility } 

In the classical theory of
electrophoresis~\cite{anderson1989,lyklema_book_vol2}, the
electrophoretic mobility $b_{ep}$ of a homogeneous particle with thin
double layers is a material constant, given by Smoluchowski's formula,
\begin{equation}
b_{ep}=b= \frac{\varepsilon_b \zeta}{\eta_b} .
\end{equation}
In particular, the electrophoretic mobility does not depend on the
background field $E_b$ or the shape or size of the particle. These are partly 
consequences of the assumption of fixed surface charge, or constant
zeta potential. For polarizable particles, the theory must be modified
to account for ICEO flows~\cite{encyclopedia_polarizable}, which
produce a shape-sensitive ICEP velocity scaling as $U \propto
\varepsilon_b R E_b^2 /\eta_b$, where $R$ is the particle
size~\cite{iceo2004a,squires2006,yariv2005}. Transverse ICEP of
metallo-dielectric Janus particles in AC fields has recently been
observed in experiments~\cite{gangwal2008} up to fairly large induced
double-layer voltages $E_bR \approx 15 kT/e$.

ICEP theories aimed at AC fields tend to assume zero total charge, but
ICEO flows can also alter the DC electrophoretic mobility of a
charged, polarizable particle. In the limit of weak fields $E_b\ll
kT/eR$, A. S. Dukhin first showed that an ideally polarizable sphere with
equilibrium zeta potential $\zeta_0$ and radius $R$ has a
field-dependent electrophoretic mobility,
\begin{equation}
b_{ep}(E_b) \sim  \frac{\varepsilon_b}{\eta_b} \left(  \zeta_0 - 
  \frac{3}{8} \frac{C_D^\prime(\zeta_0)}{C_D(\zeta_0)} (E_bR)^2
  + \ldots \right)
\label{eq:dukhin}
\end{equation}
if the diffuse-layer differential capacitance is voltage
dependent~\cite{ASdukhin1993}. (Note that field-depedent mobility is a general phenomenon~\cite{ASdukhin2005} that can also arise for fixed-charge particles due to surface conduction~\cite{shilov2003} or convection~\cite{mishchuk2002}.) This general correction has only been
applied in the context of PB theory, Eq.~ (\ref{eq:cdpb}), which
predicts decreased mobility, $\Delta b_{ep} < 0$ since $dC_D/d\psi>0$
for $\zeta_0>0$. It has also recently  been derived as the small field limit of a general PB analysis for thin double layers by Yariv~\cite{yariv2008a}.

\begin{figure}
\begin{center}
 \includegraphics[width=4.5in]{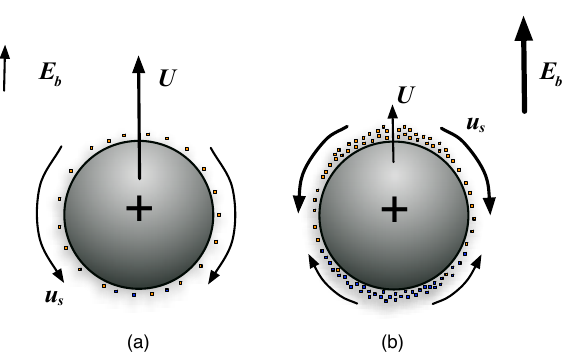}\\
(c) \includegraphics[width=3.3in]{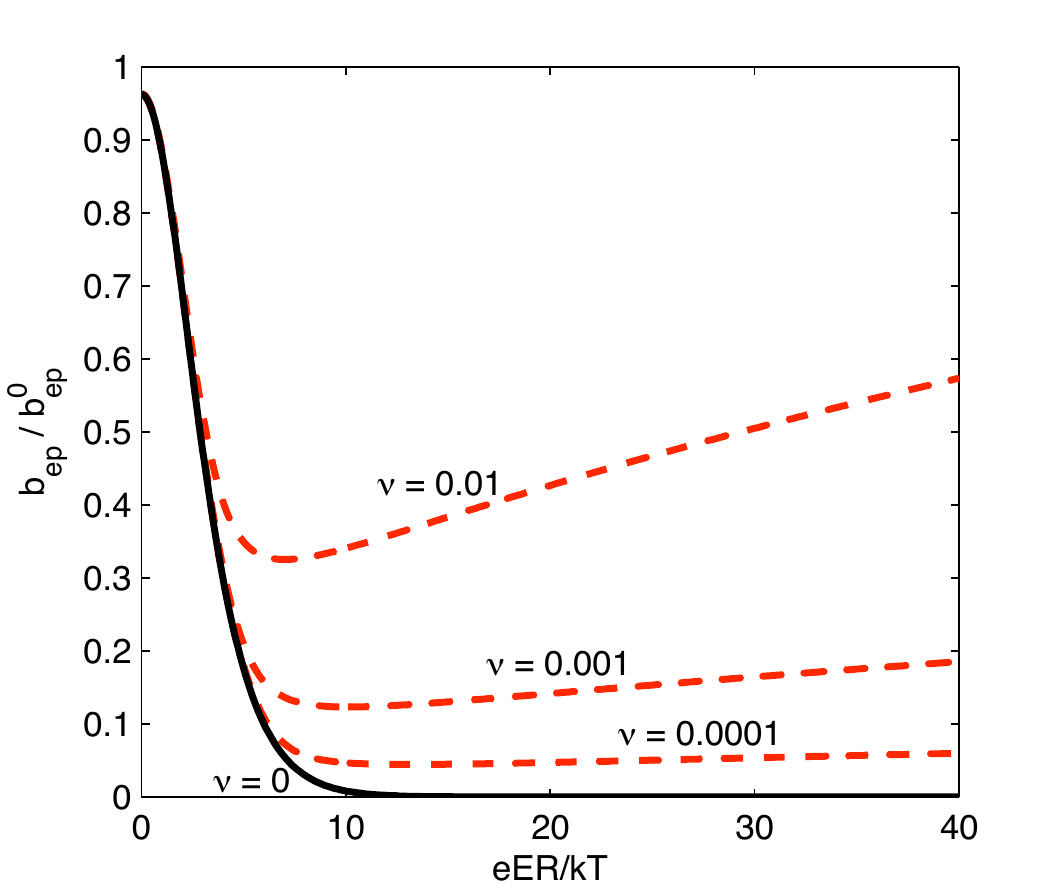}
\caption{\label{fig:mobility} Field-depenent electrophoretic velocity
  $U$ of an ideally polarizable, charged sphere of radius $R$ with
  thin double layers in a background field $E_b$. (a) In small fields,
  the mobility $b_{ep}=U/E_b$ is set by the uniformly distributed
  double-layer charge. (b) In large fields, $E_b \gg kT/eR$, the
  dipolar induced charge overwhelms the pre-existing charge and alters
  $b_{ep}$, if cations and anions do not condense at the same density
  and must redistribute to conserve total charge. (c) In PB theory,
  the unphysical collapse of point-like ions to the surface causes
  exponential decay of $b_{ep}(E_b)$ via Eq.~ (\ref{eq:bpb});
  finite-size effects in Bikerman's MPB model (Fig. \ref{fig:C})
  prevent this decay and lead to the opposite trend: increase of
  mobility in large fields via Eq.~(\ref{eq:bmpb}).}
\end{center}
\end{figure}

The basic physics of this nonlinear effect is illustrated in Fig.~\ref{fig:mobility}(a-b). If the double-layer voltage varies enough to cause spatial variations in its differential capacitance, then counterions aggregate with varying density (per area) around the surface of the particle upon polarization by the applied field, and this nonlinearity breaks symmetry  in polarity with respect to the mean voltage. For example, if the positively charged part of the diffuse layer (relative to the mean charge) is less dense (e.g. due to larger or less charged cations than anions),  it will cover more of the surface than the negatively charged part; cations are then more likely to dominate in regions of large tangential field near the equator and thus make an enhanced contribution to the electrophoretic mobility of the particle, regardless of its true surface charge. Other effects can also be important (see below), but this one is particularly sensitive to MPB models for the double layer.

Dukhin's formula (\ref{eq:dukhin}) can be derived  from the general weakly nonlinear formalism of Refs.
~\cite{encyclopedia_polarizable,squires2006} for ideally polarizable particles with thin double layers  
(yielding the same result as  Ref.~\cite{ASdukhin1993}). In the low-frequency or DC limit,  the background field $E_b$ causes nonuniform polarization of the double layer around the particle to screen the bulk electric field $\Eb = -\nabla\phi$, which thus solves Laplace's equation $\nabla^2\phi=0$ with the effective boundary condition, $\nhat\cdot\Eb=0$. If we let $\phi_0$ denote the {\it induced} potential of the particle, relative to the background applied potential, then $\Delta\phi(\rb) = \phi_0 - \phi(\rb)$ is the non-uniform voltage across the double layer, which enters the electro-osmotic slip formula, either the HS formula (\ref{eq:HS}) or one of its generalizations below. For a sphere with HS slip, the electrophoretic mobility is simply $b_{ep} = \varepsilon_b \phi_0/\eta_b$. 

The crucial step is to determine the particle's potential $\phi_0$, after polarization of the double layers, which generally differs from its initial equilibrium value, $\zeta_0$, due to nonlinearity of the charging process. (This phenomenon, first noted by Dukhin~\cite{ASdukhin1993}, was overlooked in recent ICEO papers~\cite{iceo2004a,iceo2004b,yariv2005,squires2006}.) Using the mathematical formalism of Ref.~\cite{squires2006}, 
the potential of the particle must adjust to maintain the same total charge $Q$, which can be related to the differential capacitance of the double layer as follows:
\begin{equation}
Q = \oint \left( \int_0^{\zeta_0} C_D(\psi)d\psi \right) dA 
= \oint \left( \int_0^{\phi_0 - \phi(\rb)} C_D(\psi)d\psi \right) dA,    \label{eq:Qint}
\end{equation}
for a given initial zeta potential $\zeta_0$, bulk polarization $\phi(\rb)$, and (total) double layer differential capacitance, $C_D(\psi)$. 
For simplicity, we assume the diffuse layer carries all the double-layer voltage, but a compact Stern layer can be easily included in Eq. (\ref{eq:Qint}) by replacing $\Delta\phi$ with $\Delta\phi/(1+\delta)$. Regardless of the particle shape, the assumption of a uniform, constant $C_D$ implies $\phi_0=\zeta_0=Q/(C_D A)$ (where $A$ is the surface area), and thus no impact of polarization on the electrophoretic mobility in the case of a sphere~\cite{squires2006}. With a nonlinear differential capacitance, however, Equation (\ref{eq:Qint}) is a nonlinear algebraic equation for $\phi_0$, and thus $b_{ep}$, in terms of $\zeta_0$ and $E_b$. In the geometry of a sphere, Dukhin's formula (\ref{eq:dukhin}) can be derived by asymptotic analysis in the limit of small fields, $E_b \ll kT/eR$ for any choice of $C_D(\psi)$ ~\cite{encyclopedia_polarizable}, and for some models exact solutions and the large-field limit can also be derived~\cite{kilic_thesis}.  

Using this mathematical formalism with our MPB models, 
we predict that steric effects in the electrolyte can
signficantly influence the mobility of polarizable particles in large applied fields 
and/or highly concentrated solutions.  
Here, we focus on new qualitative phenomena predicted by the theory. (More details can be found in Ref.~\cite{kilic_thesis}.) From
Fig.~\ref{fig:C} and Eq.~(\ref{eq:cdnu}), we see that the mobility of
a highly charged particle $|\zeta_0|\gg kT/e$ can increase
with the field squared in Dukhin's formula (\ref{eq:dukhin}) since
$dC_D/d\psi<0$, which is the opposite prediction of PB theory. The
mobility is also clearly sensitive to the ionic species through
$C_D$. 

The discrepancy with PB theory becomes more dramatic in a large
applied field, $E_b \gg kT/eR$, even if the particle is not highly
charged $|\zeta_0|\approx kT/e$. Previous authors have only considered
weak fields~\cite{ASdukhin2005,ASdukhin1993}, so it was apparently not
noticed until Refs.~\cite{yariv2008a,kilic_thesis} that PB theory (\ref{eq:cdpb}) leads to a surprising
prediction, shown in Fig.~\ref{fig:mobility}(c): The mobility of a charged, ideally
  polarizable, spherical particle vanishes exponentially in the limit
  $E_b \gg kT/eR$ ($=$100 V/cm for $R= 2.5 \mu$m),
\begin{equation}
  b_{ep}^{PB} \sim \frac{3 \varepsilon_b}{\eta_b}  
  \sinh\left(\frac{ze\zeta_0}{2kT}\right)  E_bR\, e^{-3 z e E_b R/4kT}, \label{eq:bpb}
\end{equation}
which is the large-voltage limit of an exact solution for the ICEP mobility of an ideally polarizable sphere with thin double layers in PB theory,
\begin{equation}
b_{ep}^{PB} = 2 \frac{kT}{ze} \sinh^{-1}\left[ \frac{3 z e E_b R}{4kT} \frac{ \sinh(ze\zeta_0/2kT) }
{\sinh(3 z e E_b R/4kT)} \right]
\end{equation}
The mechanism for this seemingly unphysical effect
is the massive overcharging of the
diffuse layer in PB theory at large voltages in Fig.~\ref{fig:C}(b),
which causes the anti-symmetric induced charge (not causing motion) to
overwhelm the symmetric pre-existing charge (giving rise to mobility). We view this prediction as another failure of PB theory, since we are not aware of any evidence that polarizable particles lose their electrophoretic mobility so strongly in such fields, which are routinely applied in electrophoresis experiments. (Note that $kT/eR = 100$ V/cm for a $5 \mu$m diameter particle at room temperature.)  
The PB prediction of vanishing mobility 
is closely tied to the unphysical pile-up of point-like ions near a highly charged surface in PB theory.

Indeed, the situation is completely different  and more physically reasonable in any mean-field theory with finite-sized ions, regardless of the model.  Using our general MPB formula
(\ref{eq:cdgen}), the mobility in large fields $E_b \gg kT/eR$ can be expressed as 
\begin{equation}
  b_{ep} \sim \frac{3 \varepsilon_b E_b R\, q(\zeta_0)}
  {2\eta_b\,  q\left( \frac{3}{2} E_b R \right)}
\end{equation}
and typically grows as $b_{ep}^\nu \propto \sqrt{E_b}$. For example, in Bikerman's
model the BF formula (\ref{eq:cdnu}) yields
\begin{equation}
b_{ep}^\nu \sim \frac{\varepsilon_b}{\eta_b} \sqrt{ \frac{3 \nu kT E_b
    R}{4ze} }.\label{eq:bmpb}
\end{equation}
As shown in Fig. \ref{fig:mobility}(c), the mobility $b_{ep}^\nu$ rapidly
decreases in weak electric fields until PB theory breaks down, and then 
gradually increases in larger fields~\cite{kilic_thesis}. Physically,
steric effects prevent overcharging of the double layer in PB theory,
thus preserving the asymmetry of the initial charge distribution.

\begin{figure}
\begin{center}
\includegraphics[width=5in]{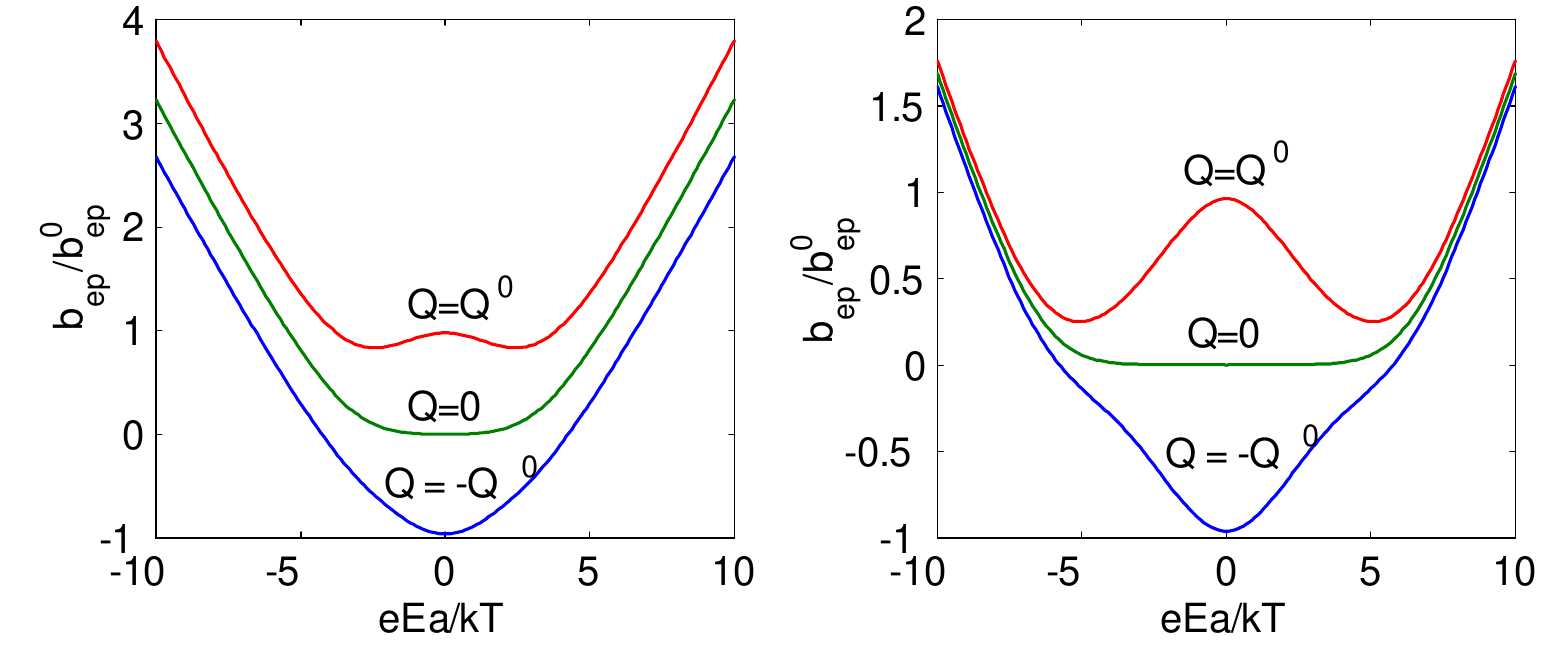}
\caption{  Electrophoretic mobility $b_{ep}$ of an ideally polarizable sphere of total charge $Q$ in an asymmetric binary $z:z$ electrolyte, scaled to $b_{ep}^0 = \varepsilon_b kT / ze \eta_b$, in the weakly nonlinear limit of thin double layers. The case of an uncharged particle $Q=0$ is compared to those of total charge $Q=\pm Q^0$ where $Q^0 = \varepsilon_b kT/ ze \lambda_D$. Using the approximation (\ref{eq:Casym}), the mobility is calculated with Bikerman's MPB model (\ref{eq:cdnu}) for an effective volume ratio $(a_-/a_+)^3=10$ in two cases:  (a) high salt concentration with $\nu_-=0.1$ and $\nu_+=0.01$ and (b) moderate salt concentration with $\nu_-=10^{-3}$ and $\nu_+=10^{-4}$.  At small electric fields and/or low salt concentrations, the size asymmetry is irrelevant, and the predictions of PB theory from Fig~\ref{fig:mobility} are apparent; at large fields and/or low concentrations, the particle acquires an apparent positive charge, due to the covering of more of the particle's area by the larger cations, regardless of its true surface charge; }
\label{fig:mob_asym}
\end{center}
\end{figure}

A related general consequence is that asymmetry in the electrolyte,
e.g. ions of different effective sizes, can affect a particle's electrophoretic mobility. Remarkably,  an uncharged particle can acquire a nonzero mobility in an asymmetric electrolyte in a large applied field ($\zeta_0=0$ but $b_{ep}\neq 0$).
An asymmetric field-dependence of the double-layer capacitance is enough to predict such exotic effects on very general grounds. 

As a first approximation for the weakly nonlinear regime using MPB theory, we simply postulate two different (homogeneous) ion sizes $a_\pm$ for positive and negative polarization of the diffuse layer,
\begin{equation}
C_D(\Psi_D) = \left\{ \begin{array}{ll}
C_D(\Psi_D; \nu = \nu^+) & \mbox{ for } \Psi_D > 0 \\
C_D(\Psi_D; \nu = \nu^-) & \mbox{ for } \Psi_D < 0 \\
\end{array} \right. \label{eq:Casym}
\end{equation}
where $\nu^\pm = 2 a_\pm^3 c_0$ (or $\Phi^\pm$ for hard sphere models).  This may seem like a crude approximation, but it is quite accurate for a binary electrolyte at large voltages since the diffuse layer mostly contains counterions of one sign; at low voltages, where all species are present in a dilute mixture, the ion sizes play no role. Inserting (\ref{eq:Casym}) into (\ref{eq:Qint}) yields the results shown in Figure~\ref{fig:mob_asym} for the (weakly nonlinear) mobility of a charged, ideally polarizable sphere in a DC field. Interestingly, at large voltages, positive or negative, the mobility tends to a  $|E|$ scaling, set by the ratio $a_+/a_-$, independent of the total charge of the particle $Q$.  We also see that an uncharged particle with $Q=0$ can still have a nonzero mobility, once nonlinear charging of the double layers sets in. 

We stress that all the calculations of this paper consider the weakly nonlinear limit of thin double layers, where the bulk concentration remains uniform and surface conduction is neglected. As such, the trends we predict for PB and MPB models are meaningful at moderate voltages, but may need to be significantly modified at large voltages for strongly nonlinear dynamics.  Moreover, even in the weakly nonlinear regime, we will now argue that at least one more physical important effect may need to be considered.

\section{ Viscoelectric effect in a concentrated solution }
\label{sec:viscosity}

\subsection{ Mean-field local-density theories }

\subsubsection{ Modified Helmholtz-Smoluchowski slip formulae }

There is a considerable literature on electrokinetic phenomena at
highly charged surfaces with large (but constant) zeta
potential~\cite{lyklema_book_vol2,hunter_book,dukhin1974,dukhin1993,delgado2007}.
In this context, it is well known that the linear electrophoretic
mobility departs from Smoluchowski's formula at large surface
potentials, $\Psi > \Psi_c$, and decreases at large voltage, due to
effects of surface conduction (large Dukhin number). Using PB theory
for thin double layers, Dukhin and Semenikhin~\cite{dukhin1974}
derived a formula for the electrophoretic mobility of a highly
charged, non-polarizable sphere, which was famously verified by
O'Brien and White~\cite{obrien1978} via numerical solutions of the
full electrokinetic equations for a dilute solution. This established
the mathematical validity of the formula, but in this article we are
questioning the {\it physical validity} of the underlying equations at
large induced voltages and/or high salt concentrations.

As described in section ~\ref{sec:expt}, recent experiments on
induced-charge electrokinetic phenomena reveal a strong decay of
electro-osmotic mobility with increasing salt concentration at highly
charged surfaces, which  cannot be explained by the standard
model, based on the HS formula (\ref{eq:HS}), even if corrected for
strongly nonlinear effects perturbing the salt
concentration. Continuum electrohydrodynamics, however, does not
require the HS formula, even for thin double layers, but instead
provides a general expression for the electro-osmotic
mobility~\cite{lyklema_book_vol2},
\begin{equation}
  b = \int_0^{\Psi_D} \frac{\varepsilon}{\eta} d\Psi \label{eq:genslip}
\end{equation}
as an integral over the potential difference $\Psi$ entering the
diffuse layer from the bulk. This allows us to derive ``modified
Helmholtz-Smoluchowski'' (MHS) formulae for $b(\Psi_D,c_0,\ldots)$
based on general microscopic electrokinetic equations.  

The standard way to interpret electrokinetic measurements is in terms
of the effective zeta potential,
\begin{equation}
\zeta_{eff} = \frac{b \eta_b }{\varepsilon_b} ,
 \label{eq:zetaeff}
\end{equation}
but we view this as simply a measure of flow generated in units of
voltage, and not a physically meaningful electrostatic potential.  One
can also use PB theory for the hypothetical ``mobile'' part of the
double layer to express the mobility in terms of an effective
``electrokinetic charge''~\cite{delgado2007,lyklema_book_vol2},
e.g. using the Gouy-Chapman solution for a symmetric binary
electrolyte,
\begin{equation}
q_{ek} = 2 \lambda_D z e c_0 \sinh\left( \frac{ze\zeta_{eff}}{2kT}
\right) \label{eq:qek}
\end{equation}
which measures the observed flow in units of charge.

It is well known that the classical theory tends to overpredict
experimentally inferred zeta potentials ($|\zeta_{eff}| < |\Psi_D|$ or
$|q_{ek}|<|q|$), and the discrepancy is often interpretted in terms of
a ``slip plane'' or ``shear plane'' separated from the surface by a
molecular distance $d\geq 0$, where the no-slip boundary condition is
applied~\cite{kirby2004,delgado2007}. If the low-voltage theory with
$\varepsilon=\varepsilon_b$ and $\eta=\eta_b$ is applied beyond this
point, then $\zeta_{eff}$ acquires physical meaning, as the potential
of the slip plane relative to the bulk. Behind the slip plane, the
fluid is assumed to be ``stagnant'' with effectively infinite
viscosity, although it may still have finite ionic surface
conductivity~\cite{delgado2007}.

As in the case of effective hydrodynamic slip ($d<0$) over hydrophobic
surfaces~\cite{vinogradova1999,bocquet2007}  the slip-plane
concept, although useful, obscures the true physics of the
interface. In particular, it has limited applicability to nonlinear
electrokinetic phenomena, where the double layer responds
non-uniformly to a large, time-dependent applied voltage.  We have
already argued this leads to signficant changes in the structure of
the double layer, and clearly it must also have an impact on its
rheology. Without microscopic models of how {\it local} physical
properties, such as viscosity and permittivity, depend on local
variables, such as ion densities or electric field, it is impossible
to predict how the effective electro-osmotic slip depends on the {\it
  global} double-layer voltage or bulk salt concentration. This is
also true in complicated geometries, such as nanochannels or porous
structures, where the concept of a flat ``slip plane'' is
not realistic.

\subsubsection{ The viscoelectric effect }

Based on Eq.~(\ref{eq:genslip}), there are good reasons to expect
reduced mobility at large voltage, $|\Psi_D| > \Psi_c$, compared to
the HS formula, $\zeta_{eff} = \Psi_D$, based only on field-dependent
properties of a polar solvent~\cite{lyklema_book_vol2}. As discussed in section~\ref{sec:dielectric},  large
normal electric fields in a highly charged double layer can decrease
the permittivity by aligning the solvent dipoles (``dielectric saturation"). The local viscosity can
also increase~\cite{vand1948,stokes_book}, through viscoelectric
thickening of a dipolar liquid with polarization transverse to the
shear direction. 

Long ago, Lyklema and
Overbeek~\cite{lyklema1961,lyklema1994} (LO) proposed the first (and to date, 
perhaps the only) microscopic electro-rheological model leading to an
MHS slip formula. They focused on the viscoelectric effect in water, which they estimated to be more important than dielectric saturation for electrokinetics (consistent with recent atomistic simulations~\cite{freund2002}, albeit at low voltage).  Physically, they argued that, for typical electric fields in aqueous double layers, dielectric saturation interferes more with the strongly cooperative rearrangements involved in viscous flow than the weakly correlated dipole alignments contributing to reduced permittivity.  To model the viscoelectric effect, LO assumed a field-squared viscosity increase in PB  theory,
\begin{equation}
\eta = \eta_b \left( 1 + f E^2 \right),   \label{eq:LO}
\end{equation}
and were able to integrate (\ref{eq:genslip}) to obtain an analytical
(although cumbersome) MHS formula. Physically, their model predicts
that $b$ saturates to a constant value at large $|\Psi_D|$, which
decays with increasing $c_0$. The saturation in the LO model, however,
is tied to the unphysical divergence of the counterion concentration
(and thus $E$) in PB theory, and thus should be revisited with volume constraints and other effects.

\begin{figure}
\begin{center}
\includegraphics[width=3.5in]{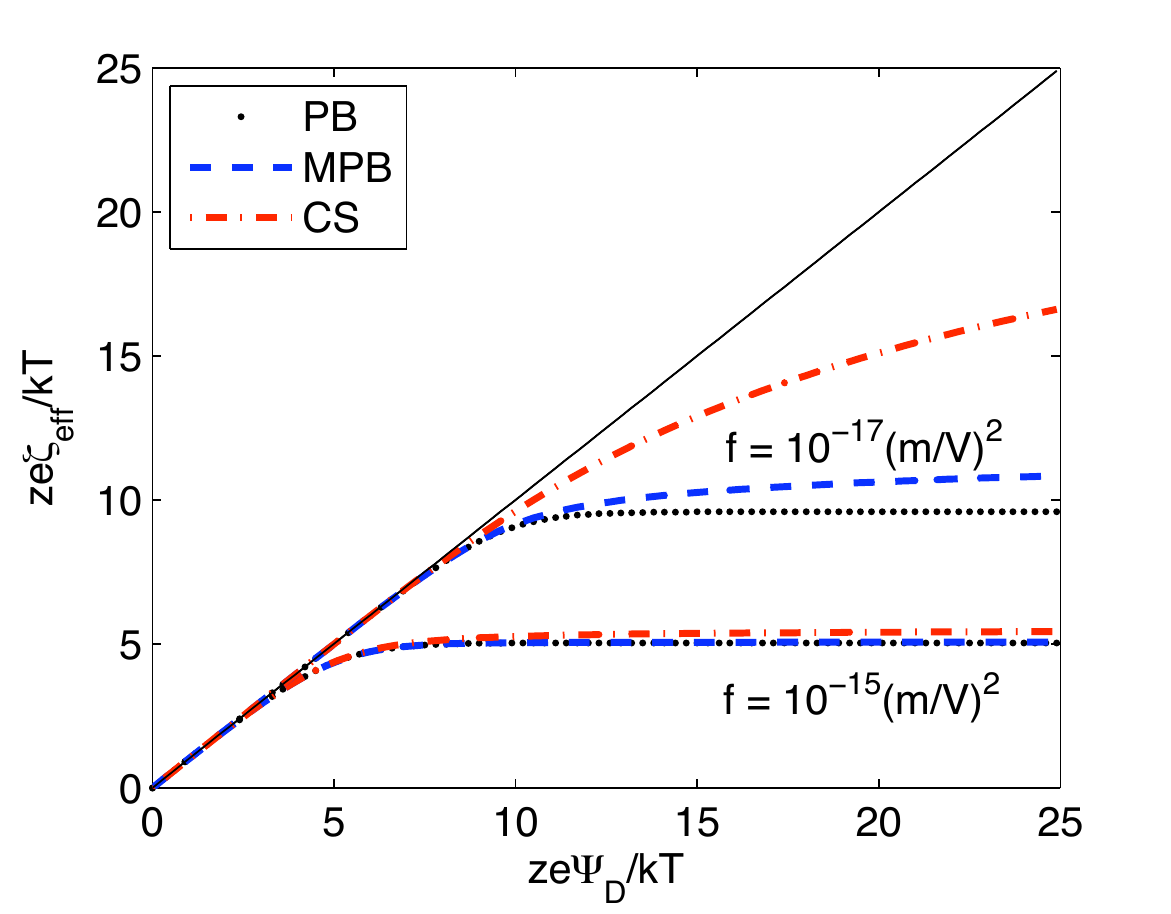}
\caption{ \label{fig:LO_MHS} Modified Helmholtz-Smoluchowski (MHS)
  slip formulae assuming a charge-independent viscoelectric effect in
  the polar solvent (\ref{eq:LO}). This example assumes a bulk
  concentration $c_0 = 1$ mM of $z:z$ electrolyte using different
  models of double-layer structure. The Lyklema-Overbeek (LO) model,
  based on Poisson-Boltzmann (PB) theory of point-like ions, is
  compared to MHS slip with the Bikerman (MPB) and Carnahan-Starling
  (CS) modified PB theories, using an effective ion size of
  $a=4$ \AA. The viscoelectric coefficient is set to the value $f =
  10^{-15} \mbox{m}^2\mbox{V}^{-2}$ suggested by LO for water as well
  as a smaller value $f = 10^{-17} \mbox{m}^2\mbox{V}^{-2}$, which
  reduces the viscoelectric effect.  }
\end{center}
\end{figure}

It is straightforward to use the LO postulate (\ref{eq:LO}) for the
viscoelectric effect in our modified PB models to obtain corresponding
MHS slip relations, although the integration of (\ref{eq:genslip})
cannot be done analytically. Numerical results for a typical case are
shown in Fig.~\ref{fig:LO_MHS}, using the value $f = 10^{-15}
\mbox{m}^2\mbox{V}^{-2}$ suggested by LO for
water~\cite{lyklema1961,lyklema1994}. It is interesting to note that
this choice makes all three models of double layer structure, PB,
Bikerman MPB, and CS MPB, yield very similiar electro-osmotic mobility
versus voltage, in spite of completely different ion density
profiles. The reason is that the viscoelectric effect sets in so
quickly with increasing voltage that the shear plane is effectively
still in the dilute, outer part of the diffuse layer, where all
theories reduce to PB. Indeed, as shown in the figure, more
differences become apparent we if choose a smaller value, $f =
10^{-17} \mbox{m}^2\mbox{V}^{-2}$. The reduction in normal electric
field due to crowding-induced expansion of the inner diffuse layer
leads to slower saturation of $\zeta_{eff}$ compared to PB theory,
especially in the CS MPB model, since it has stronger steric effects
than the Bikerman MPB model.

What this exercise shows is that the inner part of a highly charged double layer is effectively frozen by field-dependent viscosity in a similar way regardless of the model for the diffuse-charge profile. The LO model attributes this effect entirely to the solvent (water), independent of the local diffuse charge density or ionic currents, but this hypothesis does not seem entirely satisfactory. An implication is that water is effectively immobilized within a few molecular layers of the surface near a charged surface, even if there is no added salt. This effect seems to contradict  recent experimental and theoretical literature on hydrodynamic slip~\cite{vinogradova1999,stone2004,lauga2005,bocquet2007}
which  has shown that shear flow in water persists down to the atomic scale with a no-slip boundary condition for smooth hydrophilic surfaces and with significant slip lengths (up to tens of nanometers) on hydrophobic surfaces~\cite{vinogradova2003,charlaix.e:2005,joly.l:2006}. If the viscosity is a property of pure water (as opposed to the local electrolytic solution), then the field dependence should also be observable in the bulk, but we are not aware of any experiments or simulations showing that bulk de-ionized water becomes rigid in fields larger than $f^{-1} = 30$V/$\mu$m. Of course, it is hard to apply such fields in the bulk at low frequency, due to capacitive screening by the double layers and Faradaic reactions, but the LO model is time-independent and should also apply at high frequency where these effects are reduced.  

Molecular dynamics (MD) simulations of electrokinetic phenomena in nanochannels generally imply a local viscosity increase close to a charged surface, although its origin and mathematical description are not well understood. An early  MD simulation of Lyklema et al.~\cite{lyklema1998} showed a stagnant monolayer of water, which could pass ions freely, with a surface conductivity comparable to the bulk, but subsequent MD simulations of electro-osmosis have shown motion of the liquid (both ions and solvent) down to the atomic scale near the wall~\cite{freund2002,qiao2003,qiao2004,qiao2005,thompson2003,joly2004,lorenz2008}, albeit with an apparent viscosity often smoothly increasing across the closest molecular layers. We are not aware of any MD simulations of electro-osmosis in a large transverse voltage ($\gg kT/e$), but it seems that the presence of crowded counterions near the surface (described in section~\ref{sec:crowding}) must affect the apparent viscosity, since the liquid no longer resembles the pure bulk solvent.

\subsubsection{ Charge-induced thickening }

\begin{figure}
\begin{center}
\includegraphics[width=3.5in]{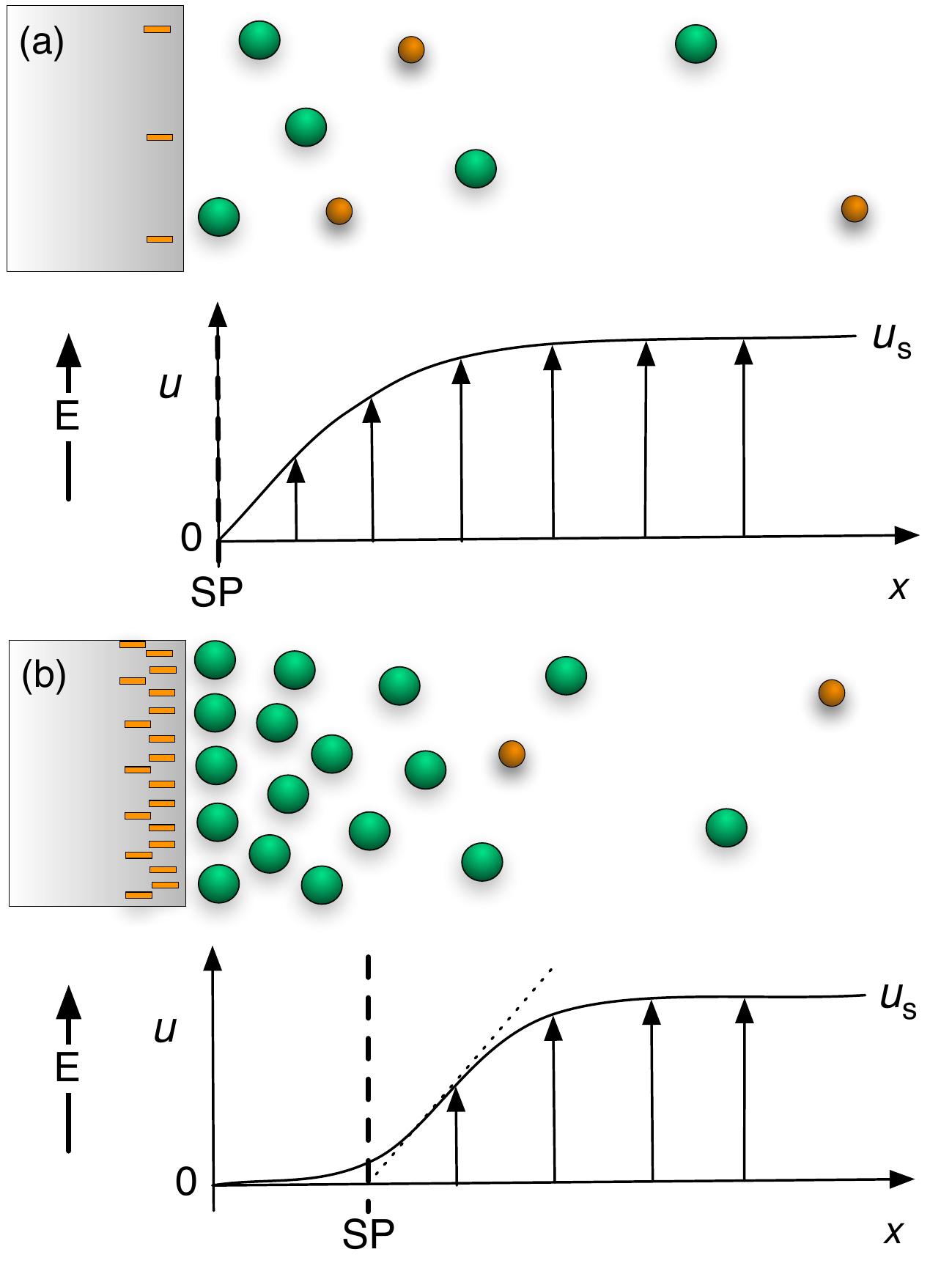}
\caption{\label{fig:slip_cartoons} Sketches of finite-sized hydrated
  ions near a polarizable surface as in Fig.~\ref{fig:cartoons},
  showing the solution velocity $u$ profile in response to a
  tangential electric field. (a) At small induced voltages, the
  no-slip boundary condition holds at the surface, and the effective
  electro-osmotic slip $u_s$ builds up exponentially across the
  diffuse layer. (b) At large voltages, crowding of hydrated ions
  increases the viscosity of the condensed layer, and the apparent
  slip plane ``SP'' (dashed line) moves away from the surface with
  increasing voltage. }
\end{center}
\end{figure}

A natural physical hypothesis, sketched in
Figure~\ref{fig:slip_cartoons}, is that the crowding of counterions in
a highly charged diffuse layer increases the local viscosity, not only
through the viscoelectric effect of the bare solvent, but also through
the presence of a large volume fraction of like-charged ions
compressed against the surface. For now, we neglect explicit field
dependence, such as (\ref{eq:LO}), to focus on effects of large charge
density. In a very crude attempt at a model, we consider an
electrolyte with solvated ions of finite size and postulate that
$\varepsilon/\eta$ diverges as a power law,
\begin{equation}
  \frac{\varepsilon}{\eta} = \frac{\varepsilon_b}{\eta_b} \left[ 1 -
    \left( \frac{|\rho|}{\rho_j^\pm} \right)^\alpha \right]^\beta   \label{eq:visc}
\end{equation}
as the local charge density approaches a critical value, $\rho_j^\pm$,
which generally must depend on the polarity $\mbox{sign}(\rho)=\pm$ in
an asymmetric electrolyte. 

\begin{figure}
\begin{center}
\includegraphics[width=2.6in]{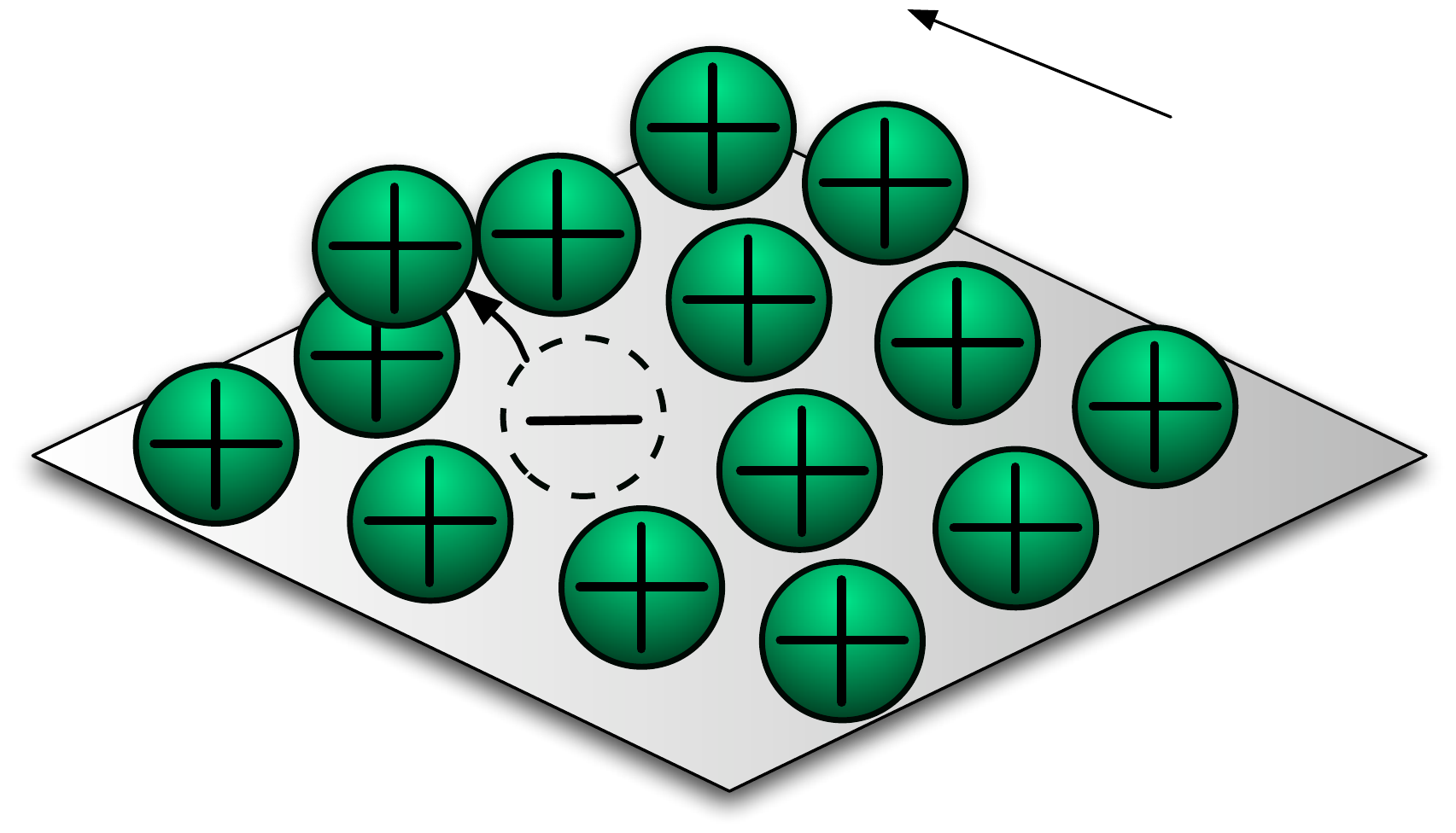}
\caption{\label{fig:hole} Sketch of a physical mechanism for charge-induced thickening due to electrostatic correlations.  A condensed layer of solvated counterions (spheres) is confined against a highly charged surface by the normal electric field. Shear of the fluid (arrow) causes an ion to leave its local equilibrium position in a Wigner-like crystal of like charges, but its motion is inhibited by a strong attraction back to its oppositely charged "correlation hole'' (dashed). }
\end{center}
\end{figure}

A natural choice is to postulate diverging viscosity (``jamming'') at the
steric limit, $\rho_j = \rho_{max}^\pm=|z_\pm|e c_{max}^\pm$. In that
case, similar exponents $\alpha$ and $\beta$ controlling the
singularity also arise in the rheology of dense granular
materials~\cite{olsson2007}. If we also assume $\alpha=\beta=1$, then
there are no new fitting parameters,
\begin{equation}
  \frac{\varepsilon}{\eta} = \frac{\varepsilon_b}{\eta_b} \left( 1 -
    \frac{|\rho|}{\rho_{max}^\pm} \right)   \label{eq:viscmax}
\end{equation}
since the steric constraints $\rho_{max}^\pm$ come from the MPB
model. In our original letter~\cite{large}, we considered only this
postulate with Bikerman's model for steric effects and thus assumed an
even simpler form,
\begin{equation}
  \frac{\varepsilon}{\eta} = \frac{\varepsilon_b}{\eta_b} \left( 1 -
   \frac{a^3 |\rho|}{ze} \right)   \label{eq:visc1}
\end{equation}
The resulting electrokinetic model is extremely simple in that it only
involves one parameter beyond dilute-solution theory, the effective
ion size, $a$. As such, it lacks flexibility to fit multiple sets of
experimental data, but we will use it in our analytical calculations
below to further understand the general consequences of steric
constraints.

The general model (\ref{eq:visc}) also allows for other types of
behavior.  With $\rho_j^\pm > \rho_{max}^\pm$, there is a finite
maximum viscosity in the condensed layer, or effectively some flow
behind the slip plane. With $\rho_j^\pm < \rho_{max}^\pm$, the model
postulates flow arrest at high charge density before the close-packing
limit is reached (and $\eta=\infty$ for $|\rho| > \rho_j^\pm$). The
new parameters $\rho_j^\pm$, $\alpha$ and $\beta$ allow some
flexibility to fit experiments or simulations, in addition to the ion
sizes $a_\pm$ from the MPB models above.

The arbitrary choice (\ref{eq:visc}) is motivated by a number of
possible physical effects:
\begin{itemize}
\item {\it Jamming against a surface.} In a bulk
  colloid~\cite{russel_book}, the maximum density corresponds to
  random close packing at the jamming point~\cite{liu1998}, where the
  shear modulus becomes finite~\cite{ohern2003}, diffusivity vanishes~\cite{gotze1992,kob1993,levin2001} and the viscosity
  diverges~\cite{olsson2007}. Molecular dynamics simulations of soft
  disks in a periodic box have recently established a viscosity
  divergence of the form (\ref{eq:visc}) with $\alpha=1$ and
  $\beta=1.7$ ~\cite{olsson2007}. In an electrolyte, strong electrostatic compression of
  solvated counterions against a (typically rough) surface may cause
  some transient local jamming of the condensed layer, and thus
  increased viscosity (and decreased mobility of individual ions) at high charge density.
\item {\it Electrostatic correlations.} Condensed counterions at large
  voltages ressemble a Wigner crystal~\cite{shklovskii1999,nguyen2001}
  (or glass) of like charges. Discrete Coulomb
  interactions~\cite{attard1996,grosberg2002,levin2002} may contribute to
  increased viscosity, e.g.  through the attraction between a
  displaced ion and its ``correlation hole'', which effectively carries an opposite charge. (See Fig.~\ref{fig:hole}.)
  We are not aware of attempts to predict the rheological response of sheared Wigner
  crystals or glasses, let alone discrete, correlated counterion
  layers, but we expect that electrostatic correlations will generally
  contribute to charge-induced thickening in electrolytes and ionic
  liquids.
\item {\it Solvent effects.} Several molecular dynamics simulations of
  linear electro-osmosis (at low voltages) have infered $\approx
  5\times$ greater viscosity within 1 nm of a flat
  surface~\cite{freund2002,qiao2003}, which grows with surface
  charge~\cite{qiao2005} and surface roughness~\cite{kim2006} (but
  decreases with hydrophobicity~\cite{joly2004}). Dielectric saturation near a highly charged surface (section ~\ref{sec:dielectric}) reduces the electro-osmotic
  mobility by lowering $\varepsilon$, but the aligning of dipoles leads to collective interactions that can also increase $\eta$, i.e. the classical viscoelectric effect, at even lower fields~\cite{lyklema1961}. In the presence of crowded, like-charged counter-ions, frustrated solvation shells and confined hydrogen-bond networks in water may
  further increase $\eta$. The latter effect may be related to
  local electrostriction, which compresses the second hydration shell
  around certain cations  (K$^+$ and Na$^+$,
  also used in ICEO experiments, Table ~\ref{table:Lambda}) ~\cite{mancinelli2007}.
\end{itemize}

Of course, this electrohydrodynamic model is rather simple. In
general, we expect that $\varepsilon$ and $\eta$ will depend
independently on both the local field $E$ and the solution
composition. We have already mentioned the different physical effects
affecting $\varepsilon$ and $\eta$ in a pure dipolar solvent, and the
situation only becomes more complicated with large ion densities. In
order to model general phenomena such as diffusio-osmosis or surface
conduction, it is necessary to provide separate functional forms for
$\varepsilon$ and $\eta$, since these variables no longer only appear
as the ratio $\varepsilon/\eta$. In that case, we would propose
viewing (\ref{eq:visc}) as a model for $\eta$ with
$\varepsilon=\varepsilon_b$ fixed, since molecular dynamics
simulations of electro-osmosis predict smaller changes in permittivity
than in viscosity near a charged surface, even within a few molecular
diameters~\cite{freund2002,qiao2003}. This is also analytically
convenient because variable $\varepsilon$ is a major complication in
MPB models, although progress can be
made as shown in section ~\ref{sec:dielectric}. Finally, our neglect of explicit
field-dependence such as (\ref{eq:LO}), in favor of charge-density
dependence (which is more closely related to our modeling of crowding
effects), does not affect the modeling of quasi-equilibrium double
layers, since the dominant normal field can be expressed in terms of
the charge-density, and vice versa, in any MPB model. In more general
models, however, it may be important to also include a field-dependent
viscoelectric effect.
  
Our basic picture of charge induced thickening is consistent with prior models 
for the increase of viscosity with salt concentration in neutral bulk
solutions \cite{vand1948,stokes_book,gering2006}. 
Traditionally, the viscosity of  electrolytes has been 
described empirically using the Dole-Jones correlations \cite{jones1929,jones1933,jenkins1995}, but 
more recent theories based on molecular models 
take a more fundamental approach. Like our 
hypothesis, these models also postulate infinite viscosity as the hard sphere solvated 
volume fraction approaches unity \cite{gering2006}. 
In a neutral bulk electrolyte, such high volume fractions are never reached due to 
limits of solubility, though a factor of ten increase in viscosity has been observed for
some systems at high salt concentration \cite{CRC}.
The effective hard-sphere diameters (see Table \ref{table:ionsize}) inferred from these viscosity 
models may also have relevance for our models, although there is an
 important difference: We neglect crowding effects in a neutral bulk
 solution and focus instead on crowding of like-charged counter-ions. 
 As discussed above, at high charge density in a large electric field,
 various physical effects could significantly  increase the  viscosity
 compared to what would be observed in a neutral solution at the same 
concentration. Moreover, solubility limits are not relevant for like-charged
 ions, so higher concentrations approaching packing limits can more easily
 be reached, as we have postulated in response to a large voltage.

Regardless of the specific form of our model, its key feature is that
the concentrated solution of counterions near a highly charged surface
is thickened with voltage, compared to the bulk, although it can still
flow slowly. This picture is consistent with molecular dynamics simulations of
electro-osmosis~\cite{freund2002,qiao2003,qiao2005,joly2004,kim2006,chakraborty2008},
which do not observe a truly stagnant layer, even at low surface
potentials. As sketched in Fig.~\ref{fig:slip_cartoons}, from a
macroscopic point of view, there is an apparent slip plane separated
from the surface, which depends on the ion density profiles and thus
external conditions of voltage and bulk solution composition. At low
voltage, near the point of zero charge, there is no change in the
local viscosity, aside from any pure solvent interactions with the
surface, e.g. due to hydrophobic or hydrophilic effects (which we
neglect). At high voltage, the crowding of counterions leads to
thickening and an apparent movement of the slip plane away from the
surface.  Without a microscopic model, it would not be possible to
predict this dependence or derive a functional form for fitting.

The picture of a thin layer near the surface with different
electro-rheological properties also appears in models of
electrokinetics with porous or soft
surfaces~\cite{lopez2003a,lopez2003b}, which build on the
Zukowski-Saville~\cite{zukowski1986a,zukowski1986b} and
Mangelsdorf-White~\cite{mangelsdorf1990,mangelsdorf1998a,mangelsdorf1998b}
theories of the ``dynamical Stern layer''. In these models, ions are
allowed to move within a flat Stern monolayer, while the diffuse layer
is described by the standard electrokinetic model. Recently,
L\'opez-Garcia, Grosse and Horno have extended these models to allow
for some fluid flow in a thicker dynamical surface layer, and this
allows more flexibility in fitting (linear) electrical and
electrokinetic measurements~\cite{lopez2007,lopez2009}. In the case of a flat
surface with an equilibrium double layer, this is similar to our
picture, if the surface layer is ascribed a higher viscosity, but in
our model there is no need to postulate a sharp plane where properties
of the solution change. Instead, by modifying the electrokinetic
equations everywhere in the solution, its electro-rheology changes
continuously as a function of local contiuum variables.

The concept of charge-induced viscosity increase may be widely
applicable in nano-scale modeling of electrolytes, in conjunction with
modified theories of the ions densities. The general modified
electrokinetic equations are summarized in
section~\ref{sec:disc}. Now, let us consider generic consequences of
this hypothesis for electrokinetic phenomena with thin double layers
and make some explicit calculations using Eq.~ (\ref{eq:visc}).

\subsection{ Implications for nonlinear electrokinetics }

\subsubsection{ Electro-osmotic slip at large voltages in concentrated solutions }

To describe electro-osmotic flows with thin double layers, we can use
our microscopic models to derive modified Helmholtz-Smoluchowski (MHS) slip formulae, which depend
nonlinearly on double-layer voltage (or surface charge) and bulk salt
concentration. Such predictions could perhaps be directly tested
experimentally by field-effect flow control of electro-osmosis in
microfluidic devices~\cite{schasfoort1999,wouden2005,wouden2006}, if a
nearly constant double-layer voltage could be maintained along a
channel and then varied systematically for different electrolytic
solutions. Alternatively, one could compare with molecular dynamics
simulations of electro-osmotic flow in nanochannels with oppositely
and highly charged walls, leading to large induced double-layer
voltages.

{\it A general MHS slip formula --- } If we assume $\varepsilon =
\varepsilon_b$ in Eq.~ (\ref{eq:viscmax}) with $\alpha=\beta=1$ (an
arbitrary choice for analytical convenience), then for any MPB model
with $|\rho_{max}^\pm|<\infty$ we can integrate Eq.~(\ref{eq:genslip})
obtain an MHS formula for the effective electro-osmotic slip outside
the double layer,
\begin{equation}
  \zeta_{eff} = \Psi_D \pm
  \frac{p_e(\Psi_D)}{\rho_j^\pm} 
= \Psi_D \pm
  \frac{q(\Psi_D)^2}{2\varepsilon_b\rho_j^\pm} 
 \label{eq:zeta_eff} 
\end{equation}
where $\pm = \mbox{sign}(q) = - \mbox{sign}(\Psi_D)$ and $p_e(\Psi_D)$
and $q(\Psi_D)$ are given by (\ref{eq:E}) and (\ref{eq:q}),
respectively. This simple and general expression reduces to the HS
formula ($\zeta_{eff} = \Psi_D$) in the limit of point-like ions
($\rho_{max}^\pm \to \infty$) and/or small voltages ($\Psi_D \ll
kT/e$). At large voltages, $|\zeta_{eff}|$ is always reduced,
depending on how the model for $\mu_i^{ex}$ imposes volume constraints
or other contributions to the excess chemical potential.

\begin{figure}
\begin{center}
 \includegraphics[width=3.5in]{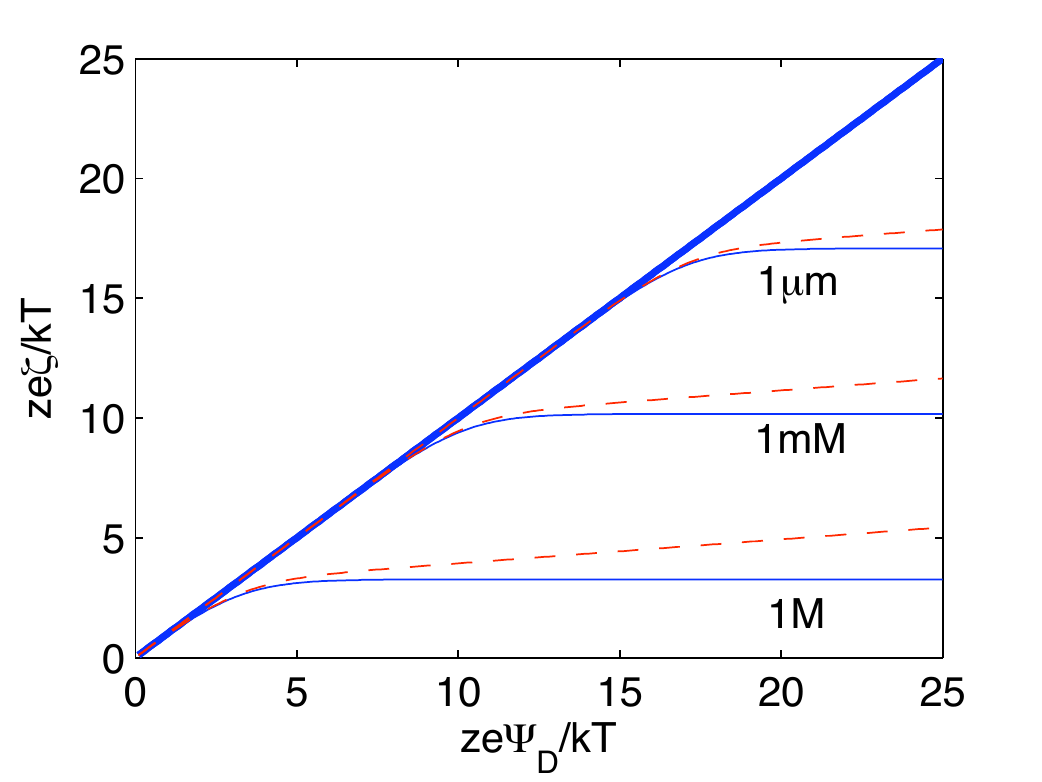}\\
 \caption{ \label{fig:B_MHS} MHS slip formula (\ref{eq:mslip}) using
   Bikerman's model with ion size $a=4$\AA and charge-induced
   thickening (\ref{eq:visc}) with $\alpha=1=\beta=1$ at
   different bulk concentrations, $c_0 = 1 \mu$M, 1 mM, and 1M. The
   viscosity is postulated to diverge either at a mean ion spacing
   $a_j=a$ (solid curves), in which case the condensed layer in
   Fig.~\ref{fig:C}(a) is effectively rigid, or at $a_j=0.9 a$ (dashed
   curves), in which case it flows with a large, but finite viscosity.
 }
\label{fig:MHSslip}
\end{center}
\end{figure}

In any MPB model, such as Bikerman's, where counterions ($\pm$) form
nearly uniform condensed layers with $c_\pm \approx c_{max}^\pm$ and
$\eta \gg \eta_b$ from Eq. (\ref{eq:visc}), the apparent zeta
potential (\ref{eq:zeta_eff}) either saturates to a constant,
\begin{equation}
  \zeta_{eff} \sim
  \Psi_c^{\pm} = -\frac{kT}{z_\pm
    e}\ln\left(\frac{c_{max}^\pm}{c_0}\right)   \label{eq:zeta_lim}
\end{equation}
if the condensed layer is stagnant ($\rho_j^\pm = \rho_{max}^\pm$) or
switches to a slower linear dependence
\begin{equation}
  \zeta_{eff} \sim \Psi_D - \left[ 1 -
    \left(\frac{a_j^\pm}{a_\pm}\right)^3 \right]
\left(\Psi_D^\pm  -
\Psi_c^{\pm}\right)  \label{eq:viscous_layer}  
\end{equation}
if the condensed layer has a finite viscosity, as shown in
Fig.~\ref{fig:B_MHS}. The former case (\ref{eq:zeta_lim}) ressembles
the logarithmic concentration dependence of equilibrium zeta potential
observed in many microfluidic systems~\cite{kirby2004}, as well as the
decay of ICEO flow noted above. The latter case allows for
intermediate behavior between strong saturation of $\zeta_{eff}\sim
\Psi_c^\pm$ and the HS limit $\zeta_{eff} = \Psi_D$.

{\it MHS slip with Bikerman's model --- } The reduction of
$\zeta_{eff}$ arises in different ways depending on the diffuse-layer
model. 
In Bikerman's model
for a symmetric electrolyte~\cite{large}, the limiting behavior is 
reached fairly suddenly. In that case the integral
(\ref{eq:visc1}) can be performed analytically to obtain a simple
formula:
\begin{equation}
  \zeta_{eff}^\nu = \Psi_D \pm \frac{kT}{ze} \,   \left(\frac{a_j}{a}\right)^3 
  \ln\left[ 1 + 4 a^3 c_0 \sinh^{2}\left( \frac{ze\Psi_D}{2kT}
    \right)\right],    \label{eq:mslip}
\end{equation}
illustrated in Fig.~\ref{fig:B_MHS}. For a rigid condensed layer,
this model predicts a simple logarithmic decay of ICEO flow with
concentration, Eq. (\ref{eq:zeta_lim}). If the condensed layer has a
large, but finite viscosity, the decay is slower and more complex via
Eq.  (\ref{eq:viscous_layer}). A general feature of even these very
simple models is that ICEO flow becomes concentration-dependent and
ion-specific at large voltages and/or high salt concentrations, 
through $z_\pm$, $a_\pm$, $a_j^\pm$,and $c_0$.

It is interesting to compare Eq.~(\ref{eq:mslip}) to the only previous
MHS formula of Lyklema and Overbeek ~\cite{lyklema1961,lyklema1994},
based on the viscoelectric effect (\ref{eq:LO}) in the context of PB dilute-solution
theory. As shown in Figures~\ref{fig:LO_MHS} and \ref{fig:B_MHS}, the
two formulae make similar predictions of saturation of the zeta
potential with voltage with $a_j=a$, but our formula (\ref{eq:mslip})
is simpler and more amenable to mathematical analysis, as illustrated
below. (In contrast, the LO formula takes several lines to write down
in closed form~\cite{lyklema1994}.) The parameter $a$ is also more constrained on
physical grounds, to be of order the hydrated ion size, than the
empirical viscoelectric constant $f$. Qualitatively, the LO formula
based on PB theory does not offer any explanation of the experimental fact that ICEO
flows depend on the particular ions, even in different solutions of
the same ionic valences, $\{z_i\}$ (such as NaCl and KCl), although we
do not claim that our Bikerman MHS formula correctly captures any specific
trends.

\begin{figure}
\begin{center}
 \includegraphics[width=3.5in]{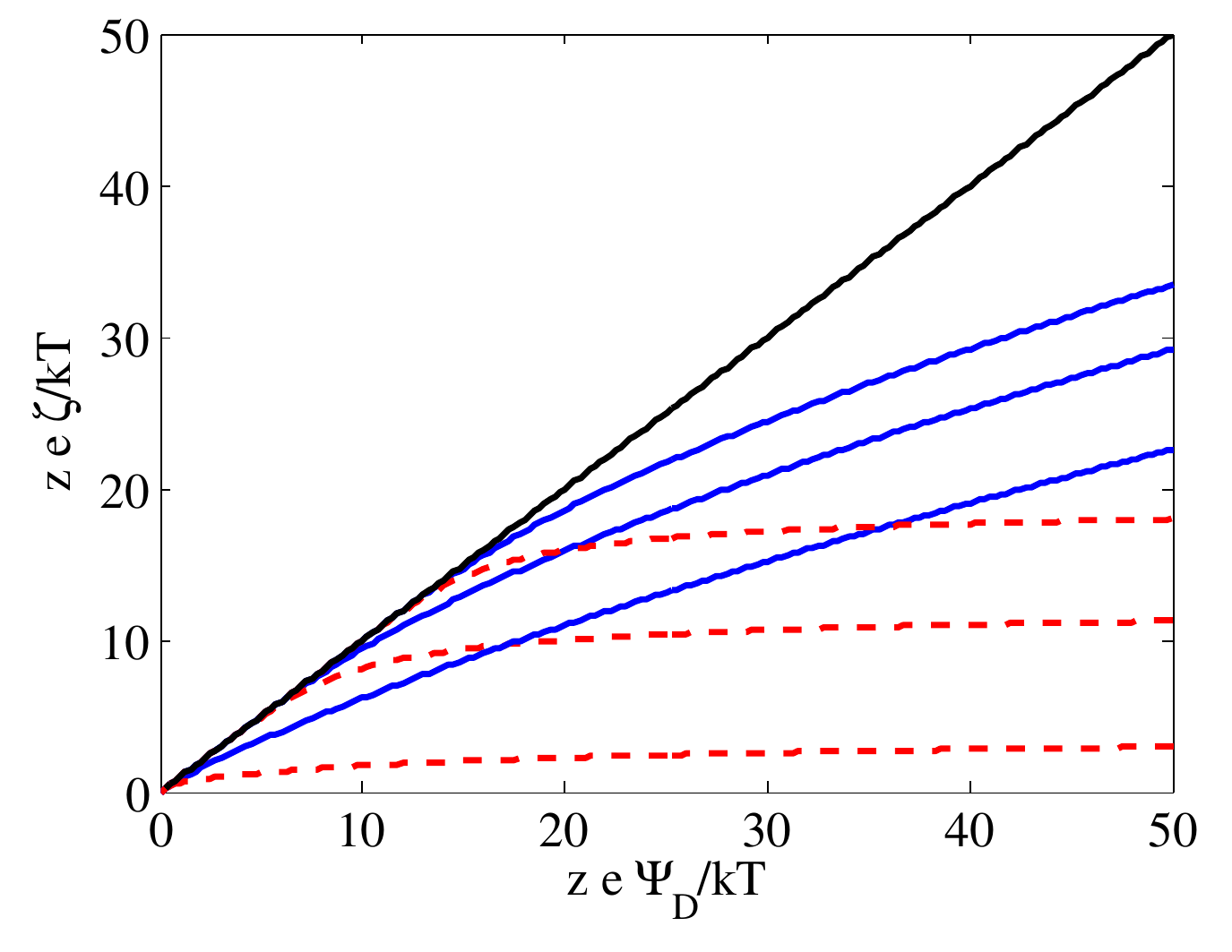}
\caption{ \label{fig:CS_MHS} Effective zeta potential $\zeta_{eff}$
  versus diffuse-layer voltage $\Psi_D$ at different bulk
  concentrations using the Carnahan-Starling MPB
  model for charged hard spheres of diameter $a=4$ \AA from
  Fig.~\ref{fig:CS}. The concentration are
  $c_0=1 \mu$M, 1 mM, and 1 M from top to bottom. The solid curves 
  use the MHS slip formula
  (\ref{eq:visc}) with $\alpha=\beta=1$ and  $\rho_j =
  \rho_{max}$, and the dashed curves change to $\beta=4$.
}
\end{center}
\end{figure}

{\it MHS slip in a hard-sphere liquid --- } As noted above,
hard-sphere liquid models show  qualitative differences
with Bikerman's lattice-based model, beyond just allowing the use of
more realistic, smaller effective ion sizes. As shown in
Fig.~\ref{fig:CS}, steric effects are stronger in a hard-sphere
liquid. The ion density  approaches close packing as the voltage is 
increased  much more slowly 
with the Carnahan-Starling model when compared to Bikerman's model, as shown in 
Fig.~\ref{fig:CS}(d). When the CS model is used to compute the effective
zeta potential with  $\rho_j^\pm= \rho_{max}^\pm$ and $\alpha=\beta=1$ 
in Eq.~(\ref{eq:visc}) as shown in  Fig.~\ref{fig:CS_MHS}, 
the effective zeta  potential does not
saturate as with Bikerman's model, and the layer continues to flow
until extremely high voltages are reached, albeit much more slowly than in the classical HS model without charge-induced thickening.

However, at this time
we have no reason to assume that  $\alpha=\beta=1$ in Eq.~(\ref{eq:visc}). 
The crucial
exponent controlling the viscosity divergence is $\beta$. If we set
$\beta=4$, then the divergence is fairly strong as shown by the dashed
curves in  Fig.~\ref{fig:CS_MHS}.  
Here, we find  a strong saturation of
$\zeta_{eff}$, similar to what is predicted by Bikerman's model
(Fig.~\ref{fig:B_MHS}), which more easily forms a condensed layer. 
A range of possible MHS slip behavior is possible with a 
given MPB model for the ion density profile, depending on the precise postulate for charge-induced thickening.

It is important to note that the CS equation was never intended to be used at high 
volume fractions and only fits the homogeneous chemical potential in the liquid state for $\Phi < 0.55$. From Eq. \ref{eq:CS}
and Fig.~\ref{fig:CS}(d) we see that the chemical potential 
diverges as $\Phi \to 1$, but the 
physical maximum 
for random close packing for bulk hard spheres 
is $\Phi\approx 0.63$ 
~\cite{liu1998,ohern2003,olsson2007}, which is exceeded in the CS MPB theory already at 1 volt across the diffuse layer. Moreover, crowding in the double layer occurs against a hard wall, which  removes geometrical degrees of freedom and thus reduce the accessible local volume fraction for random close packing of hard spheres~\cite{rycroft2006}, and induces correlations not captured by the LDA~\cite{antypov2005}.  It is therefore not clear
what the proper value of $\rho_{max}^\pm$ should be to control our postulated viscosity divergence. The 
development of accurate models of the local rheology of highly charged
double layers should ideally be guided by molecular theories and
simulations. We simply give a range of examples to show what kind of
qualitative behavior can be predicted by various simple MPB/MHS
models, which are suitable for macroscopic theory and simulation of
electrokinetic phenomena.

\begin{figure}
\begin{center}
\includegraphics[width=5in]{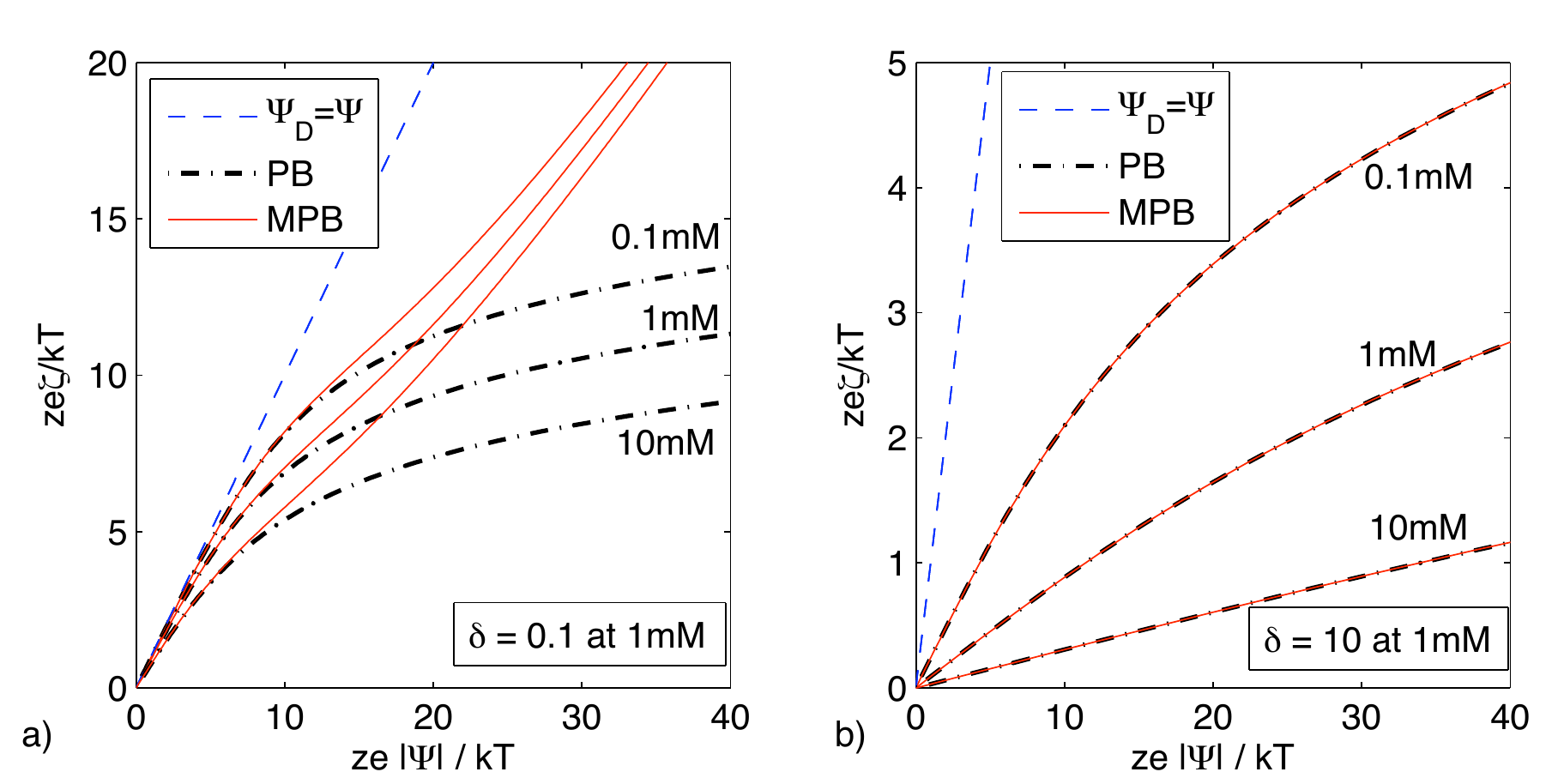}
\caption{  Compact layer effects in traditional HS slip theory assuming a thin dielectric coating between the surface and the diffuse layer, whose importance is controlled by the parameter $\delta = \lambda_S/\lambda_D(c_0) = C_D(c_0)/C_S$ defined in Eq.~(\ref{eq:Laminv}). Both PB (dot-dash curves) and Bikerman's MPB (solid curves) models are considered for the diffuse layer. The zeta potential, equal to the diffuse-layer voltage $\zeta= \Psi_D$, is plotted versus the total double layer voltage $\Psi$ at different values of the bulk salt concentration $c_0$ (labeled) for $\delta=0.1$ in (a) and $\delta=10$ in (b). }
\label{fig:sternHS}
\end{center}
\end{figure}

{\it Comparison to compact-layer models. } For completeness, we briefly discuss how the general liquid-state models we develop above compare to the traditional approach of dividing the double layer into two parts, a flowing diffuse layer and a rigid compact layer, and assuming a constant viscosity and permittivity everywhere, leading to the HS formula. First we consider the classical Gouy-Chapman-Stern model which postulates an uncharged dielectric monolayer of solvent of constant effective thickness $\lambda_S$ in contact with a diffuse layer of point-like ions obeying PB theory. The nonlinear charge-voltage relation~\cite{bazant2004},
\begin{equation}
\Psi = \Psi_D + 2 \frac{kT \lambda_D}{ze \lambda_S} \sinh\left(\frac{ze\Psi_D}{2kT}\right),
\end{equation}
then implies that only a logarithmically small diffuse-layer voltage $\Psi_D$ contributes to the zeta potential at large total voltage $\Psi$ applied to the double layer,
\begin{equation}
\zeta = \Psi_D \sim \frac{kT}{ze}  \ln\left(\frac{\varepsilon_b \Psi^2}{2kT\lambda_S^2c_0}\right)  \left[ 1- 
\frac{kT}{ze\Psi} + \ldots \right].
\end{equation} 
This HS model also predicts that the zeta potential decays logarithmically with concentration and somewhat resembles our MHS models, as shown in Figure~\ref{fig:sternHS}. 

Although the thin-dielectric compact-layer hypothesis leads to some reasonable predictions, it is not fully satisfactory either. As noted above, the saturation of zeta depends on the pileup of point-like ions, albeit reduced by transferring most of the voltage to the compact layer.  This can be avoided by using an MPB model with steric constraints, such as Bikerman's model, together with a Stern layer and HS slip. As shown in Fig.~\ref{fig:sternHS}(a), this tends to reduce the Stern-layer effect, since the diffuse layer is able to carry more voltage due to its reduced capacitance. With increasing $\delta$, however, the Stern layer always carries most of the voltage, and the difference between PB and MPB models on HS slip with the Stern model is eventually lost, as shown in Fig.~\ref{fig:sternHS}(b) for $\delta=10$. 

This exercise shows that some concentration and voltage effects on slip can be captured by the classical Stern model and HS slip formula, but we are left with the same general criticisms made above for charging dynamics. The model places most of the large voltage across a region for which no detailed physical model is assumed, without attempting to account for liquid-state properties, salt concentration, surface roughness, liquid-surface interactions, etc. It is also not clear that a hypothetical Stern monolayer of solvent could withstand several Volts, while leaving a dilute diffuse layer with a small voltage, and some model for dynamical effects in response to applied voltages should be required. 

We have already seen in section~\ref{sec:expt} that the classical model is unable to predict all the trends in nonlinear electrokinetics.  Except in the case of a true dielectric coating (e.g a native oxide on a metal electrode), it seems more physically realistic to discard the classical concept of the compact layer (defined as the inner region where PB theory does not apply) and instead ascribe all of the double layer response to the dynamics of the liquid state, except perhaps for the dielectric response of a true Stern monolayer solvating the charged surface. An appropriate modified theory of the liquid can approximate the traditional properties of the compact layer through strong molecular interactions, as sketched in Fig.~\ref{fig:slip_cartoons}. At least in this work, we have shown that it is possible to describe a variety of nonlinear effects in charging dynamics and electro-osmotic flow at large voltages and/or large salt concentrations without resorting to lumping the errors from dilute solution theory in a hypothetical compact layer outside the continuum model. Perhaps a more accurate  theory would combine the ideas of this paper for the liquid phase with boundary conditions representing a compact interfacial phase, along the lines of the dynamical Stern layer model~\cite{zukowski1986a,mangelsdorf1990,lopez2007,lopez2009}.

\subsubsection{ Ion-specific electrophoretic mobility } 

We have already noted that induced-charge electrokinetic phenomena are
sensitive to the solution composition in our models, via both the
nonlinear capacitance and the effective zeta potential. There are some
surprising, general consequences, which are well illustrated by ICEP
of an uncharged metallic sphere. It has been predicted using
low-voltage models~\cite{iceo2004a,squires2006,yariv2005} and
observed~\cite{murtsovkin1990,gangwal2008} that asymmetric polarizable
particles in a uniform field have an ICEP velocity scaling as $E_b^2$,
but it is widely believed that linear velocity scaling, $U = b_{ep}
E_b$, implies nonzero total charge. For non-polarizable particles with
thin double layers, this is the case, unless the particle has both
asymmetric shape and a non-uniform charge
density~\cite{long1996,long1998}. For polarizable particles, our
models predict that a nonzero mobility $b_{ep}$ can result simply from
{\it broken symmetry in the electrolyte}, even for a perfectly
symmetric particle.

\begin{figure}
\begin{center}
\includegraphics[width=4in]{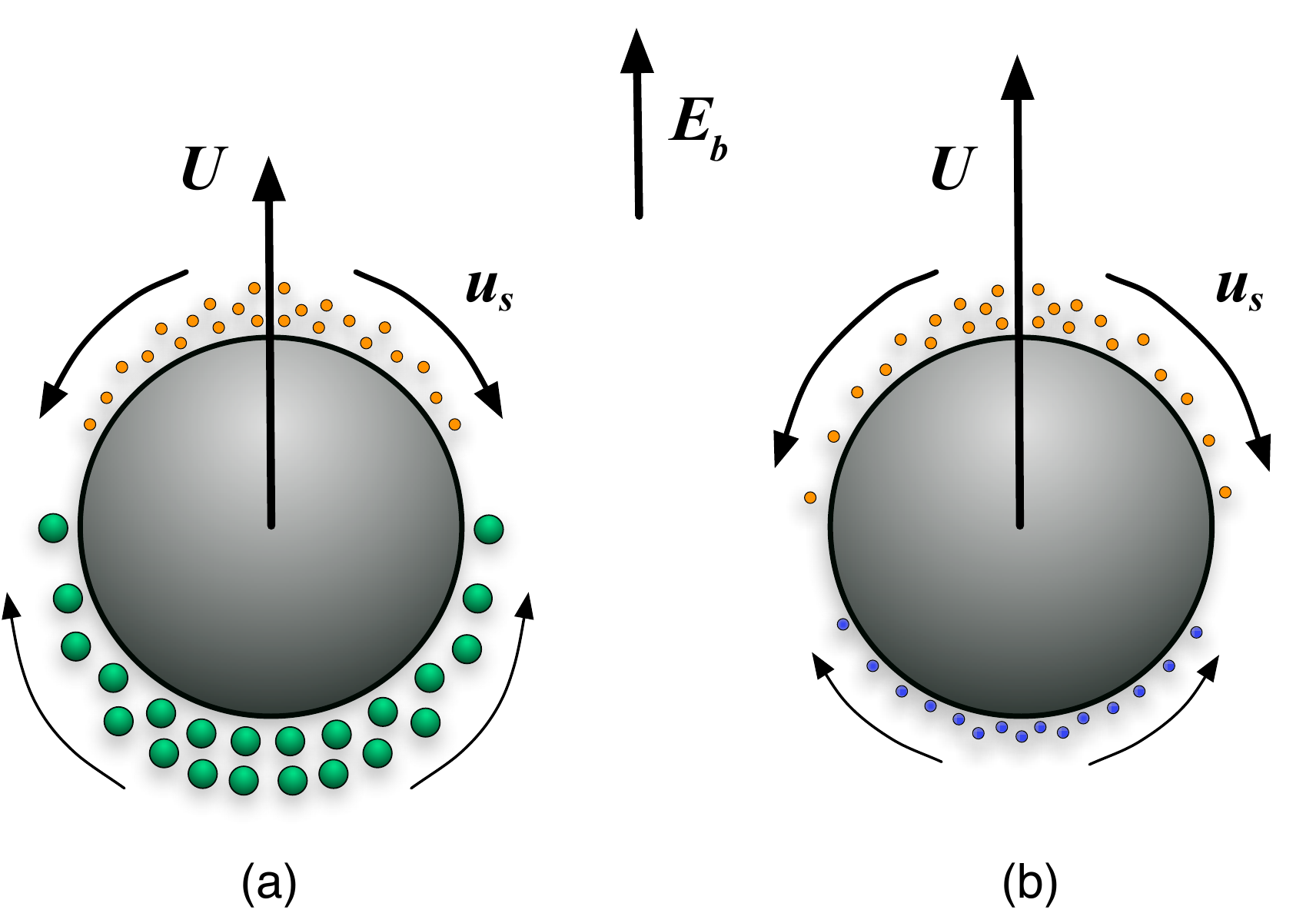}
\caption{ \label{fig:sphere} Mechanisms for DC electrophoretic motion
  $U$ of an uncharged metallic sphere in an asymmetric electrolyte
  (\ref{eq:b0}) due to saturated ICEO flow (\ref{eq:zeta_lim}) in a
  large field $E_b \gg kT/eR$. (a) Larger cations (below) in a $z:z$
  electrolyte ($a_+>a_-$, $z_+=|z_-|$) pack at lower density than
  smaller anions (above) and thus cover more of the sphere, but
  produce less slip due to greater crowding. (b) Divalent cations
  (below) cover less area and also produce less slip than monovalent
  anions of the same size ($a_+=a_-$, $z_+ > |z_-|$). In both
  examples, the sphere has an {\it apparent} positive charge ($U >
  0$).  }
\end{center}
\end{figure}

Consider an ideally polarizable, uncharged sphere of radius $R$ in an
asymmetric binary electrolyte with ions of unequal effective sizes
$a_\pm$ and charges $z_\pm e$, subject to a uniform DC background
field $E_b$. The first effect to consider is the shift in the
potential $\phi_0(E_b)$ of the particle (relative to the applied
background potential) due to its asymmetric nonlinear capacitance,
since the induced charge (which must integrate to zero) is more dense
on one side than the other, as shown in Fig.~\ref{fig:sphere}. As
noted above, this already yields a nonzero mobility with the HS
formula, $b_{ep} = \varepsilon_b \phi_0 / \eta_b$.

The effect of zeta saturation (\ref{eq:zeta_lim}) provides a different
dependence on the solution composition, which dominates in large
fields, $E_b \gg kT/eR$. Since $|\phi_0(E_b)| \ll E_bR$, the change in
polarity of the induced charge occurs near the equator, around which
there is only a narrow region with $|\zeta_{eff}|<|\Psi_c^{\pm}|$. In
this limit, therefore, we approximate one hemisphere with uniform $\zeta_{eff} =
\Psi_c^+$ and the other with $\Psi_c^-$, which yields the
ion-dependent mobility
\begin{equation}
  b_{ep} \sim \frac{1}{2} \frac{\varepsilon_b}{\eta_b} \frac{kT}{e}
  \ln \left[ \frac{ (a_+^3 c_0)^{1/z_+}}{(a_-^3 c_0)^{1/|z_-|}}\right]  \label{eq:b0}
\end{equation}
The apparent zeta potential from a DC electrophoresis measurement
$\zeta_{ep} = \eta_b b_{ep} / \varepsilon$, tends to a constant of
order $\approx kT/e$, independent of $E_b$ and $R$. In a $z:z$
electrolyte, the limiting value 
\begin{equation}
\zeta_{ep} \sim \frac{3}{2} \frac{kT}{ze} \ln \frac{a_+}{a_-}
\end{equation}
is set by the
size ratio $a_+/a_-$ and does not depend on the bulk concentration $c_0$. With equal sizes
$a_+=a_-=a$,  the limiting apparent zeta potential
\begin{equation}
\zeta_{ep} \sim \frac{1}{2}\frac{kT}{\bar{z} e}\ln (a^3 c_0)
\end{equation}
is set by harmonic mean of the valences, $\bar{z}=z_+z_-/(z_+ + z_-)$, if $z_+ +
z_- \neq 0$. 

\begin{figure}
\begin{center}
\includegraphics[width=5in]{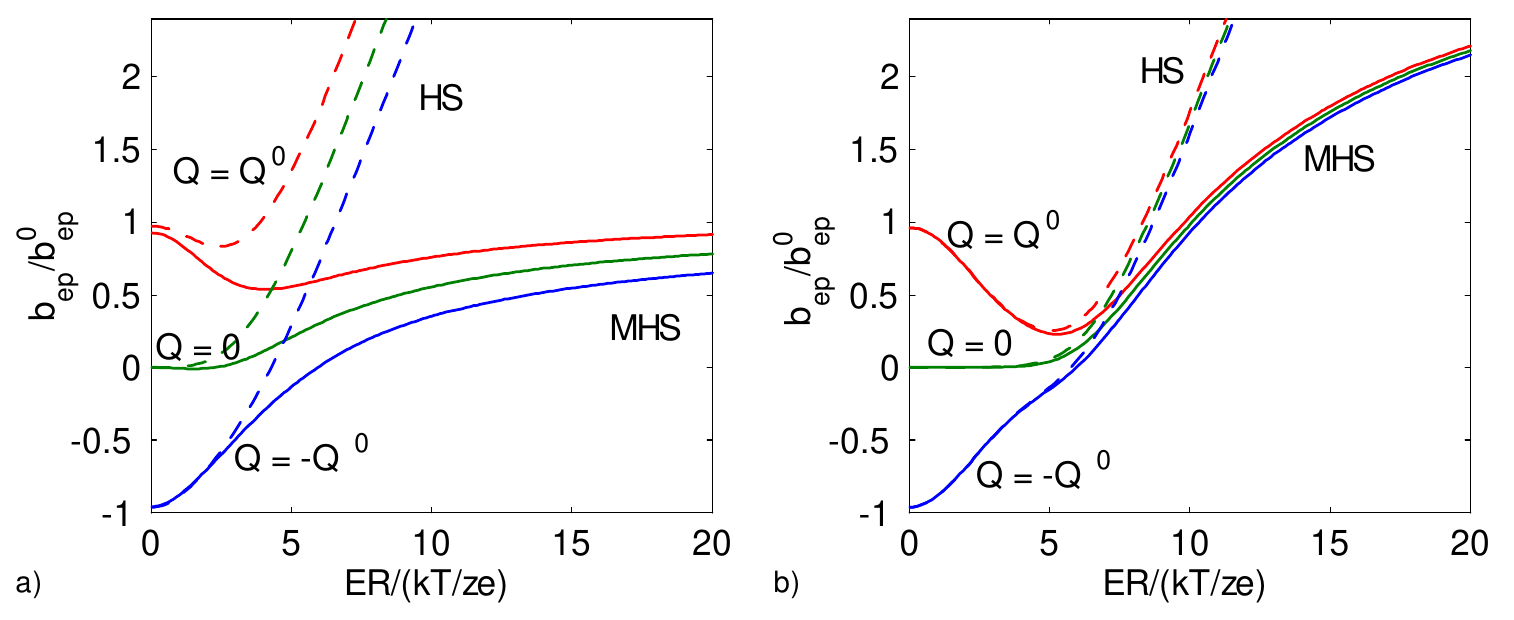}
\caption{  Electrophoretic mobility of an ideally polarizable sphere of total charge $Q$ in an asymmetric electrolyte with $(a_-/a_+)^3 = 10$ from Fig.~\ref{fig:mob_asym} with HS slip (dashed curves) compared to the same calculations redone with MHS  slip (\ref{eq:mslip}) for (a) high salt concentration $\nu_-=0.1$ and $\nu_+=0.01$ and (b) moderate salt concentration $\nu_-=10^{-3}$ and $\nu_+=10^{-4}$. Charge-induced thickening has the opposite effect of nonlinear capacitance, since it gives more weight to smaller and/or more highly charged counterions in determining the electrophoretic mobility. This significantly reduces the apparent charge of the particle at large fields and/or high salt concentrations, but not enough to change its sign. }
\label{fig:mob_asym_mhs}
\end{center}
\end{figure}

An interesting feature of this nonlinearity is that the limiting mobility is set by properties of the electrolyte and is independent of the true charge of the particle. As sketched in Fig.~\ref{fig:sphere}, the induced viscosity increase alone causes the neutral sphere to have an apparent charge whose sign is that
of the ions which condense at a lower potential (larger $z$ and/or
larger $a$). For consistency, however, we should also include the nonlinear capacitance effect, which can act in the opposite direction, making the apparent charge that of the ions which pack less densely (smaller $z$ and/or smaller $a$). As shown in Fig.~\ref{fig:mob_asym_mhs}, it turns out that the nonlinear capacitance effect is stronger and determines the sign of the apparent charge of the particle in large fields and/or high salt concentrations. Nevertheless, the charge-induced thickening effect significantly reduces the mobility in this regime and introduces a strong decay with salt concentration.

By now, it should be clear that ICEP of charged, asymmetric, polarizable particles can
have a very complicated dependence on the solution chemistry in large electric fields and/or high salt concentrations.  We must stress again that we do not include strongly nonlinear effects such as surface conduction and salt adsorption by the double layers, so the predictions of this paper only pertain to moderate voltages and thin double layers in the weakly nonlinear regime. Nevertheless, we already see some interesting new qualitative features predicted by the modified models.  Our examples also show that the electrophoretic
mobility of a homogeneous polarizable particle need not provide a
reliable measure of its total charge, contrary to common wisdom.

\subsubsection{  Concentration dependence of AC electro-osmosis  }

\begin{figure}
\begin{center}
\includegraphics[width=3.5in]{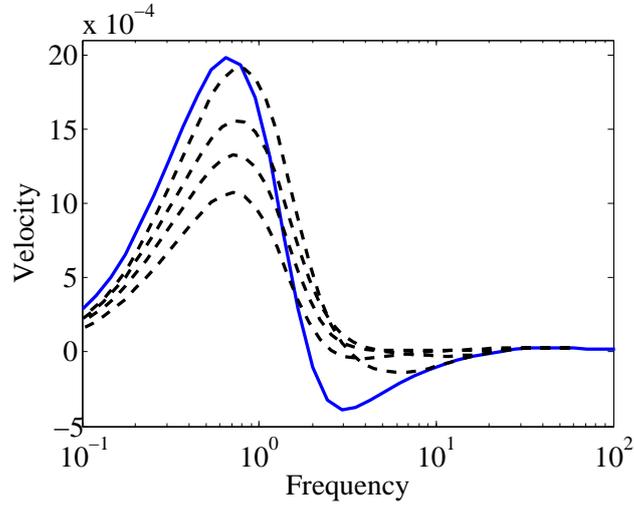}
\caption{
Frequency response of ACEO pumping with MHS slip and Bikerman's model of 
steric effects. Data are shown in dimensionless form where frequency is scaled
by the RC time of the equivalent circuit and the velocity is 
scaled by $\varepsilon V^2/\eta L_0$ \cite{olesen2006}. The values of $\nu=2 c_o a^3$ 
used in  (\ref{eq:mslip}) are $0,~ 10^{-10},~ 10^{-6},~ 10^{-4},~ \mathrm{and}~ 10^{-2}$, 
from top to bottom. 
Even at the extremely low values of $\nu$, this model of charge induced thickening
essentially removes the prediction of reverse flow.   
For the steric effects, $\nu=0.01$, in the Bikerman model was used for all cases. 
}
\label{fig:ve_aceo}
\end{center}
\end{figure}

Next we revisit the weakly nonlinear analysis of ACEO pumping by adding the effect of  charge-induced thickening via the MHS slip formula (\ref{eq:mslip}) in a symmetric electrolyte. 
In Fig. \ref{fig:ve_aceo} we show predictions of an  ACEO pump including steric effects 
(Bikerman) and the simplest MHS, Equation \ref{eq:mslip} with $a_j/a=1$. 
In interpreting these
 data, it is useful  to remember that our simple model predicts the effective
 zeta potential to be the same as the voltage across the double layer up to a 
critical voltage whereafter the effective zeta potential saturates, see Fig. \ref{fig:MHSslip}. 
At high frequency, there is insufficient
 time for the double layers to fully charge and therefore the cutoff voltage
 is not reached and the slip is uneffected. 
At low frequency and high volatge, the double layers fully charge and the 
saturated zeta-potential severly limits the flow. 
Thus the MHS model acts as a high-pass filter for  electroosmotic slip.
 This effect can be seen in Fig. \ref{fig:ve_aceo} as the ion size decreases (thus the cutoff voltage  increases). 
The upper  dashed curve, given the physical interpretation of the model would  
require an ion size  of 0.1 angstroms at 0.1 mM concentration. 
For more realistic ion sizes (at 0.1 mM and $a=3$ angstroms, $\nu=7\times 10^{-6}$) 
we find that the viscoelectric effect essentially  
eliminates  the prediction flow reversal from Bikerman's model. 
To date we have not successfully predicted flow decay with concentration 
and high frequency reversal in ACEO with a single, unified model though 
work continues in this direction.

\begin{figure}
\centering
a) \includegraphics[width=2.4in]{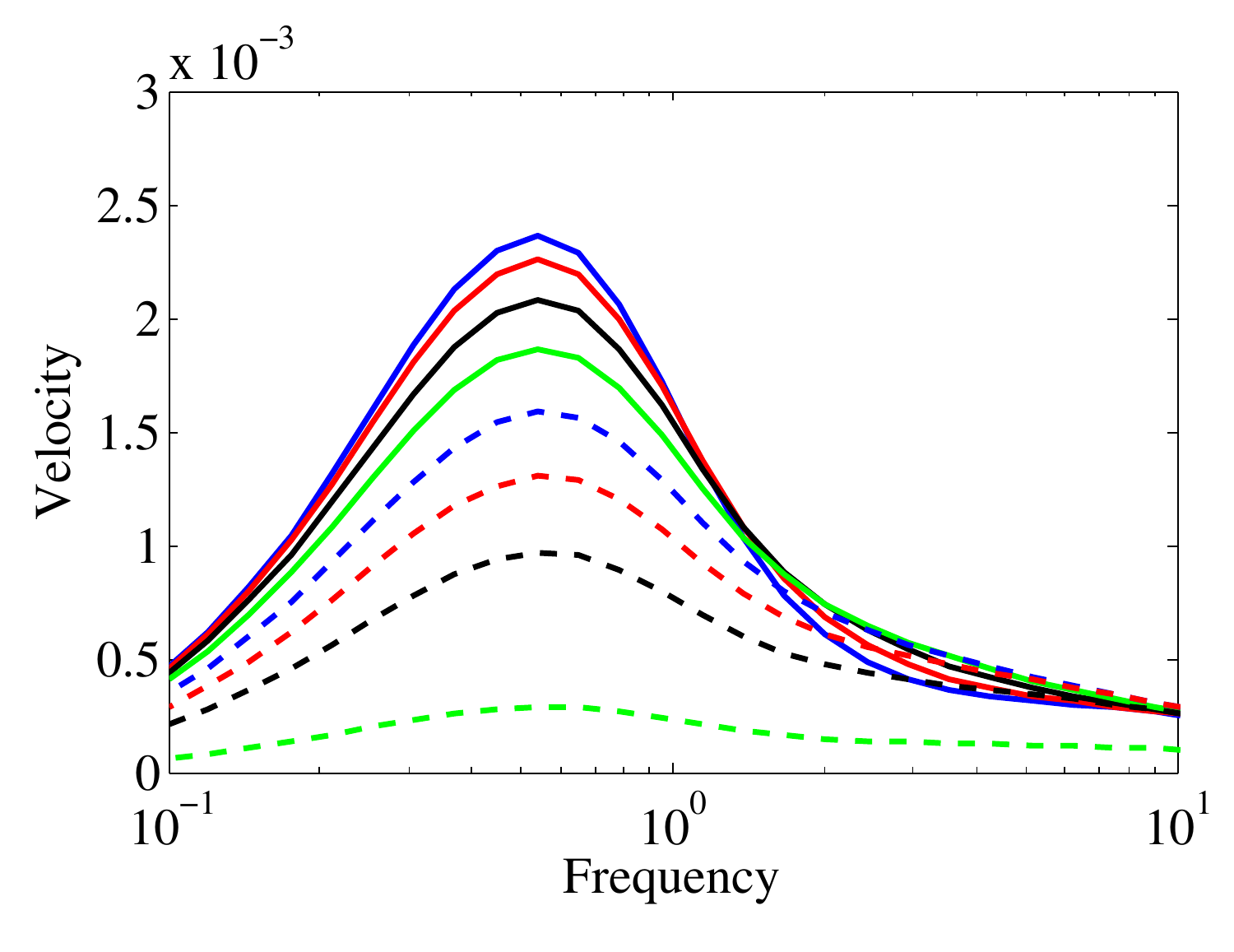}
b) \includegraphics[width=2.4in]{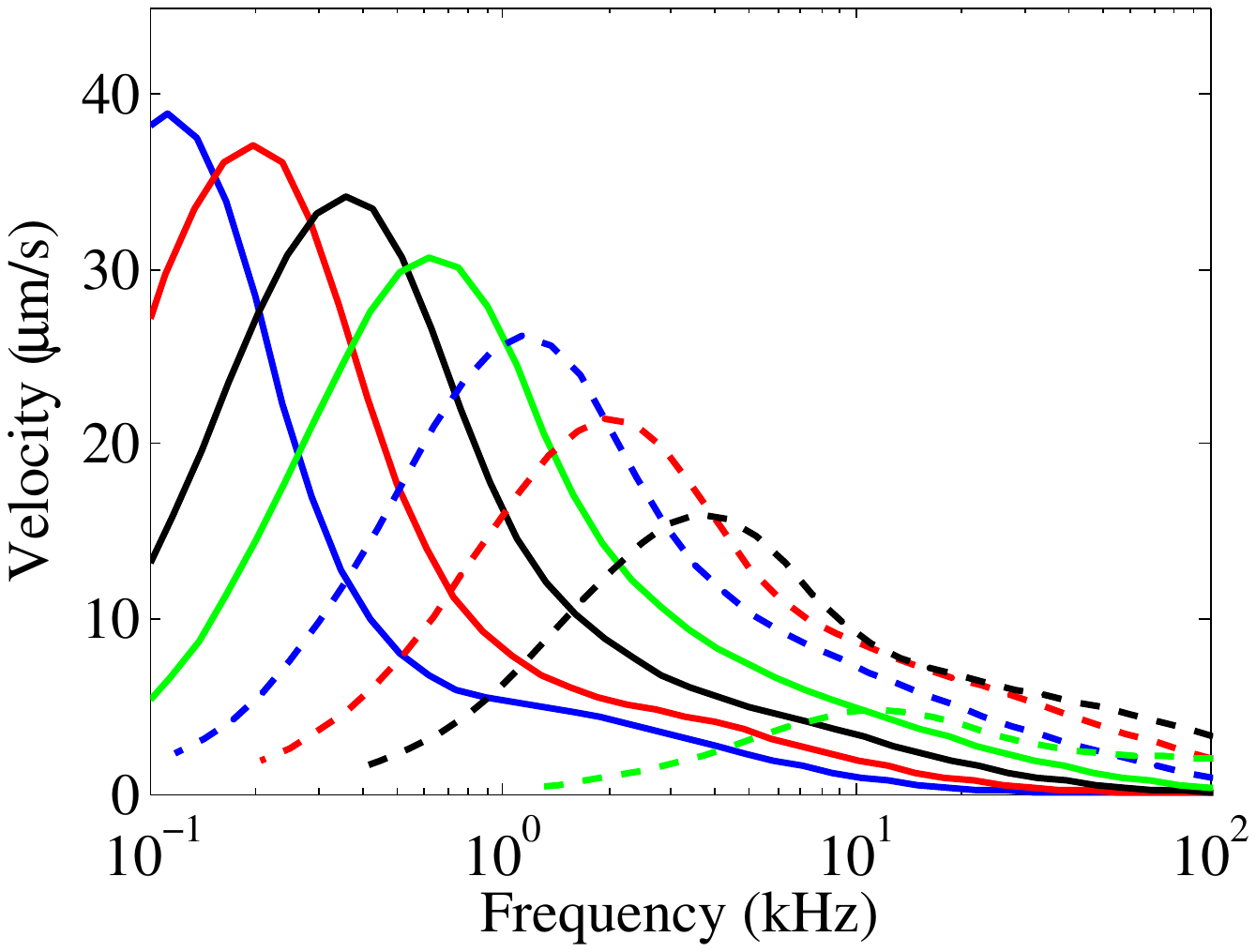}
\caption{
Predicted frequency response of the ACEO pump  of Urbanski 
\textit{ et al} ~\cite{urbanski2006} from Fig.~\ref{fig:expt}(a) with an 
MPB double-layer model also accounting for MHS slip with charge-induced thickening. 
In (a) we show the dimensionless frequency response of the pump as we change concentration 
(C = 0.001, 0.003, 0.01, 0.03, 0.1 0.3, 1, and 10 mM from top to bottom). 
In (b) the same data are plotted in dimensional form. 
The model of charge-induced thickening uses a constant ion size of 4
nm.  The response  is computed with the Bikerman model using a constant 
value of $\nu=0.01$, though no high-frequency flow reversal is predicted at this voltage.  
}
\label{fig:jp_aceo_ve}
\end{figure}

Figure \ref{fig:jp_aceo_ve} shows  results applying the MHS model, (\ref{eq:mslip}), 
to the  ACEO pumps of Urbanski
\textit{ et al} ~\cite{urbanski2006}. 
We see a decay in the maximum flow velocity with 
concentration that is reminiscent of experiments, Fig. \ref{fig:Exp_data}, when we
view the data in dimensionless form. 
Figure  \ref{fig:jp_aceo_ve} (a) 
shows the dimensionless frequency response as we change concentration as in  the experiments
using a fixed ion size of 4 nm. 
As with the predictions of flow reversal in ACEO, it seems that the ion size that best fits the experiments
is an order of magnitude larger than we would expect.
The simulated  data show the promise of a simple charge-induced thickening model
to predict decay of flow with concentration. 
It is interesting to note that at the relatively low voltage of these
 experiments (3 Vpp $\approx $ 1 Vrms) there is no flow reversal predicted 
even when the viscoelectric effect is relatively weak, contrary to the 
experimental observations. 

We show the data in dimensional form in  Figure  \ref{fig:jp_aceo_ve} (b)  for a direct comparison
to experiments. We again  see one of the key discrepancies
which occurs in all models be they based in classical electrokinetic theory or 
based on the modified models presented here. All the models 
assuming blocking electrodes predict that the features in the frequency
response are strongly concentration dependent, while the data show relative insensitivity to 
concentration. As discussed in Section \ref{sec:highfreqrev}, we hypothesize this 
discrepancy is due to neglect of Faradaic reactions though this remains an area for
future study.

These simple models are capable of reproducing at least in 
qualitative way the general experimental trends of ACEO. The Standard Model which does not
account for charge induced thickening does not predict a strong  flow decay as
concentration is increased as all ACEO  experiments have shown.  We should also emphasize that our models are surely oversimplified, and various neglected effects, such as specific solvent-mediated forces or electrostatic correlations can effectively increase the range of crowding effects (to allow the use of realistic ion sizes) and change their form in ways that may improve the ability to fit experimental data.

\section{ Mathematical modeling of electrokinetics in a concentrated solution }
\label{sec:disc}

\subsection{ Nanoscale physics }
 
The fundamental difficulty in modeling all electrokinetic phenomena is
that complex molecular-scale phenomena at the electrified interface
give rise to macroscopic fluid motion.  The modified models above
attempt to take into account some new physics -- steric effects for finite-sized ions and 
charge-dependent visco-electric effects - but these are just
simple first steps away from dilute-solution theory in
electrokinetics. In this section, we develop a general theoretical framework for modeling electrokinetic phenomena in concentrated solutions (including large applied voltages in dilute solutions). 

We begin by  discussing various neglected effects in our models above 
that might need to be incorporated into the general theory, some of which were
anticipated (qualitatively) by J. J. Bikerman decades
ago~\cite{bikerman_book_1948,bikerman_book_1970}. In some sense, what we are attempting is to model part of the
``compact layer'' in microscopic detail using the {\it same} continuum
equations that describe the ``diffuse layer'' and the bulk
electrolyte. The hypothetical partitioning of the interface between diffuse and compact parts has
become entrenched in electrokinetics and had many successes, at least
in describing linear phenomena with non-polarizable surfaces of fixed
charge. In such cases, it usually suffices to define an effective slip
plane, which marks the sharp transition from a dilute solution to a
stagnant~\cite{kirby2004,delgado2007} or
dynamical~\cite{zukowski1986a,mangelsdorf1990,lopez2007,lopez2009} compact
layer, whose fixed position can be fit to electrokinetic
measurements. At large induced voltages or concentrated solutions, however, we believe it is necessary to
describe the nanoscale rheology of the liquid in more detail, since it is otherwise
not clear how to shift the effective shear plane with voltage or
concentration, as sketched in
Fig.~\ref{fig:slip_cartoons}.  As in prior work, it may still be useful to maintain the theoretical construct of a separate "compact layer" via effective boundary conditions on the continuum region, but crowding effects, which vary with the local electric field and ionic concentrations, should also be included in modified electrokinetic equations.

The following are some nanoscale physical effects, other than volume constraints and viscoelectric effects (in the local density approximation), that we
have neglected or included only heuristically, which may be important in nonlinear electrokinetics and other situations discussed below. In some cases, simple continuum models are available but have not yet been applied to electrokinetic phenomena as part of a coherent theoretical framework.  Setting the stage for such modeling is the goal of this section.
\begin{itemize}

\item {\it Electrostatic correlations.} To our knowledge, all mathematical models in electrokinetics are based on the mean-field approximation, where each ion only feels an electrostatic force from the mean charge density of all the other ions In reality, ions are discrete charges that exert correlated forces on each other, which become especially important with increasing valence~\cite{attard1996,vlachy1999}. The breakdown of the mean-field approximation for multivalent ions can lead to counterion condensation on the surface, effectively leading to a correlated two-dimensional liquid (or glass) resembling a Wigner crystal of like charges in the limit of strong coupling~\cite{grosberg2002,levin2002}.  

In addition to this effect, a promising direction for continuum modeling (discussed below) may be to build on MPB equation of Ref.~\cite{santangelo2006}, which describes the effective restoring force acting on an ion that tries to fluctuate from its local electrostatic equilibrium position (in a one-component plasma~\cite{levin2002}), only to be drawn back toward its correlation hole.  We conjecture that this effect may not be so important at very high charge densities in large applied voltages, where simple crowding becomes dominant and flattens out any oscillations in the charge-density profiles. Some evidence comes from recent molecular simulations of the double-layer capacitance of ion liquids~\cite{federov2008b}, which verify the square-root scaling of MPB theory with steric effects discuss above. 

On the other hand, electrostatic correlations may be crucial for the dynamics of highly charged double layers under mechanical shear stress, which to our knowledge has never been studied. We have conjectured above that the correlation hole interaction may effectively enhance the viscosity of the solution and lead to charge-induced thickening.  Electrostatic correlations may also effectively increase  the critical length for crowding effects, and we have noted elsewhere that the Bjerrum length $\ell_B = e^2/4\pi\varepsilon_b kT$ is at the same scale as the effective ion size required in our steric MPB models of ACEO, especially if corrected for reduced permittivity in large field~\cite{kilic2007a,kilic2007b,storey2008}.  The electrostatic correlation length $\lambda_c$ (defined below) is approximately $\lambda_c = z^2 \ell_B$ for a $z:z$ electrolyte, so ion-ion correlations are particularly important for  multivalent ions.  

The relative importance of corrections to mean-field theory due to electrostatic ion-ion correlations is controlled by  the dimensionless parameter,
\begin{eqnarray}
\delta_c&=&\frac{\lambda_c}{\lambda_D} \nonumber \\
&=& \frac{\ell_B z^2}{\lambda_D} =\frac{(ze)^2(2c_0)^{1/2}}{4\pi(\varepsilon_bkT)^{3/2}} 
\ \ \ \mbox{($z:z$ electrolyte)}
\end{eqnarray}
which grows with bulk salt concentration like $\sqrt{c_0}$, as the Debye screening length shrinks. The Bjerrum length in bulk water is $\ell_B \approx 7$ \AA, so we expect strong correlation effects on electrochemical transport and electrokinetics when $\lambda_D > 7 z^2$ \AA which is 7 \AA, 1.4 nm, and 2.8 nm for monovalent, divalent, and trivalent ions, respectively. The condition of ``intermediate coupling" $\delta_c=O(1)$ is met in many concentrated aqueous solutions, so it may be necessary to include correlation effects in theories of nonlinear electrokinetics and double-layer charging dynamics. Below, we discuss a possible modification of Poisson's equation~\cite{santangelo2006}, which could provide a starting point.

\item {\it Specific ion-ion interactions. } In addition to entropic effects such as hard-sphere repulsion and long-range electrostatic forces, ions can interact via more complicated short-range forces, due to direct molecular interactions or solvent-mediated effective forces. DiCaprio et al.~\cite{dicaprio2004} recently expressed the excess free energy density of an electrolyte as the sum of a hard-sphere entropic contribution (e.g. the CS model above, or its generalization to polydisperse mixtures) plus the first,  quadratic terms in a Taylor expansion in the ionic concentrations. In our theoretical framework, the latter corresponds to an additional linear, enthalpic contribution to the excess chemical potential
\begin{equation}
\mu_i^{ex} = kT \sum_j a_{ij} c_j \ \   \mbox{ (specific ion-ion 
  interactions)}
\label{eq:dicaprio}
\end{equation}
where the coupling coefficients $a_{ij}$ (with units of volume) are assumed to be concentration independent.  These terms effectively "soften" the hard-sphere interactions by introducing further short-range forces. For example, if $a_{ii}  > 0$ (or $< 0$), then like-charged ions experience an additional repulsion (or attraction) within a volume $a_{ii}$, and this increases (or decreases) the effective hard-sphere radius for crowded counterions. (More complicated concentration-dependent enthalpic terms have also been postulated for ions intercalated in crystalline solids in rechargeable battery electrodes~\cite{singh2008}. ) Specific interactions may also contribute to dynamical friction coefficients between different ionic species, e.g. leading to off-diagonal elements in the mobility tensor (see below), which could be important in large ac fields~\cite{olesen2009} where oppositely charged ions must quickly pass each other upon polarization reversal.

\item {\it Ion-surface correlations.} Just as ions in the liquid phase can have short-range interactions with each other, beyond the mean-field electrostatic force and hard-sphere repulsion,  so too can they have specific interactions with molecules comprising a phase boundary, such as a solid wall. The simplest of these result from hard-sphere ordering and solvent-mediated forces which produce correlations near the flat wall. Molecular
  simulations taking into account the solvent (beyond the primitive
  model) reveal strong density oscillations near a
  surface~\cite{marcelja2000,freund2002,qiao2003}, and we have already discussed how non-local WDA theories can capture hard-sphere aspects~\cite{tarazona1985,tarazona1987,zhang1992,tang1992a,zhang1993,tang1994,antypov2005,levin2002}. 
 Solvent-induced layering can also be described by local continuum MPB models, by
  simply adding an effective external potential $V_w(x)$ to the excess
  chemical potential~\cite{qiao2003,qiao2004,joly2004},
\begin{equation}
  \mu_i^{ex} = V_w(x) = -kT \ln \left(\frac{\rho_s(x)}{\rho_b}\right) 
\label{eq:wall}
\end{equation}
which can be expressed in terms of the statistical density profile
(pair correlation function) $\rho_s(x)$ of the solvent molecules near
the wall, relative to their bulk density $\rho_b$
~\cite{marcelja2000}. Contributions to the effective ion-wall potential $V_w(x)$ can also arise from other effects, such as the polarizability of ions~\cite{levin2009} and ``cavitation forces" due to disruption of hydrogen bonding networks in water~\cite{hummer1996,huang2008}; these effects have been studied for water/gas interfaces, but could also be important at solid metal surfaces. 

Stern's original model of a solvation monolayer
separating ions from the surface~\cite{stern1924} can be viewed the simplest model of type (\ref{eq:wall}) to account for excluded volume,
since he essentially postulated an infinite chemical potential
$V_w(x)$ for ions in the monolayer. Of course, it is more realistic to
allow for smooth oscillations in the ion-wall correlation function
over several molecular diameters, but at least Stern's model is easily
incorporated into a boundary condition~\cite{bazant2004,bazant2005} (see below). 

\item {\it Surface heterogeneity. } Discrete surface charges contribute to electrostatic correlations~\cite{grosberg2002}, but chemical heterogeneities can also play an important role in the structure of the double layer and electrokinetic phenomena~\cite{lyklema2005cr,duval2004}. In the context of PB theory, the
  effect of surface roughness on the differential capacitance of the
  double layer depends on the correlation length of the roughess
  relative to the Debye length~\cite{daikhin1996,daikhin1997}. With
  finite-size ions, roughness could have a much stronger effect, not
  only on capacitance, but also on slip generation. Molecular dynamics
  simulations of electro-osmosis over atomically rough surfaces have
  revealed departures from PB theory~\cite{kim2006}, but large induced
  voltages have yet to be studied. Given the arguments of the previous
  section, the charge-induced viscosity increase is likely to be
  enhanced by roughness, since the electrostatic compression of
  hydrated ions against molecular scale asperities should thicken the
  fluid and make the apparent ion size seem larger.

\item {\it Specific adsorption of ions. } Another effect we have neglected is the specific adsorption of ions, which break free of solvation and come into direct molecular contact with a surface, as already included by Stern~\cite{stern1924} in his model for the isolated equilibrium diffuse layer based on a Langmuir adsorption isotherm. Much more refined surface adsorption models have been developed since then, especially to describe the colloid chemistry of oxidic materials~\cite{hiemstra1989}. Because the specific adsorbed of ions, and thus the surface charge, is a direct function of the ion concentration within the diffuse layer directly next to the interface, the close approach of two colloidal particles and the overlap of their respective diffuse layers will lead to surface charge modulation, so-called ``charge regulation"~\cite{biesheuvel2004b,lyklema2005cr}. For identical interfaces approaching one another, the surface charge will be reduced, and in the limit of touching Stern (adsorption) planes, the surface charge will become zero. Charge regulation of different interfaces (hetero-interaction) leads to the sum of the two interfacial charge densities to approach zero when the interaction distance is reduced, an effect which can be so strong that for amphoteric materials (i.e., those that can be both positive and negatively charged) it will lead to the inversion of the surface charge on one of the interfaces. Simultaneously, the force-distance curve can be highly non-monotonic with for instance repulsion at contact and at sufficient separation, but with an electrostatic attraction at intermediate separation~\cite{biesheuvel2004b}. 

Ion adsorption effects are begining to be considered in electrokinetics and may be particularly important in nonlinear electrokinetics, due to the large driving force for desolvation in large applied voltages. Effects of specific ion adsorption on the equilibrium surface charge of metal electrodes as function of surface potential~\cite{duval2001} and on the  electro-osmotic flow in a microchannel between electrodes have been considered by Duval~\cite{duval2004b}. Adsorption is an important mechanism of ion specificity of electrokinetics with hydrophobic surfaces~\cite{huang2007,huang2008}. In the present context of induced-charge electrokinetics,  we have already highlighted the recent work of Suh and Kang~\cite{suh2008,suh2009} incorporating surface adsorption of ions in models of ACEO flow. 

In electrochemistry, the effect of specific ion adsorption is widely invoked to explain the increase of differential capacitance of the double layer at high voltage observed in many experimental situations, since the distance between plates of the equivalent capacitor effectively shrinks from the ``outer Helmholtz plane" to the ``inner Helmholtz plane" (Fig.~\ref{fig:IHP}), and a number phenomenological models are available~\cite{bockris_book,damaskin1995}.  Some results combining specific adsorption with MPB models of the diffuse layer are in Ref.~\cite{kilic_thesis}. Of course, at very large voltages with steric effects included, the differential capacitance at  blocking electrodes must eventually decrease, once the IHP and OHP both become saturated with ions, as we have argued above. This regime of universal square-root decay is often inaccessible at low frequency due to Faradaic reactions, but our modeling of ACEO flow reversal above suggests that it can be probed at high frequency.

\begin{figure*}
\begin{center}
\includegraphics[width=3.5in]{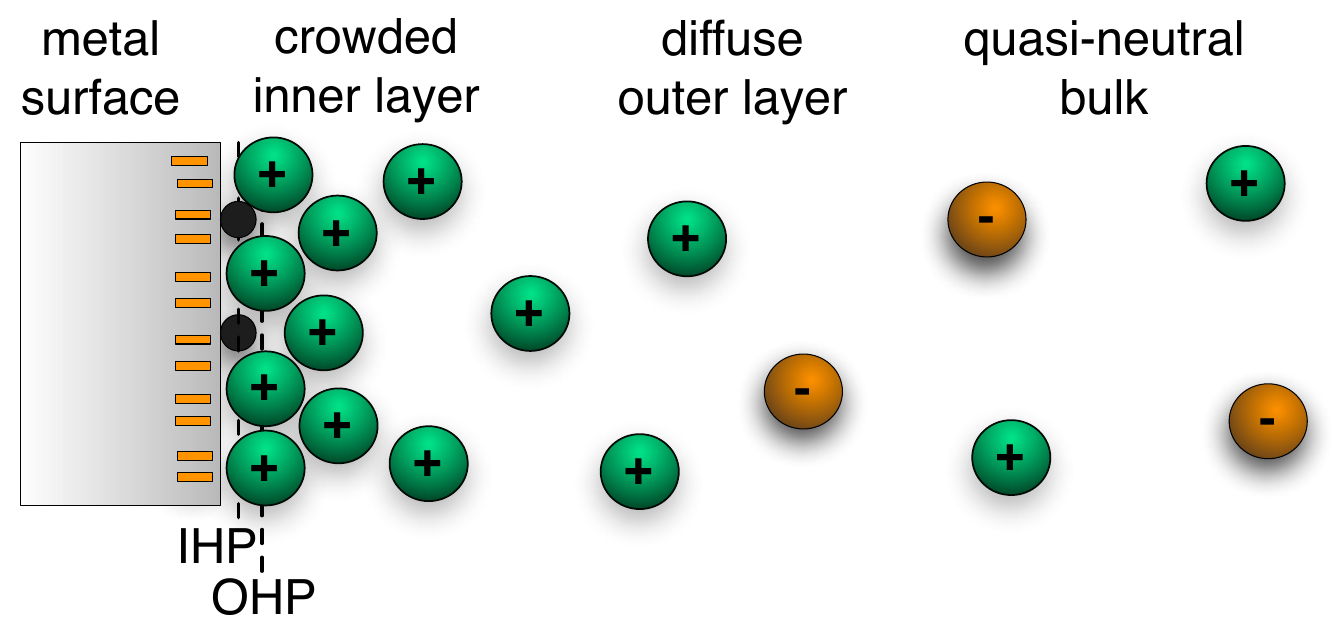}
\caption{\label{fig:IHP} 
Sketch of the double layer near a blocking electrode at high voltage. Solvated counterions (green) are crowded in the inner region and smoothly transition across the outer diffuse region to a dilute solution with solvated anions (orange). An ion can break free from its solvation shell and adsorb on the surface (black), thereby moving from the outer Helmholtz plane (OHP) to the inner Helmholtz plane (IHP) 
}
\end{center}
\end{figure*}

\item {\it Normal current and Faradaic reactions. } Faradaic electron-transfer reactions can affect local surface potentials and ion concentrations, and thus also electro-osmotic flows over electrodes. This effect has recently been studied by Duval et al.~\cite{duval2004b,duval2003faradaic_theory,duval2003faradaic_expt,duval2006faradaic}  in the context of nonlinear streaming potentials over electrodes in the presence of a redox couple in solution.  Reversible redox couples, in particular, can greatly reduce streaming potentials over electrodes due to conduction that results from bipolar electrolysis~\cite{duval2003faradaic_expt}, and this effect can be isolated using an indifferent supporting electrolyte~\cite{duval2006faradaic}. These studies are based on the standard Butler-Volmer equation, with reactions driven by the full double layer voltage (rather than just the compact part~\cite{biesheuvel2009galvanic}) and a Frumkin correction based on dilute-solution theory, so some modifications may be required for large applied voltages (see below). 

Most models for induced-charge electrokinetics assume blocking surfaces in order to focus on capacitive double-layer charging and simple ICEO flow. As noted in section 2, however, there is growing evidence that Faradaic  reactions play a major role, especially at low frequency and high voltage (as in Fig. \ref{fig:ACEOexpt}). This leads to normal currents, which can perturb the equilibrium structure of the double layer.  Lacoste et al.~\cite{lacoste2009} have recently noted that reverse ICEO flows can arise even at low voltages in a mean-field theory of biological membranes passing a normal ionic current, in the Helmholtz limit of a thick dielectric layer ($\delta\to\infty$). Since this limit corresponds to a large ``correction factor" in the Standard Model as inferred from most experiments (see section 2), it may be that non-equilibrium double-layer structure in the presence of Faradaic reactions is involved in the low-frequency, high-voltage flow reversal in ACEO and TWEO, in addition to effects of diffusion-layer electroconvection~\cite{gonzalez2008}. Below, we note that the mathematical description of Faradaic reactions may also need to be modified for large voltages or concentrated solutions. 

\end{itemize}
In the next section, we summarize a general mathematical framework for the dynamics of electrolytes and ionic liquids, which naturally allows the incorporation of some of these effects.

\subsection{ Modified electrokinetic equations }
\label{sec:general}

\subsubsection{ Continuum modeling approaches }

Until now we have focused on situations with thin double layers and
integrated over the double-layer structure to obtain modified
effective boundary conditions on the quasi-neutral bulk fluid, namely
the MPB differential capacitance in section~\ref{sec:crowding} and MHS
slip in section~\ref{sec:viscosity}. In this section, we summarize the
modified continuum electrokinetic equations corresponding to these
models, which could be applied to arbitrary geometries with thin or
thick double layers in nonlinear electrokinetics. A few examples of this approach been developed for linear electrokinetics~\cite{cervera2001,cervera2003} and electrochemical dynamics~\cite{kilic2007b,soestbergen2008}. Such modified continuum equations aim to capture more of the essential physics of nanoscale electrokinetics, 
while remaining much simpler and widely applicable than brute-force molecular
dynamics~\cite{lyklema1998,freund2002,qiao2003,thompson2003,joly2004,lorenz2008} or more complex    
statistical
approaches~\cite{gillespie2002,roth2005,liu2008}, and could have broad applicability beyond the problems considered here. 

We have made the case for modified electrokinetic equations based on experimental and theoretical arguments in nonlinear electrokinetics, but similar issues also arise in other fields. We have already noted that Bikerman's model has recently been adapted to model the double layer in ionic liquids~\cite{kornyshev2007,federov2008,federov2008b}, but mainly to predict the differential capacitance for use in RC circuit models. 
 We are not aware of any attempt to describe electrokinetics or non-equiilibrium dynamics of ionic liquids, so our general approach below may have some relevance, in the limit of very high salt concentration, approaching a molten salt. Steric effects in polyelectrolytes have also been described by Bikerman's model~\cite{gonzalez2004} as well as the CS and BMCSL hard-sphere models~\cite{devos2008,biesheuvel2008poly}, and this is another interesting area to consider electrokinetics with crowding effects.

Even in the more familiar context of electrolytes, there has been a recent explosion of interest in  nanofluidics~\cite{eijkel2005,schoch2008}, since the high surface-to-volume ratio of nanochannels amplifies the importance of transport phenomena occurring in confined geometries, effectively inside double layer. In recent years, the classical electrokinetic equations for a dilute solution have been used extensively (and exclusively) to model nanochannel phenomena such as 
ion selectivity~\cite{zheng2003,pennathur2005a,pennathur2005b,baldessari2006} and mechanical-to-electrical power conversion~\cite{morrison1965,yao2003,daiguji2004,heyden2006,heyden2007,pennathur2007}, but quantitative agreement with experiments often requires fitting "extra" compact-layer properties. Modified electrokinetic equations with additional physics might improve theoretical predictions without relying as much on adjusting boundary conditions. 

As noted in the introduction,  Liu et el.~\cite{liu2008} have recently considered correlations and crowding effects in steady nanochannel transport via a continuum MPB theory, based on more complex statistical thermodynamical models than those considered here~\cite{outhwaite1980,outhwaite1983,bhuiyan2004}. This approach incorporates more physics than we have considered above, but this comes at the expense of sacrificing some generality and   mathematical simplicity, since it assumes equilibrium charge profiles and requires numerical integration of coupled integral equations to determine the self-consistent charge density in Poisson's equation.  Nonlinear integral equations for ion profiles or correlation functions also result from other statistical approaches~\cite{hansen_book}, such as the hyper-netted chain and related models~\cite{vlachy1999,attard1996}, and statistical density functional theory~\cite{gillespie2002,gillespie2003,gillespie2005}. For hard-sphere models, perhaps the simplest theories of this type are based on the weighted-density approximation~\cite{antypov2005}, as discussed above. In spite of many successes,  however, all of these methods require significant effort to solve numerically, even in simple situations, while their underlying physics remains relatively simple (e.g. charged hard spheres). As such, it may be more fruitful to apply modern computers to molecular dynamics simulations with more realistic interatomic forces than to solve non-local continuum models numerically. 

Twenty years ago, in the context of fitting double-layer capacitance data, Macdonald concluded that ``integral-equation statistics treatments are too complicated and too limited... to be of practical usefulness", while lattice-gas models may be ``entirely adequate to describe the diffuse layer" ~\cite{macdonald1987}. Although computational advances have made integral-equation approaches more feasible today, we believe that simple mathematical descriptions are still useful, if not necessary, to model non-equilibrium dynamical phenomena. Therefore, we focus here on local continuum models with finite-size ions, following Cervera et al.~\cite{cervera2001,cervera2003}. We will also mention a simple non-local WDA, which could serve as a first correction to the LDA in the general theory.

\subsubsection{ Electrochemical transport }

We now present a general modeling framework based on electrochemical potentials, which  applies to non-equilibrium situations and includes the simple cases considered above.  In principle, one could start with the full theory of non-equilibrium thermodynamics of multicomponent systems~\cite{degroot_book,taylor_book}, but we develop a simpler phenomenological theory for electrokinetics in an isothermal concentrated solution. Our general starting point is to postulate a MF-LDA continuum model for the electrochemical potential of an ion of species $i$ (possibly including a solvation shell, depending on the model), decomposed into ideal, electrostatic, and excess contributions as follows:
\begin{equation}
\mu_i = kT \ln c_i + z_i e \phi + \mu_i^{ex}(x,\{c_j\},\phi) \label{eq:muidef}
\end{equation}
The thermodynamic meaning of $\mu_i$ is the Gibbs free energy difference upon adding a particle of species $i$ (and replacing other particles or empty space) within a continuum element, viewed as a local open system in quasi-equilibrium with the reservoir of nearby elements. Its gradient $-\nabla\mu_i$ acts as a mean ``thermodynamic force" on each particle, driving the system toward local equilibrium. In (\ref{eq:muidef}), we have defined $\mu_i^{ex}$ as the excess chemical potential of ion $i$
relative to a dilute solution, expressed in terms of local continuum
variables, such as the position $x$
(e.g. distance from a surface), ionic concentrations $\{c_j\}$ and their gradients (to approximate non-local contributions),  electrostatic potential $\phi$ and 
field $E = -\nabla \phi$, etc.  The fundamental significance of electrochemical potentials is emphasized by Newman~\cite{newman_book}, who also questions validity of the mean electrostatic potential $\phi$ at the molecular level in a multi-component concentrated solution. Nevertheless, it is necessary to separate long-range electrostatic forces from short-range chemical interactions to develop continuum equations for electrokinetics, so we proceed with the phenomenological decomposition in (\ref{eq:muidef}). From this perspective, we can view $\mu_i^{ex} = kT \ln f_i$ as defining the chemical activity coefficient $f_i$ in terms of the mean-field approximation for $\phi$.

In principle, the excess chemical potential $\mu_i^{ex}$ can be derived from microscopic statistical models or by fitting to molecular dynamics simulations or experiments.  
For example, throughout this article, we have focused on two simple MF-LDA models of $\mu_i^{ex}$ for steric effects of excluded volume for solvated ions, namely Bikerman's lattice-based
model (\ref{eq:bike}) and the Carnahan-Starling hard-sphere-liquid
model (\ref{eq:CS}). Above in this section, we noted some other possible contributions. 
Simple MF-LDA models are also available for
specific solvent-mediated ion-wall interactions (\ref{eq:wall}) and
ion-ion interactions (\ref{eq:dicaprio}). 

More complicated effects of statistical correlations can also be included in $\mu_i^{ex}$ by making it a non-local functional of the ion density-profiles. The simplest prescription of this type is the weighted density approximation (WDA), where $\mu^{ex}$ for a bulk homogeneous liquid is evaluated with local reference concentrations $\bar{c}_i$ in place of $c_i$, which are obtained by non-local averaging of the nearby inhomogeneous concentrations,
\begin{equation}
\bar{c}_i({\bf r}) = \int w(|{\bf r}-{\bf r}'|)c_i({\bf r}')d{\bf r}'.
\end{equation}
Several methods to construct the weight function $w(r)$ are available for general liquids~\cite{tarazona1985,tarazona1987,curtin1985,groot1987,levin2002} and have been applied in equilibrium MF-WDA electrolyte simulations~\cite{tang1992a,zhang1992,zhang1993,tang1994,antypov2005}. For hard-sphere liquids, the simplest and most natural choice for $w(r)$ is  the hard-sphere indicator function, $w=3/(4\pi a_i^3)$ for $r<a_i$ and $w=0$ for $r>a_i$, which turns out to be quite successful, so it would be interesting to try to include this particular WDA in electrokinetic models, as a first correction to the LDA. This WDA leads to realistic density oscillations near a surface without the need to fit an empirical wall potential (\ref{eq:wall}) in the LDA. On the other hand, LDA models are much simpler to implement numerically and allow analytical progress, while any non-local continuum model may be intractable for dynamical problems or complicated geometries.

In non-equilibrium thermodynamics~\cite{degroot_book,taylor_book,newman_book}, the mass flux densities $\Fb_i$ are obtained from the phenomenological hypothesis of linear response:
\begin{equation}
\Fb_i = c_i \ub  - \sum_j L_{ij} c_j \nabla \mu_j   \label{eq:Lij}
\end{equation}
where $\fb_j = - c_j \nabla \mu_j$ is the  thermodynamic force density (force/volume) 
acting on species $j$ and  $L_{ij}$ is the (symmetric, positive definite) Onsager mobility tensor converting these forces into mean drift velocities in the frame moving with the mass-averaged velocity $\ub$. The mobility tensor is related to the diffusivity tensor by the Einstein relation, $D_{ij} = L_{ij} kT$, and is usually assumed to be  
be diagonal, $D_{ij} = D_i \delta_{ij}$, but this can only be justified for a dilute solution. In a highly concentrated solution, the mobility tensor (or its inverse, the friction tensor) may have significant off-diagonal elements~\cite{degroot_book,taylor_book}.  

The Onsager mobility in (\ref{eq:Lij}) is generally take to be a constant (i.e. linear response), but in highly concentrated and dissipative liquids, there  may be nonlinear concentration dependence. Various models for glassy relaxation~\cite{kob1993,gotze1992} and granular flow~\cite{levin2001} have postulated a power-law decay of the mobility, 
\begin{equation}
L_{ij} = L_{ij}^0 \left( 1 - \frac{\Phi}{\Phi_c} \right)^p
\end{equation}
as the particle volume fraction $\Phi$ approaches the jamming or glass transition $\Phi_c$, respectively.  The crowding and compression of counterions against a highly charged surface by a large, time-dependent normal electric field may cause a similar, temporary decrease in mobility very close to the surface. Such a nonlinear effect on ion transport would be consistent with the charge-induced viscosity increase proposed above.

With the fluxes defined by (\ref{eq:Lij}), the differential form of mass conservation is 
\begin{equation}
\frac{\partial c_i}{\partial t} + \nabla\cdot\Fb_i = r_i \label{eq:ioneqs}
\end{equation} 
where $r_i$ is the reaction rate density for
production (or removal) of ion $i$, which is usually (but not
always~\cite{macdonald1954_b}) set to zero for electrolytes. With these further assumptions, the modified Nernst-Planck equations take the general form~\cite{kilic2007b},
\begin{equation}
\frac{\partial c_i}{\partial t} + \nabla\cdot(c_i \ub) = \nabla \cdot \left\{ D_i \left[  \nabla c_i +  c_i \nabla 
\left( \frac{z_i e \phi + \mu_i^{ex} }{kT} \right) \right] \right\}
\end{equation} 

\subsubsection{ Electrostatics }

The system of modified PNP equations~\cite{kilic2007b} is usually closed by making the mean-field approximation, in which the electrostatic potential self-consistently satisfies Poisson's equation
\begin{equation}
  -\nabla\cdot(\varepsilon \nabla \phi) = \rho = \sum_i z_i e c_i \label{eq:poiseq}
\end{equation}
where $\rho$ is the mean charge density.  The permittivity
$\varepsilon$ of a polar solvent like water is usually taken as a constant  in (\ref{eq:poiseq}), but numerous models exist for field-dependent permittivity $\varepsilon(|\nabla\phi|)$, such as (\ref{eq:sat}) or (\ref{eq:epsE}), discussed in section~\ref{sec:dielectric}. The classical effect of dielectric saturation reduces the permittivity at large fields due to the alignment of solvent dipoles~\cite{bockris_book,grahame1950,macdonald1962,macdonald1987}, although an increase in dipole density near a surface may have the opposite effect~\cite{abrashkin2007}. The permittivity can  also vary with temperature,  due to Joule heating or reactions, but here we only consider isothermal systems.  

We are not aware of any attempts to go beyond the mean-field approximation (\ref{eq:poiseq}) in dynamical problems of ion transport or electrokinetics. This would seem to require a simple continuum treatment of correlation effects, ideally leading to a general modification of (\ref{eq:poiseq}). Recently, Santangelo~\cite{santangelo2006} derived a simple modified PB equation accounting for ion-ion electrostatic correlations in a one-component plasma~\cite{levin2002} near a charged wall in the relevant regime of ``intermediate coupling", $\delta_c=\lambda_c/\lambda_D=O(1)$, which suggests modifying Poisson's equation with an additional term,
\begin{equation}
\left( \lambda_c^2 \nabla^2 - 1\right) \nabla\cdot(\varepsilon\nabla \phi) = \rho
\label{eq:mpois}
\end{equation}
where $\lambda_c$ is the electrostatic ion-ion correlation length, set by the balance of thermal energy and Coulomb energy in the dielectric medium. Physically, the extra term roughly accounts for interactions between an ion and its correlation hole during thermal fluctuations. The higher derivative of the correction term  introduces the possibility of oscillations in the ion densities at the scale of the correlation length. (It also requires additional boundary conditions, discussed below.)  The relative importance of the correction term in (\ref{eq:mpois}) is measured by the dimensionless parameter $\delta_c=\lambda_c/\lambda_D$ introduced above, which takes the form $\delta_c=z^2\ell_B/\lambda_D$ for a $z:z$ electrolyte. Since $\delta_c=O(1)$ for concentrated aqueous solutions (and increases if local permittivity decreases), correlation effects could be important in electrokinetics, and Santangelo's equation may provide a useful starting point for analysis. For ionic liquids, correlation effects are even more important, since the diffuse layer shrinks to the molecular scale ($\delta_c \gg 1$). 
 
\subsubsection{ Electrochemical hydrodynamics }

To determine the mass-averaged solution velocity, $\ub$, we enforce the conservation of linear momentum 
\begin{equation}
  \rho_m \frac{\partial \ub}{\partial t}  + \nabla\cdot\Tb = \fb 
\label{eq:momeq}
\end{equation}
where $\rho_m$ is the total solution mass density (which can be set to its bulk value in most cases), $\Tb$ is the hydrodynamic stress tensor (arising from mechanical friction), and $\fb = \sum_i \fb_i$ is the  thermodynamic force density (acting on all the ions, independent of fluid flow). The nonlinear inertial convection term $\ub\cdot\nabla\ub$ could be added to (\ref{eq:momeq}), but it is typically negligible in nonlinear electrokinetics due to a very small Reynolds number. For the stress tensor, the  first approximation is the Newtonian form, $\Tb = p \Ib -  \Tb^{(v)}$ with contributions from the dynamic pressure $p$ (to satisfy incompressibility, $\nabla\cdot\ub=0$) and the viscous stress tensor,
\begin{equation}
T_{ij}^{(v)} =  \eta \left( \frac{\partial u_i}{\partial
      x_j} +  \frac{\partial u_j}{\partial x_i} \right). \label{eq:Tv}
\end{equation}      
As explained in section~\ref{sec:viscosity}, there may be significant
local changes in viscosity near a highly charged surface. For example,
the following empirical form combines the viscoelectric effect
(\ref{eq:LO}) and charge-induced thickening (\ref{eq:visc}),
\begin{equation}
  \eta = \eta_b \left[ 1 -
    \left(\frac{\rho}{\rho_j^\pm}\right)^2 \right]^{-\beta}
  \left( 1 + f E^2 \right)^\gamma .    \label{eq:visc_general}
\end{equation}
For simplicity, one could typically set $\gamma=0$, since $E$ tends to
grow with $\rho$ in a similar way, via the MPB equations, as noted
above. Alternatively, one could set $\beta=0$, but that removes any
explicit dependence on the effective ion sizes or other modified
interactions.

The thermodynamic force $\fb$, which acts as a source of momentum in (\ref{eq:momeq}), 
can be simplified, if we assume small departures from local thermal  equilibrium. We also neglect heat transfer and assume isothermal conditions. In that case, the Gibbs free energy density, $g = \sum_i c_i \mu_i$,  varies as 
\begin{equation}
\delta g = \delta p_0 + \sum_i \mu_i \, \delta c_i + \rho \, \delta \phi = \sum_i (c_i \delta \mu_i + \mu_i \, \delta c_i )
\end{equation}
where $p_0$ is the hydrostatic pressure.  Taking the variation between adjacent continuum elements, we obtain
\begin{equation}
-\fb = \sum_i c_i \nabla \mu_i = \nabla p_0 + \rho \nabla \phi  \label{eq:gd}
\end{equation}
which is a form of the Gibbs-Duhem relation~\cite{degroot_book,newman_book}, adapted for an isothermal charged system. Note that $p_0$ includes the osmotic pressure that balances concentration gradients, which takes the form, $kT \sum_i c_i$, in a dilute solution upon inserting (\ref{eq:mu_ideal}) into (\ref{eq:gd}). In a concentrated solution, the  osmotic pressure can take a more complicated, possibly non-algebraic form, but its gradient should still uphold the local  Gibbs-Duhem relation(\ref{eq:gd})  near thermal equilibrium.  

Since we assume incompressible flow, we can insert (\ref{eq:gd}) into (\ref{eq:momeq}) and absorb $p_0$ into the dynamical pressure $p$. In this way, we arrive at the familiar form of the unsteady Stokes equation,
\begin{equation}
  \rho_m \frac{\partial \ub}{\partial t}  + \nabla\cdot\Tb =  \fb_e 
\label{eq:stokes}
\end{equation}
with an electrostatic force density, $\fb_e = -\rho \nabla \phi$.  The unsteady term $\partial \ub/\partial t$ in (\ref{eq:stokes}) is often overlooked, but it can be important in nonlinear electrokinetics, e.g. for oscillating momentum boundary layers and vortex shedding in response to AC forcing~\cite{iceo2004b,olesen_thesis}.
 
The Stokes equation (\ref{eq:stokes}) is the standard  description of fluid mechanics at low Reynolds number, which is normally applied in to a dilute solution, but we see that it also holds for an isothermal, concentrated solution near equilibrium, regardless of the form of $\mu^{ex}_i$. In their theory of non-equilbrium thermodynamics, De Groot and Mazur~\cite{degroot_book} instead assert (\ref{eq:stokes}) as the fundamental expression of momentum conservation, where the "external" or "long-ranged" force $\fb_e$ acts as the source of momentum flux in a continuum element, whose internal "short-ranged" forces are described by $\Tb$. Here, we show the equivalence of starting with (\ref{eq:momeq}) based on the full thermodynamic force $\fb$ for a concentrated solution and deriving (\ref{eq:stokes}) as the quasi-equilibrium limit, where only the external (electrostatic) force $\fb_e$ produces momentum. In this limit, which is consistent with the assumption of linear response (\ref{eq:Lij}) for the mass fluxes, all other the "chemical" interactions in $\fb-\fb_e$ only contribute to the (osmotic) pressure.

Non-equilibrium thermodynamics can be extended to account for the electrical polarizability of a  concentrated solution~\cite{degroot_book}. For a linear dielectric medium with variable permittivity $\varepsilon$, the electrostatic force density can be expressed in the familiar form, $\fb_e = - \nabla\cdot\Tb^{(e)}$, where  
\begin{equation}
 T_{ij}^{(e)} = \varepsilon
  \left( E_i E_j - \frac{1}{2} |E|^2 \delta_{ij} \right)  
\end{equation}
is the Maxwell stress tensor~\cite{jackson_book}. As noted above, in polar solvents, the permittivity should generally decrease in large fields. Various phenomenological models for $\varepsilon(E)$ can be incorporated into the theory of electrokinetics for concentrated solutions, but they complicate analysis and can introduce seemingly unphysical oscillations or singularities in the concentration profiles~\cite{kilic_thesis} and are perhaps best avoided, or included only heuristically in the boundary conditions.

\subsection{ Modified boundary conditions }

\subsubsection{ Electrostatic boundary conditions }

For Poisson's equation in the mean-field approximation (\ref{eq:poiseq}), the
electrostatic boundary conditions at a dielectric surface require
continuity of the tangential electric field $\Eb_t$ and equate the jump in
normal electric displacement $\varepsilon \Eb$ across the interface to
the free charge $q_S$ (which is related to the equilibrium zeta
potential)~\cite{jackson_book}. For a low-dielectric surface with a fixed surface charge density, the internal electric field can often be neglected, yielding the
standard boundary condition,
\begin{equation}
\varepsilon\, \nhat \cdot \nabla \phi= q_S. \label{eq:insbc}
\end{equation} 
Alternatively, for a metal surface, one can simply fix 
the potential $\phi = \phi_0$ or allow for a thin dielectric layer (or compact Stern layer) on
the surface through the mixed boundary condition~\cite{bazant2005,yossifon2006,yossifon2007,lacoste2009},
\begin{equation}
\Delta\phi_S = \phi - \phi_0 = \lambda_S \nhat\cdot\nabla \phi - \frac{q_S}{C_S}, \label{eq:sbc}
\end{equation}
where $\lambda_S=\varepsilon h_S / \varepsilon_S$ is an effective
thickness of the layer, equal to the true thickness $h_S$ by the ratio
of permittivities of the solution $\varepsilon$ and the layer
$\varepsilon_S$, and $C_S=\varepsilon_S/h_S$ is its capacitance. The boundary condition can also be generalized for voltage-dependent surface capacitance, which makes the surface-layer voltage drop $\Delta\phi_S$ a nonlinear function of the normal electric field~\cite{bazant2005}. When applying (\ref{eq:sbc}) to a metal electrode, one can set $q_S=0$ to model the Stern layer as a thin dielectric coating of solvent molecules~\cite{macdonald1962}, while specific adsorption of ions would lead to $q_S\neq 0$.

The preceding boundary conditions can be imposed on the mean-field Poisson equation (\ref{eq:poiseq}), but the modified equation for electrostatic correlations (\ref{eq:mpois}) introduces a fourth derivative term and requires one more boundary condition on each surface. Charge conservation requires the following boundary condition 
\begin{equation}
\nhat\cdot \left[ \left(\lambda_c^2\nabla^2 - 1 \right)\varepsilon \nabla\phi \right] = q_S
\end{equation}
where brackets indicate the jump across the boundary.  For consistency with the derivation of (\ref{eq:mpois}), Santangelo sets the second term (jump in mean dielectric displacement) to zero and thus equates the surface charge to the jump in the curvature of the field~\cite{santangelo2006}.  For an insulating surface of fixed charge density, this would imply replacing (\ref{eq:insbc}) with two boundary conditions
\begin{equation}
\nhat \cdot \nabla\phi = 0 \ \ \mbox{ and } \ \ 
 \nhat\cdot  \lambda_c^2\nabla^2 \left(\varepsilon \nabla\phi\right) = q_S
\end{equation}
For a metal surface with a compact Stern layer modeled as a thin dielectric coating (with a uniform electric field), we would replace (\ref{eq:sbc}) with
\begin{equation}
\phi - \phi_0 = \lambda_S \nhat\cdot\nabla \phi \ \ \mbox{ and } \ \ 
 \nhat\cdot  \lambda_c^2\nabla^2 \left(\varepsilon \nabla\phi\right) = q_S.
\end{equation}
This mathematical model provides an interesting opportunity for analysis of correlation effects in electrochemical dynamics and electrokinetics, although it is only a first approximation.

\subsubsection{ Electrochemical boundary conditions }

Standard
boundary conditions for the concentration fields equate normal ionic fluxes with surface reaction
rates 
\begin{equation}
\nhat\cdot\Fb_i=R_i(\{c_i\},\phi,\{\mu_i\},\ldots),
\end{equation}
which vanish for inert ions ($R_i=0$). The flux boundary condition can also be generalized for a dynamical Stern layer, which supports tangential ionic
fluxes~\cite{zukowski1986a,mangelsdorf1990,lopez2007} or adsorption of ions~\cite{mangelsdorf1998a,mangelsdorf1998b,suh2008,suh2009}, although some
effects of this type are already captured by the modified
electrokinetic equations, as noted above. The reaction rate $R_i$ may describe surface adsorption, in which case there is an auxiliary equation for the surface concentration, 
\begin{equation}
\frac{\partial c_s}{\partial t}  + \nabla_s\cdot {\bf F}_s = R_i,
\end{equation}
where the second term allows for surface diffusion. If the kinetics of this reaction are fast (large Damkoller number) and surface transport is slow (small Dukhin-Bikerman number) compared to bulk transport, then the reaction is in quasi-equilibrium. In that case, the surface concentration $c_s$ is given by an adsorption isotherm, which equates the surface and nearby liquid chemical potentials, $\mu_s=\mu_i$. For example, the popular Langmuir isotherm follows from a lattice-gas model of the surface adsorption sites, $\mu_s = kT \ln[ c_s / (c_{max}-c_s)]$, with dilute solution theory for $\mu_i$.  The reaction rate $R_i$ may also describe Faradaic electron-transfer reactions, such as electrodeposition (which also involves an adsorption step, and resulting motion of the metal surface) or redox reactions (which alter the charge of ions remaining in the liquid region).

The proper mathematical description of electrochemical reaction kinetics is complex and not fully understood~\cite{bockris_book}. The standard model in electrochemistry is the Butler-Volmer equation,  usually applied across the entire double layer under conditions of electroneutrality~\cite{newman_book}. Applying an analogous expression at the molecular level has better theoretical justification and introduces the "Frumkin correction" for diffuse-layer voltage variations~\cite{frumkin1933}. See Ref.~\cite{biesheuvel2009galvanic} for a recent review.  For example, for the redox reaction $R \leftrightarrow O + ne^-$, this model asserts  Arrhenius kinetics for the forward (anodic) and backward (cathodic) reaction rates,
\begin{equation}
R= k_a c_R e^{-\alpha_O n e\Delta\Phi_S/kT} - k_c c_O e^{\alpha_R ne\Delta\phi_S/kT}, \label{eq:bv}
\end{equation}
where $k_a$ and $k_c$ are rate constants for the anodic and cathodic reactions, $c_R$ and $c_O$ are concentrations of species $R$ and $O$, and $\alpha_R$ and $\alpha_O$ are transfer coefficients defined below ($\alpha_R+\alpha_O=1$) . The bias voltage $\Delta\phi_S$ can be interpreted as the Stern-layer voltage in models of the type we have considered here~\cite{chu2005,bonnefont2001,biesheuvel2009fuelcell,biesheuvel2009galvanic}. This approach has been used to model ACEO~\cite{olesen2006} and TWEO~\cite{gonzalez2008} at reacting electrode arrays in dilute solutions. It is straightforward to include nonlinear differential capacitance of the Stern layer, $C_S(\Delta\phi_S$), as well~\cite{chu2005}, but more significant modifications may be needed for concentrated solutions and large voltages. 

For consistency with our theoretical framework based on non-equilibrium thermodynamics, the reaction rate should properly be expressed in terms of the electrochemical potentials~\cite{10.95},
\begin{equation}
R= k_0 \left( e^{(\mu_R - \mu^{ex}_{TS})/kT} - e^{(\mu_O - \mu^{ex}_{TS})/kT} \right), \label{eq:react}
\end{equation}
where $\mu_R$ and $\mu_O$ are the complete electrochemical potentials of the reaction complex in the reduced and oxidized states, and $\mu^{ex}_{TS}$ is the excess electrochemical potential in the transition state, and $k_0$ is an arbitrary rate constant (which can be set by shifting $\mu^{ex}_{TS}$). The Butler-Volmer equation (\ref{eq:bv}) follows from dilute-solution theory ($\mu_R^{ex} = \mu_O^{ex}=0$) and a purely electrostatic model for the activation barrier which is a linear combination of the electrostatic energy of the reduced and oxidized states, weighted by the transfer coefficients:
\begin{equation}
\mu_{TS}^{ex} = E_a + \alpha_R q_R \phi + \alpha_O (q_O\phi - ne\phi_0) 
\end{equation}
 where $E_a$ is a composition-independent activation energy barrier, absorbed into the rate constants $k_a$ and $k_c$, and $q_R$ and $q_O=q_R+ne$ are the charges of the reduced and oxided states.  The general expression (\ref{eq:react}) can be derived from statistical transition-state theory in a hypothetical local open system, and it contains a variety of possible non-electrostatic influences on the reaction rate, via the excess contributions to the chemical potentials. This approach was recently introduced in the context of ion intercalation in rechargeable-battery materials, where the chemical potential in the electrode, and thus the reaction rate, depends on gradients in the ion concentration~\cite{singh2008}. Steric effects were also recently considered in the context of fuel-cell membranes~\cite{biesheuvel2009fuelcell}, and a form of (\ref{eq:react}) was effectively applied. In nonlinear electrokinetics, it may also be necessary to consider more general forms of the reaction rate in (\ref{eq:react}), whenever the voltage is large enough to invalidate the dilute-solution approximation close to the surface.

\subsubsection{ Hydrodynamic boundary conditions }

Until recently, almost all theoretical studies in  
electrokinetics have assumed the no-slip boundary condition for the liquid velocity, $\ub = \Ub$, where $\Ub$ is the velocity of the surface.  With the emergence of microfluidics~\cite{stone2004}, the phenomenon of hydrodynamic slip has been studied extensively in simple, Newtownian fluids~\cite{vinogradova1999,bocquet2007,lauga2005} and interpreted in terms of the Navier boundary
condition~\cite{navier1823},
\begin{equation}
\Delta \ub = \ub - \Ub = b \, \nhat \cdot \nabla \ub, \label{eq:navier}
\end{equation}
where the slip $\Delta \ub$ is
proportional to the shear strain rate via the
slip length $b$.  Flow past smooth hydrophilic surfaces has been shown
to be consistent with the no-slip hypothesis, but $b$ can reach tens
of nanometres for hydrophobic
surfaces~\cite{vinogradova2003,charlaix.e:2005,joly.l:2006} or even several microns over super-hydrophobic textured surfaces with trapped nanobubbles~\cite{vinogradova.oi:1995b,cottin_bizonne.c:2003.a,ou2005,joseph.p:2006,choi.ch:2006}. 

The study of electrokinetic phenomena in the presence of slip was perhaps first pursued by the group of N. V. Churaev~\cite{kiseleva1979,muller1986}. For electro-osmotic flow in a microchannel, Kiseleva et al.~\cite{kiseleva1979} considered the effect of exponentially varying viscosity $\eta(x)$ near a wall, increasing toward a hydrophilic surface or decreasing toward a hydrophobic surface,  and Muller et al. ~\cite{muller1986} studied the impact of the slip boundary condition (\ref{eq:navier}), which enhances the flow by a factor $(1+b/\lambda_D)$ at low voltage~\cite{bocquet2007}. This enhancement of electro-osmosis was recently analyzed and demonstrated by Joly et al.~\cite{joly2004} via molecular dynamics simulations and extended to diffusio-osmosis by Ajdari and Bocquet~\cite{ajdari2006}. (The permittivity may also vary near a hydrophobic surface, and this also can affect particle interactions~\cite{mishchuk2008hydrophobic}.) The possibility of slip-enhanced (linear) electro-osmotic flows has generated considerable excitement in nanofluidics~\cite{pennathur2007,tandon2008}, but so far it has only been analyzed with the classical electrokinetic equations and the simple, purely viscous  boundary condition (\ref{eq:navier}). 

We suggest using a modified Navier slip boundary condition~\cite{tensor}, 
\begin{equation}
\Delta \ub = \Mb \, (\Tb\cdot\nhat), \label{eq:navier2}
\end{equation}
where $\Tb\cdot\nhat$ is the total normal traction on the surface due to short-range forces (force/area) and $\Mb$ is an interfacial mobility tensor (velocity$\times$area/force), which is non-diagonal for anisotropic surfaces. For an isotropic, impermeable surface, the mobility matrix is diagonal, $\Mb = M \Ib$, with zero elements for normal flow, and then   Eq.~(\ref{eq:navier}) is recovered from (\ref{eq:navier2}) with $b= M \eta$. 

Using these models, it would be interesting to study the competition of viscoelectric effects (\ref{eq:visc_general}) and hydrodynamic slip (\ref{eq:navier}) or (\ref{eq:navier2}) in nonlinear electrokinetic phenomena at polarizable, hydrophobic surfaces in large applied voltages and/or concentrated solutions. The simple low-voltage amplification factor $(1+b/\lambda_D(c_0))$ associated with (\ref{eq:navier})  increases with concentration and becomes appreciable when the diffuse-layer thickness becomes smaller than the slip length, so hydrophobic surfaces may counteract the effect of increasing viscosity very close to the surface and allow induced-charge electro-kinetic phenomena to be observed at higher concentrations and voltages.  On the other hand, these predictions depend sensitively on the location of the slip plane where the hydrodynamic boundary condition is imposed, and how it relates to the compact-layer plane (OHP, Stern plane, reaction plane, etc.) where the electrochemical boundary conditions are imposed, especially in the concentrated-solutions models. For example, we have seen that charge-induced thickening reduces the flow in the diffuse layer, but this can be counteracted by hydrodynamic slip if the slip plane lies closer to the surface than the thickened region, due to the amplification of viscous stress on the slip  plane. We pose the effect of hydrodynamic slip in a highly charged double layer on a metal surface as an interesting open question for future work.

\subsection{ Thin double layers and diffusion layers }

The modified electrokinetic equations and boundary conditions above may be useful in modeling
nanoscale electrokinetic phenomena, e.g. taking into account steric
effects of finite ion sizes, but at larger scales, where the double
layers become thin, matched asymptotic expansions can be used to systematically
integrate out the diffuse layer and derive effective boundary
conditions on the quasineutral bulk.  First, we briefly summarize the results in the typical situation where the voltage is not strong enough to drive the double layer out of equilibrium or fully deplete the bulk salt concentration, due to diffusion limitation. In that case, the ion transport equations
(\ref{eq:ioneqs}) remain unchanged in the bulk, but Poisson's equation
(\ref{eq:poiseq}) is replaced by the condition of electroneutrality,
$\sum_i z_i e c_i = 0$.  The fluid equations are also unchanged, and
bulk viscosity variations can usually be neglected.

As noted above, chemical potentials are approximately constant (or "quasi-equilibrium" holds) in the
normal direction across a thin double layer.  For the ionic
concentrations, the boundary conditions then take the form of surface
conservation laws~\cite{chu2007},
\begin{equation}
  \frac{\partial \Gamma_i}{\partial t} + \nabla_s \cdot \Fb_i^{(s)} =
  \nhat\cdot\Fb_i - R_i,  \label{eq:surf}
\end{equation}
where $\Gamma_i(\Psi_D,\{c_i\})$ is the excess concentration of
species $i$ per unit area, $\nabla_s \cdot
\Fb_i^{(s)}(\Psi_D,\{c_i\})$ is the surface divergence of the
integrated tangential flux in the diffuse layer, $\nhat\cdot\Fb_i$ is
the normal flux from the bulk, and $R_i(\Psi_D,\{c_i\})$ is the
Faradaic reaction rate density at the surface, evaluated in terms of
the bulk variables, which thus includes the Frumkin correction. See Ref.~\cite{chu2007} for expressions for $\Gamma_i$ and $\Fb_i^{(s)}$ using
Bikerman's model, neglecting convective fluxes, and Refs.~\cite{bazant2005,bonnefont2001,richardson2007,biesheuvel2009galvanic} for expressions for $R_i$ for Faradaic reactions in a dilute solution, neglecting tangential transport. Equation
(\ref{eq:surf}) generalizes the RC boundary condition (\ref{eq:RC}).

Integrating over the diffuse-layer also yields a ``first kind''
effective slip boundary condition for the bulk fluid velocity,
\begin{equation}
\ub_s = b^{(eo)} \, \Eb_t - \sum_i b_i^{(do)} \, kT \nabla_t \ln \frac{ c_i}{c_0}, \label{eq:do}
\end{equation}
where the first term describes electro-osmosis driven by the bulk
tangential field with $b^{(eo)}(\Psi_D,\{c_i\})$ given by (\ref{eq:genslip})
and various approximations above. The second term
describes diffusio-osmosis in response to tangential bulk salt
concentration
gradients~\cite{deryaguin1961,dukhin1974,prieve1984,anderson1989,rubinstein2001}. The
diffusio-osmotic mobilities $b_i^{(do)}(\Psi_D,\{c_i\})$ can be
systematically derived from the full transport equations above in the limit of thin double layers,
following the asymptotic analysis of Prieve et al.~\cite{prieve1984}
or Rubinstein and Zaltzman~\cite{rubinstein2001,zaltzman2007}. Simple expressions can be derived for dilute solution theory, such as 
\begin{equation}
\ub_s = \frac{\varepsilon_b}{\eta_b} \left[ \zeta \Eb_t - 4 \left( \frac{kT}{ze} \right)^2 \ln \cosh \left( \frac{ze\zeta}{4kT}\right) \nabla_t \ln \frac{c}{c_0} \right]
\end{equation}
for a dilute $z:z$ electrolyte, where $c=c_+ = c_-$ is the quasi-neutral bulk salt concentration. (Equivalent forms are in Refs.~\cite{prieve1984,rubinstein2001}, and this corrects a missing factor of 4 in Eq. (111) of Ref.~\cite{khair2008}.). More cumbersome expressions, which are not expressible in terms of elementary functions, result from modified equations with volume constraints
~\cite{kilic_thesis}. Similar asymptotic methods can also express the effective slip boundary condition in terms of bulk electrochemical potential gradients, $\{\nabla_t \mu_i$\}.

In our many examples of induced-charge electrokinetic phenomena with
blocking surfaces, we have assumed ``weakly nonlinear''
dynamics~\cite{bazant2004,chu2006}, where the bulk concentration is
not significantly perturbed. Under ``strongly nonlinear'' dynamics at large voltages, even 
with blocking electrodes~\cite{bazant2004,beunis2008,verschueren2008,marescaux2009,olesen2009,suh2008,suh2009}, strong bulk concentration gradients
can develop, and the other terms in Equations (\ref{eq:surf}) and
(\ref{eq:do}) become
important~\cite{dukhin1974,rubinstein2001,zaltzman2007}.  Although steric
effects generally reduce the importance of surface conduction (smaller
Dukhin-Bikerman number) compared to dilute solution theory since
there are not nearly as many ions in the double
layer~\cite{kilic2007a}, diffusio-osmosis, concentration
polarization, and diffusion-layer electro-convection~\cite{gonzalez2008} are affected less and could be significant. In addition to Faradaic reactions, other sources of normal ion flux, such as salt adsorption by the diffuse-charge layers and Langmuir adsorption of ions on the surface, can also produce time-depedent  diffusion layers (thicker than the diffuse charge layer, but thinner than the bulk region) oscillating at twice the frequency of the AC forcing, although these layers are weakly charged and may not have a large effect on the effective fluid slip~\cite{suh2008,suh2009,olesen2009}. All of the problems we have considered above should be revisited in the strongly nonlinear regime to better understand the predictions of the modified electrokinetic equations, but this is beyond the scope of the paper.

As noted in section 2, the effects of Faradaic reactions or other mechanisms for normal ionic flux are still poorly understand in nonlinear electrokinetics, even for thin double layers. Normal currents can disturb the quasi-equilibrium structure of the double layer, even at small currents (in the Helmholtz limit $\delta\to 0$), and lead to seemingly reverse ICEO flows ~\cite{lacoste2009}. Concentration gradients can also
develop due to normal ionic fluxes under diffusion limitation.  If the bulk salt concentration approaches
zero, at a limiting current, the quasi-equilibrium diffuse layer
expands into a non-equilibrium space charge layer and drives
second-kind electro-osmotic flows~\cite{dukhin1993} and hydrodynamic
instability~\cite{rubinstein2000,zaltzman2007}. The classical
electrokinetic equations suffice in that case, since the decreasing
concentration only helps to validate dilute solution theory.  In the
case of induced-charge electrokinetic phenomena, however, the strongly
nonlinear regime is just beginning to be explored and may require
modeling with modified electrokinetic equations.

 \section{ Conclusion}

\label{sec:conc}

We have provided a critical review of recent work in nonlinear ``induced-charge'' electrokinetics, comparing theory to experiment for the first time across a wide range of phenomena to extract general trends. In doing so, we were naturally led to question the theoretical foundations of the field and develop modified equations for electrokinetics in concentrated solutions at large voltages. These equations may find applications in diverse areas of electrochemistry and fluid mechanics. A crucial aspect of this effort was to survey microscopic models of electrolytes from different fields, where convection and charge relaxation are neglected. Our experience shows the importance of integrating knowledge across scientific communities. For example,  we managed to find {\it seven} independent formulations of Bikerman's model (1942) from 1947 to 1997 and two rediscoveries of Freise's capacitance formula (1952) in 2007, each leading to separate literatures without any cross references (section ~\ref{sec:history}).

We have argued that many new nonlinear 
phenomena can arise at large induced voltages, and we focused on three that could play a major role in induced-charge electrokinetics: (i)
Crowding of counterions against a blocking surface (Fig.~\ref{fig:cartoons}) decreases the differential capacitance
(Fig. ~\ref{fig:C}), which may explain high frequency flow reversal in
ACEO pumps (Fig. ~\ref{fig:rev}) and imply ion-specific mobility of
polarizable particles in large fields (Fig. ~\ref{fig:mobility}); (ii) a charge-induced 
viscosity increase upon ion crowding (Fig.~\ref{fig:slip_cartoons}) reduces the effective 
zeta potential (Fig. ~\ref{fig:MHSslip}), which implies flow decay with
increasing concentration and an additional source of ion-specificity
(Fig. \ref{fig:sphere}); (iii) Each of these effects is enhanced by dielectric saturation of the solution in large electric fields (Fig. ~\ref{fig:Cd_epsE}). To illustrate these phenomena, we
have derived extensive analytical formulae based on simple models in the mean-field and local-density approximations (MF-LDA), including lattice-gas and hard-sphere models for steric effects of finite ion sizes,  as well as various postulates of charge-induced thickening leading to modified electro-osmotic slip formulae. 

We have also developed a  theoretical framework for electrokinetics based on non-equilibrium thermodynamics in a concentrated solution (section ~\ref{sec:disc}). Although motivated by induced-charge electrokinetics, these general equations and boundary conditions could find applications in many other areas involving nanoscale electrochemical transport.  Our examples of MF-LDA models focusing on the effects above  
should be refined and extended in future work, e.g. to account for
various solvent effects, specific adsorption of ions,
and Faradaic reactions, and we have reviewed some of the relevant literature for guidance. Especially for multivalent
ions, it may be necessary to go beyond the MF approximation to account for electrostatic
correlations, which we alter  the inner double-layer structure and could contribute to charge-induced thickening. A proper description of volume constraints may require going beyond the LDA, especially very close to a surface, to account for short-range correlations and related density oscillations. 

Legend has it that Wolfgang Pauli once said,``if God made materials, then surfaces are the work of the Devil''. At end of this study, it is tempting to draw the same conclusion, given the complexity of possible  phenomena occurring at large surface potentials and our inability to devise a single model to predict all the experimental data. We remain optimistic, however, that simple, predictive models will follow from improved nanoscale understanding of the double layer.

Our study raises an important general question for continuum modeling, ``Where should the continuum region end and give way to a boundary condition?". This question has received much attention recently in the context of hydrodynamic slip, but the situation is much more complicated for electrochemical relaxation and electrokinetic phenomena. The pioneering papers of Stern~\cite{stern1924} and Bikerman~\cite{bikerman1942} introduced two opposing, general perspectives, which provide the historical context for our work.  Stern first proposed describing the outer ``diffuse part" of the double layer with dilute solution theory, while lumping any discrepancies into an empirical boundary condition on the inner ``compact layer''. Bikerman then showed that a similar finite-size cutoff of dilute-solution theory could instead by obtained by modifying the bulk equations for a concentrated solution, without ajdusting the boundary condition. By the latter half of the century, Stern's approach became widely adopted through simple empirical models for the compact layer, and Bikerman's paper was essentially forgotten. 

We have argued that Bikerman's  perspective should be revisited, especially for extreme conditions of large applied voltage,  salt concentration, and/or frequency, since otherwise it is not clear how compact-layer boundary conditions should change to describe a dynamical region of nonlinear response near the surface. Bikerman's perspective also applies more easily to nontrivial nanoscale geometries, where the classical concepts of ``Stern plane" and ``slip plane" are less well defined.  We have shown that simple continuum equations for ion crowding, dielectric saturation, and viscoelectric response all predict that the compact layer and slip plane both effectively advance into the solution with increasing surface charge, but without the need to define these empirical concepts. On the other hand, Stern's perspective remains attractive for mathematical modeling, since it preserves simple equations for the fluid domain and lumps complicated molecular details into boundary conditions, so perhaps a combined approach is needed. Certainly, true surface effects, such as specific adsorption and Faradaic reactions, require effective boundary conditions, consistent with any modifications of the bulk equations (section~\ref{sec:general}).

In our opinion, it remains a grand challenge to describe double-layer structure, electrochemical relaxation, and ICEO flow at a polarizable surface over vast parameter ranges, from 10 mV to 10 
V in voltage, 0 to 100 kHz in frequency, and 1 $\mu$M to 1 M in ionic strength, using a simple -- but not over-simplified -- mathematical model, amenable to analytical results (in idealized limits) and numerical simulations (for typical experimental situations). The upper extremes of these conditions correspond to a new regime for the theory of
electrokinetic phenomena, where counterions become crowded during time-dependent relaxation and flow near a highly
charged surface. Nanoscale experiments and molecular dynamics simulations
will be crucial to further develop and validate various theoretical postulates. If properly validated, we believe that modified  
mathematical models can advance  
our understanding and better predict experimental observations.

\ 

This research was supported by the National Science Foundation, under
Contract DMS-0707641 (MZB). AA acknowledges the hospitality of MIT and
financial help from ANR grant Nanodrive. MZB acknowledges the
hospitality of ESPCI and support from the Paris Sciences Chair. MZB
and MSK also thank A. Ramos for helpful discussions, made
possible by the MIT-Spain program. We are also grateful to   
P. M. Biesheuvel, Y. Levin, J. R. Macdonald, and the anonymous referees for comments and important references.

\bibliographystyle{elsarticle-num}
\bibliography{elec24,slip2}

\end{document}